\documentclass[prd, 12pt,a4paper,groupedaddress,preprintnumbers,nofootinbib,floatfix]{revtex4-1}
\pdfoutput=1

\usepackage[top=2.9cm, bottom=2.1cm, left=2cm, right=2cm]{geometry}
\usepackage[english]{babel}
\usepackage{amsmath,amssymb,amsfonts, dsfont}
\usepackage{graphicx}
\usepackage{color}
\usepackage{hyperref}
\usepackage{subfigure}
\usepackage{dcolumn}
\usepackage{bm}
\usepackage{float}
\usepackage{mathtools}
\usepackage{amsthm}
\usepackage{bbm}
\usepackage{pxfonts}
\usepackage[vcentermath]{youngtab}
\usepackage[all]{xy}
\usepackage{pstricks}
\usepackage{accents}
\usepackage{cases}
\usepackage{comment}
\usepackage{placeins}
\usepackage{xspace}
\usepackage{cancel} 
\usepackage{slashed}
\usepackage{soul}
\usepackage{feynmp}
\definecolor{orange}{cmyk}{0,0.5,1,0}
\definecolor{rossoCP3}{cmyk}{0,.88,.77,.40}
\definecolor{graa}{rgb}{0.8,0.8,0.8}
\definecolor{blaa}{rgb}{0.2,0.2,0.6}
\hypersetup{
	colorlinks, 
	bookmarksopen, 
	bookmarksnumbered,
	citecolor=blaa, 		
	linkcolor=rossoCP3,	
	urlcolor=rossoCP3,			
	    }

\newcommand{\be}{\begin{equation}}
\newcommand{\ee}{\end{equation}}
\newcommand{\pd}{\overset{\text{\small$\leftrightarrow$}}{\partial^\mu}}

\newcommand{\parenbar}[2][4]{%
  \mkern#1mu
  \sbox0{$#2$}%
  \makebox[0pt][r]{\raisebox{\ht0}{$\scriptscriptstyle($}}%
  \overline{\mkern-#1mu#2\mkern-1mu}%
  \makebox[0pt][l]{\raisebox{\ht0}{$\scriptscriptstyle)$}}%
  \mkern1mu
}

\usepackage{xcolor}

\begin{document}

 
\title{\texorpdfstring{\Large\color{rossoCP3}   Theory and Phenomenology \\ of \\ The Elementary Goldstone Higgs}{}}
\author{Helene {\sc Gertov}}
\email{gertov@cp3.dias.sdu.dk} 
\author{Aurora {\sc Meroni}}
\email{meroni@cp3.dias.sdu.dk} 
\author{Emiliano {\sc Molinaro}}
\email{molinaro@cp3.dias.sdu.dk} 
\affiliation{{\color{rossoCP3} {CP}$^{ \bf 3}${-Origins}} \& the Danish Institute for Advanced Study {\color{rossoCP3}\rm{Danish IAS}},  University of Southern Denmark, Campusvej 55, DK-5230 Odense M, Denmark.}
\author{Francesco {\sc Sannino}}
\email{sannino@cp3.dias.sdu.dk}
\affiliation{{\color{rossoCP3} {CP}$^{ \bf 3}${-Origins}} \& the Danish Institute for Advanced Study {\color{rossoCP3}\rm{Danish IAS}},  University of Southern Denmark, Campusvej 55, DK-5230 Odense M, Denmark.}

\begin{abstract}
We show, via a careful analytical and numerical analysis, that a pseudo Goldstone nature of the Higgs is naturally embodied  by an elementary realization that also serves as ultraviolet completion. Renormalizability married to perturbation theory allows to precisely determine the quantum corrections of the theory while permitting to explore  the underlying parameter space. By characterising the available parameter space of the extended Higgs sector we discover that the preferred electroweak alignment angle is centred around $\theta \simeq 0.02$, corresponding to the Higgs chiral symmetry breaking scale $ f \simeq 14~$TeV.  The latter is almost 60 times higher than the Standard Model electroweak scale. However,  due to the perturbative nature of the theory, the  spectrum of the enlarged Higgs sector remains in the few TeV energy range. We also analyse precision constraints and the relevant phenomenological aspects of the theory.  
\\[.3cm]
{\footnotesize  \it Preprint: CP$^3$-Origins-2015-030 DNRF90 \& DIAS-2015-30}
\end{abstract}
\maketitle
\newpage
\tableofcontents
\newpage

\section{The Elementary Goldstone Higgs Boson}
 
The discovery of the Higgs boson \cite{Aad:2012tfa, Chatrchyan:2012ufa}   is one of the triumphs of the Standard Model (SM) of particle interactions  \cite{Agashe:2014kda}. Of course, several puzzles remain unexplained such as the nature of Dark Matter, neutrino masses and mixing as well as the cosmological matter-antimatter asymmetry of the Universe.  Solutions to these puzzles can be envisioned, that can address one or several of these problems simultaneously. 

The SM Higgs sector, by construction, identifies the electroweak (EW) scale, $v_{\rm EW}\simeq 246$~GeV, with the vacuum expectation value (vev) of the Higgs field. This scale is linked to the only dimensional parameter of the theory, that is the Higgs mass term operator. Indeed, the relevant predictions of the SM gauge sector are 
\begin{equation}\label{SMrelations1}
m_W^2 \; = \;\frac{g^2}{4} v^2_{\rm EW} = \frac{g^2}{8} \frac{m_h^2}{\lambda} \,,\quad\quad
m_Z^2 \; = \; \frac{1}{4}\left( g^2+g'^2 \right) v^2_{\rm EW} \; = \frac{m_W^2}{\cos^2\theta_W}\,,
\end{equation}
where $m_W$, $m_Z$ and $m_h$ are the physical masses of the $W$, $Z$ and Higgs boson, respectively, $\theta_W$ is the weak mixing angle  and $\lambda$ is the SM Higgs self-coupling.  Any other mass scale is then generated from $v_{\rm EW}$ up to dimensionless couplings that have to fit experiments. 

In this paper we explore and further establish a different paradigm, that is the one that allows to disentangle the vacuum expectation of the {\it elementary} Higgs sector from the EW scale \cite{Alanne:2014kea}.  
In this setup the Higgs sector symmetry is larger than the minimally required symmetry needed to break (spontaneously) the EW gauge symmetry. And the physical Higgs can emerge as a pseudo Nambu Goldstone Boson (pNGB). Once the SM gauge and fermion sectors are embedded in the larger symmetry one discovers that {\it calculable} radiative corrections induce  the proper breaking of the EW symmetry and naturally aligns the vacuum in the pNGB Higgs direction.  In this way the EW scale is only radiatively induced and, as we shall show, it is order of magnitudes smaller than the scale of the Higgs sector in isolation. 

This setup is profoundly different from the composite Goldstone Higgs scenario according to which the Higgs sector is composed of a new strongly interacting theory, 
and therefore it is not automatically UV complete~\footnote{By UV completion  we mean that the model not only constitutes the fundamental theory making the Higgs sector but can also accommodate the SM fermion masses in a simple manner. It is well known that giving masses to the SM fermions is typically much more involved in the composite pNGB construction. }. In contrast, the present realisation is, by construction, UV complete and under perturbative control.

Other attempts at constructing models where the Higgs boson is a pNGB  have appeared in the  literature. 
Examples are the little-Higgs models  (see, e.g., \cite{Schmaltz:2005ky} for a review), that  feature the Higgs as a pNGB of a spontaneously broken approximate global symmetry. 
The collective breaking ensures that no quadratically divergent contributions 
to the Higgs potential arise at one loop.  This mechanism can stabilise the model up to $\Lambda\lesssim$ 10 TeV.  Little Higgs models are, however, typically effective field theories valid up to a cutoff scale $\Lambda\sim 4\pi f$.

In this work we use as template an Higgs sector leading to the  $SU(4)\to Sp(4)$ pattern of chiral symmetry breaking  \cite{Alanne:2014kea} that was introduced for composite dynamics in \cite{Appelquist:1999dq,Duan:2000dy,Ryttov:2008xe}.   
Via an in depth analytical and numerical analysis we shall demonstrate that a pNGB nature of the Higgs is naturally embodied  within the elementary realisation.  Via this UV complete model we shall show that renormalizability alongside perturbation theory allows to precisely determine the quantum corrections of the theory. By investigating the available parameter space of the extended Higgs sector we discover that the preferred electroweak alignment angle is centred around $\theta \simeq 0.02 $, corresponding to the Higgs chiral symmetry breaking scale of $ f \simeq 14~$TeV.  This value is almost 60 times higher than the SM electroweak scale. Due, however, to the perturbative nature of the theory the  new spectrum of the enlarged Higgs sector  is in the few TeV energy range. It is important to note that it is the intrinsic structure of the quantum corrections the culprit for this interesting result.  This is very different from the composite case.  The reason being that for the composite case the final vacuum alignment is dictated by cut-off contributions that, de facto, do not align the vacuum in the pNGB direction and therefore further require the introduction of new operators rendering the pNGB nature fine-tuned \cite{Cacciapaglia:2014uja}. Furthermore in the composite case the new resonances are of the order of $4\pi f$ and therefore typically much harder to discover at present and future colliders.

The structure of the paper is the following. We review the tree-level spectrum and vacuum properties in Section~\ref{tree}, and the EW symmetry and associated radiative corrections in Section~\ref{radiative}. The quantum corrected spectrum and relevant couplings are  discussed in Section~\ref{pheno} together with their phenomenology. We go beyond the initial investigation of the phenomenological consequences of the model presented in \cite{Alanne:2014kea} and perform a more detailed study of the parameter space of the theory by scanning over the four independent parameters of the model. We use as constraints the masses of the electroweak gauge bosons, the mass of the Higgs and the requirement that we have a local minimum in the potential. In the original work \cite{Alanne:2014kea}, for simplicity, all the masses of the massive scalars were taken to be identical. This constraint has been lifted here allowing for several, even charged, scalars to be sufficiently light to be within the LHC run 2 reach. 

We also provide a thorough study of the electroweak precision measurements  to further gain insight on the viable parameter space of the theory.  
This study  allows for an indirect way to investigate deviations from the SM emerging from this theory both  
at the second three-year LHC run and  by the  next collider generation ---ILC (E$_{CM}$ $\lesssim$ 1TeV)\cite{Baer:2013cma},  CLIC (E$_{CM}$ $\lesssim$3 TeV) \cite{Linssen:2012hp}
or a large circular $e^+ e^-$ collider with E$_{CM}$  $\lesssim$  350 GeV \cite{Koratzinos:2013ncw} and/or a $pp$ collider with E$_{CM}$  $\lesssim$ 100 TeV.
The program of the envisioned future colliders, in fact, aims at a more precise determination of the Higgs couplings for which our model is the ideal {\it Guinea pig}.   

\section{The Minimal Model: $SU(4) \to Sp(4)$}
\label{tree}

As a general framework, we  identify the Elementary Goldstone Higgs (EGH) as one of the Goldstone bosons which live in the coset of the spontaneously broken global symmetry 
of the scalar sector. The latter is an enlarged symmetry group that contains the $SU_L(2)\times SU_R(2)$ (global) chiral symmetry of the SM Higgs sector. 
The possible patterns of chiral symmetry breaking, then, must  lead to the $SU_V(2)$ custodial symmetry of the tree-level Higgs potential, which guarantees (at leading order) 
the second relation in (\ref{SMrelations1}) or, equivalently, the constraint
\begin{equation}
	\rho \; \equiv \;  \frac{m_W^2}{m_Z^2\, cos^2\theta_W} \;=\;1 \,.
\end{equation}
Typically, in this class of custodially symmetric models, the Higgs boson does acquire a mass and becomes a pNGB 
due to the embedding of the Yukawa and EW gauge interactions, which explicitly break the global symmetry of the scalar sector. In this framework,
the radiative corrections responsible for the Higgs mass can be precisely computed within perturbation theory, as we will see below.

A complete classification of the patterns of chiral symmetry breaking leading to a Goldstone Higgs boson can be found , e.g., in \cite{Cacciapaglia:2014uja} where unified scenarios of composite (Goldstone) Higgs dynamics were studied. The minimal viable framework  is in this case
\begin{equation}\label{minchibreak}
	SO(6)\sim SU(4) \;\to\;  Sp(4)\sim SO(5)
\end{equation}
where the breaking of the chiral symmetry  is triggered by the vacuum expectation value of the antisymmetric  6-dimensional (pseudo-real) representation of $SO(6)\sim SU(4)$, with 5 Goldstone bosons in the coset. 
The latter can be decomposed into a $(2,\,2)+(1,\,1)$ representation of the $SO(4)$ subgroup of $SO(5)$. 
Other (non-minimal) chiral symmetry breaking patterns that allow to embed a SM Higgs doublet are \cite{Cacciapaglia:2014uja}: $i)$ $SU(5)\to SO(5)$ with 14 Goldstone bosons,
decomposed into the $(3,\,3)+(2,\,2)+(1,\,1)$ representations of $SO(4)$; $ii)$ $SU(6)\to Sp(6)$, which contains 2 Higgs doublets and  6 singlets.

In this work we focus on the EGH scenario with the simplest  pattern of chiral symmetry breaking given in eq.~(\ref{minchibreak}), where
  the EGH arises as one of the 5 Goldstone bosons belonging to the coset $SU(4)/Sp(4)$.   
  In this case, 
  the most general vacuum of the theory, $E_\theta$, can be expressed as the linear combination \cite{Alanne:2014kea}
  \be E_\theta =\cos\theta \,E_B +  \sin\theta\, E_H= -E^T_\theta\,,\label{eq:E}\ee
where $0\leq\theta\leq\pi/2$ and the two independent  vacua $E_B$ and $E_H$ are defined as
\be  E_B =
\begin{pmatrix}
i\sigma_2 & 0\\
0 & -i \sigma_2
\end{pmatrix}, \quad  E_H=
\begin{pmatrix}
0 & 1\\
-1& 0
\end{pmatrix}\,.\label{vacua}
\ee
In the context of Composite (Goldstone) Higgs scenarios,  $E_B$  ($E_H$)  
is the vacuum of the theory that preserves (explicitly breaks) the EW symmetry and therefore can be used to construct  Composite Higgs (Technicolor) models 
(see \cite{Cacciapaglia:2014uja} for a detailed discussion).
   
   The vacuum $E_\theta$ satisfies the relations
    \begin{equation}
	\label{eq:sp4algebra}
	S^a_\theta E_\theta+E_\theta\,S^{a\,T}_\theta=0,\qquad a=1,\dots,10\,,
    \end{equation}
    where $S^a_\theta$ are the 10  unbroken generators of $SU(4)$, which obey to the symplectic algebra of $Sp(4)$.~\footnote{The representations of  these generators can be found in  Appendix~\ref{app:generators}.}~ After EW symmetry breaking, the vacuum remains invariant under $U_{em}(1)$, that is~\cite{Alanne:2014kea}
       \begin{equation} \label{condE2}
    E_\theta \,Q_{em} - Q_{em}\, E_\theta^T =0, \quad Q_{em} = T_3 + Y = \frac{1}{2\sqrt{2}}\begin{pmatrix}
	\sigma_3 & 0 \\
	0 & -\sigma_3^T
	\end{pmatrix}\,,
    \end{equation}
    where $Q_{em}$ is  the electromagnetic charge operator. 

The scalar sector of the theory strictly depends on the choice of the vacuum $E_\theta$. As we will see in the following, the 
alignment angle $\theta$ is completely determined by the radiative corrections and the requirement that the model reproduces the phenomenological success of the Standard Model. {This framework} is very different from the composite (Goldstone) Higgs scenario because there the different structure of the radiative corrections induced by the EW and top mass alone prefer the Technicolor limit rather than the composite Goldstone Higgs realisation.

\subsection{Scalar sector}\label{sec:scalars}

In the minimal scenario depicted above, the  scalar sector  can be constructed out of the vacuum $E_\theta$, making use of the  two-index antisymmetric irrep 
$M\sim \mathbf{6}$ of $SU(4)$ 
\be \begin{split}
M   
     & =\left[\frac{1}{2} \left( \sigma + i\, \Theta\right) + \sqrt{2}\, ( \Pi_i+i \,\tilde \Pi_i) \,X^i_\theta \right] E_\theta\,,  \\ \end{split}
\label{eq:higgssector}
\ee
where  $X^i_\theta$  ($i = 1, \ldots, 5$)  are the broken generators associated to the breaking of  $SU(4)$ to $Sp(4)$, reported 
in Appendix~\ref{app:generators} 
The explicit expression of $M$ is 
\be M=\frac{1}{\sqrt{2}}\begin{pmatrix} 
 0 & -S^\ast + i \tilde S^\ast & \Pi_0^\ast + i \tilde\Pi_0 & \Pi^+  -i \tilde\Pi^+  \\
 S^\ast - i \tilde S^\ast & 0 & -\Pi^- + i \tilde\Pi^- & \Pi_0 - i \tilde\Pi_0 \\
 -\Pi_0^\ast- i \tilde\Pi_0 &\Pi^- - i \tilde\Pi^- & 0 & S - i \tilde S \\
 -\Pi^+ +i \tilde\Pi^+  & -\Pi_0 + i \tilde\Pi_0  & -S + i \tilde S & 0 \\
\end{pmatrix}\label{eq:MM}\ee
where we define
\be \Pi^{\pm}=\frac{\left(\Pi_2 \mp i \Pi_1 \right)}{\sqrt{2}},\qquad \tilde\Pi^{\pm}=\frac{\left(\tilde\Pi_2 \mp i \tilde\Pi_1 \right)}{\sqrt{2}},\ee
and we use the shorthand notation
\be \Pi_0= \sqrt{2} \left(\Pi _4 \cos\theta +\sigma  \sin\theta -i \Pi _3\right), \qquad  \tilde\Pi_0= \sqrt{2}  \left(-\Theta  \sin\theta +\tilde \Pi_4 \cos\theta -i \tilde\Pi_3\right),\ee
\be S=\sqrt{2}  \left(\Pi _4 \sin \theta -\sigma  \cos\theta +i \Pi _5\right), \qquad  \tilde S=\sqrt{2}  \left(\Theta  \cos\theta +\tilde\Pi_4 \sin\theta +i \tilde\Pi_5\right).\ee
Accordingly, the  $SU(4)$ invariant (tree-level) scalar potential reads~\footnote{In the following we will assume that all the couplings in eq.~(\ref{eq:pot}) are real.}
\begin{equation}
	\label{eq:pot}
	\begin{split}
	    V_M=&\frac{1}{2}m_M^2 Tr[M^{\dagger} M]+\left( c_M Pf(M)+\mathrm{h.c.}\right)+\frac{\lambda}{4}Tr [M^{\dagger}M]^2\\
	    &+\lambda_1 Tr  [M^{\dagger}MM^{\dagger}M]
		    -2\left(\lambda_2 Pf(M)^2+h.c \right)+\left(\frac{\lambda_3}{2}Tr [M^{\dagger}M] Pf(M)+h.c. \right) \,,
\end{split}\end{equation}
where $Pf(M)$ is by definition the Pfaffian of $M$, i.e.~$Pf(M)=\frac{1}{8}\, \epsilon_{i j k l} M_{ij} M_{kl}$. ~The explicit expression of $V_M$ 
as a function of the fields introduced in eq.~(\ref{eq:pot}) is reported in Appendix~\ref{app:stability}, where the stability conditions are also discussed. Note that before adding the operators including the $Pf(M)$ the symmetry is $U(4)$. 

Because of the $SU(4)$ symmetry we can choose  the vacuum of the theory to be aligned in the $\sigma$ direction as follows:
  \be \left. \frac{\partial V_M}{\partial\sigma}   \right |_{\sigma=f} =0 \quad \Rightarrow\quad  \langle \sigma^2\rangle\equiv f^2= \frac{c_M-m_M^2}{4\,\lambda_{11}}\,,\label{vev-tree}\ee
where $c_M>m_{M}^2$ and  $\lambda_{11}$ is a positive effective coupling defined as~\footnote{As shown in Appendix~\ref{app:stability}, the positivity of $\lambda_{11}$ guarantees the stability of the scalar potential.}
\be  \lambda_{11}=\frac 1 4 (\lambda+ \lambda_1-\lambda_{2}-\lambda_3)\,.\label{La11} \ee 
Notice that, before adding the EW interactions, the tree-level scalar potential in eq.~(\ref{eq:pot}) is independent of the parameter $\theta$. Therefore, the theory at tree-level has an infinite number of degenerate vacua, of which the solution  $\theta=0$, that is $E_{0}=E_B$, preserves the EW symmetry. As we will see in the next section, quantum corrections to the scalar potential arising from the electroweak and Yukawa sectors lift the degeneracy of the vacua and a global minimum
arises selecting  a non-zero value of $\theta$. In turn this guaranties the breaking of the EW symmetry and, concurrently, the generation of a mass term for the Higgs boson at the quantum level.

At tree-level the  mass eigenstates of the theory are obtained by diagonalizing  the scalar mass matrix
\be {\cal M}^2 (\Phi)\big|_{\sigma=f}\;\equiv\;\partial_{\phi_i}  \partial_{\phi_j} V_M \big|_{\sigma=f}\,, \label{Masstree} \ee
where $\Phi$ denotes the collection of all the scalar fields in $M$, eq.~(\ref{eq:higgssector}). Thus, due to the vacuum structure of the theory,
one can show that  the five bosons $\Pi_i$  
correspond to the eigenstates of the matrix (\ref{Masstree}) with zero eigenvalues, i.e. they are the Nambu Goldstone Bosons (NGB)s  associated to the spontaneous breaking of the $SU(4)$ symmetry in the scalar sector. They can be rearranged in terms of the $(2,2)$ and $(1,1)$ representations of $SO(4)\sim SU(2)_L\times SU(2)_R$, namely
\begin{equation}
\left(	\begin{array}{c} \Pi_2\,-\,i\,\Pi_1\\ \Pi_4\,-\,i\,\Pi_3 \end{array}\right) \,\,,\quad\quad \Pi_5\,.
\end{equation}
Notably, the first  representation has the same quantum numbers as the SM Higgs doublet. Therefore, we can identify the fields $\Pi_{1,2,3}$ with  the longitudinal polarisation of the $W$ and $Z$  gauge bosons (see subsection~\ref{sec:gauge}), whereas the EGH is given (at tree-level) by   $\Pi_4$.
As remarked above and discussed in more detail in the following section, 
the radiative corrections to (\ref{eq:pot}) allow to generate a mass term for the Higgs boson, which in this case arises as a linear combination of
the fluctuations of the  $\sigma$ and $\Pi_4$ fields around the vacuum.

On the other hand, the scalar fields  $\sigma$, $\Theta$ and
$\tilde{\Pi}_i$ ($i=1,\dots,5$)  acquire a non-zero mass at tree-level given by
\begin{equation}\label{treemasses}
\begin{split}
 m_{\sigma}^2 &\equiv M_\sigma^2\,\,,\, \quad m_{\Theta}^2 \equiv M_\Theta^2\,\,,\, \quad m_{\tilde\Pi_i}^2\equiv M_\Theta^2+2\lambda_f f^2\,,\\
\end{split}
\end{equation}
with 
\be \lambda_f \;\equiv \;\lambda_1 -\lambda_{2}\; > \; - \frac{M_\Theta^2}{2\, f^2}\,. \label{lambdaF}\ee
In terms of the parameters in eq.~(\ref{treemasses})
 we can recast the dimensional terms $c_M$, $m_M^2$ in $V_M$ and linear combination $\lambda_{11}$ in (\ref{La11}) as
\be   c_M= \frac 1 2 \left(M_\Theta^2 - f^2(4 \lambda_{11} -\tilde\lambda )\right)\,, \quad m_M^2=\frac 1 2 \left(M_\Theta^2 - f^2 (4 \lambda_{11} -\tilde\lambda )- M_\sigma^2\right)\,, \quad \lambda_{11}= \frac{M_\sigma^2}{8 f^2} \,,\label{eq:ren}\ee
with
\be \tilde \lambda\; = \;4\, \lambda_{11}\,+\,\lambda_3 \,+\,4\, \lambda_2 \,. \label{tildeL}\ee

Finally, we notice that the $\Pi_5$ can acquire mass at tree-level by introducing a small breaking of the $SU(4)$ symmetry by adding the following operator to the potential  in eq.~(\ref{eq:pot}) 
\be V_{DM}= \frac{\mu_M^2}{8} Tr\left[ E_A M \right] Tr\left[ E_A M \right]^\ast =\frac{1}{2} \mu_M^2\left(\Pi_5^2 +\tilde \Pi_5^2 \right), 
\qquad \mbox{with} \quad 
E_A= \begin{pmatrix}
i \,\sigma_2 & 0 \\
0 & i \,\sigma_2
\end{pmatrix}\,. \label{eq:DMmass}
\ee
As shown in \cite{Alanne:2014kea}, in this case $\Pi_5$  is a stable massive particle - due to the presence of an accidental $Z_2$ symmetry -
and provides a viable Dark Matter candidate. 
Accordingly, the full tree-level scalar potential of the theory is 
\be V= V_M +V_{DM}\,. \label{VPhi}\ee
The minimum of  $V$ is still aligned in the $\sigma$ direction, but now there are new massive excitations for $\mu_M\neq 0$, that is
\begin{equation}
\begin{split}\label{treemasses2}
m_{\tilde \Pi_5}^2 & \equiv M_\Theta^2+2\lambda_f f^2+ \mu_M^2\,\,,\,    \quad m_{\Pi_5}^2 \equiv \mu_M^2\,.
\end{split}
\end{equation}

All in all, once the symmetry breaking scale $f$ is fixed, the scalar sector of the theory can be described in terms of only five independent parameters: $M_\sigma$, $M_\Theta$, $\mu_M$, $\lambda_f$  and $\tilde\lambda$.

\subsection{Gauge sector}\label{sec:gauge}
In order to embed the EW gauge sector of the SM into the larger group $SU(4)$, we gauge 
the $SU(2)_L\times U(1)_Y$ part of the chiral symmetry group  $SU(2)_L\times SU(2)_R \subset SU(4)$.
In this way, the  scalar degrees of freedom are minimally coupled to the EW gauge bosons via
the covariant derivative of $M$ 
	\be
	    \label{eq:covM}
	    D_{\mu}M=\partial_{\mu} M- i \left(G_{\mu}M+M \,G_{\mu}^T\right)\,, \quad \mbox{with} \quad   G_{\mu}=gW^i_{\mu}T_L^i +g' B_{\mu}T_Y\,,
	\ee
where the $SU(2)_L$ generators $T_L^i$  ($i=1,2,3$)  and the hypercharge generator $T_Y=T_R^3$ are given in Appendix~\ref{app:generators}. 
The kinetic and EW gauge interaction Lagrangian of the scalar sector reads 
\be  \mathcal{L}_{gauge}= \frac{1}{2} \text{Tr}\left[D_{\mu}M^{\dagger}D^{\mu}M\right] \,, \label{eq:L}\ee
which explicitly breaks the global $SU(4)$ symmetry.
For any non vanishing $\theta$  the EW gauge group breaks spontaneously and  
the weak gauge bosons acquire non-zero masses through the Higgs-Brout-Englert mechanism that read
	\begin{equation}
	    \label{eq:WBosMasses}
	    m_W^2=\frac{1}{4}g^2f^2\sin^2\theta, \quad\text{and}\quad m_Z^2=\frac{1}{4}(g^2+g'^2)f^2\sin^2\theta\,.
	\end{equation}  
Comparing these expressions with the corresponding SM predictions reported in eq.~(\ref{SMrelations1})
we see that $f$ and $\theta$ must satisfy the phenomenological constraint
\be f\sin \theta\;=\;v_{\rm EW}\;\simeq\; 246 \mbox{ GeV} \,.\label{eq:ew}\ee

It has been established in \cite{Alanne:2014kea}  that a non-vanishing $\theta$ indeed occurs when radiative corrections are taken into account. We will further demonstrate, in the following section, that a small value of $\theta$ is naturally preferred by the theory once the full parameter space is properly investigated.

\subsection{Yukawa sector}\label{sec:Yukawa}
%
We construct the Yukawa sector of the  theory introducing EW gauge invariant operators that explicitly break  the $SU(4)$ global symmetry and
correctly  reproduce the SM fermion masses and mixing. 
 Following this reasoning, we formally accommodate each one of the SM fermion families in the fundamental irrep of $SU(4)$, namely
\be  \mathbf{L}_{\alpha}=\begin{pmatrix} L, &  \tilde \nu, & \tilde \ell
\end{pmatrix}_{\alpha L}^T\sim \mathbf{4}, \qquad  
\mathbf{Q}_{i}=\begin{pmatrix} Q, &  \tilde q^u, & \tilde q^d
\end{pmatrix}_{i \,L}^T \sim \mathbf{4},\label{eq:ql}
\ee
where $\alpha=e,\mu,\tau$ and $i=1,  2, 3$ are generation indices  and the tilde indicates the charge conjugate fields of the RH fermions, that is, for instance, 
$\tilde\nu_{\alpha L}\equiv (\nu_{\alpha R})^c$, $\tilde \ell_{\alpha L}\equiv (\ell_{\alpha R})^c$, $L_{\alpha L}\equiv (\nu_{\alpha L}, \ell_{\alpha L})^T$ and similarly for the quark fields. Notice that a RH neutrino  $\nu_{\alpha R}$ for each family must be introduced in order to define $\mathbf{L}_{\alpha}$ to transform according to the fundamental irrep of $SU(4)$.

Given the embedding of quarks and leptons in $SU(4)$, we now construct a Yukawa mass term for the SM fermions. 
 For this we make use of $SU(4)$ spurion fields \cite{Galloway:2010bp} $P_a$ and $\overline{P}_a$, with  $a=1,2$ is an $SU(2)_L$ index. They transform   as  $\parenbar[1]{P}_a \rightarrow (u^\dagger)^T \,\parenbar[1]{P}_a \,u^\dagger$, with $u \in SU(4)$. In this case we have
\be \begin{split}
 P_1= 
\frac{1}{\sqrt{2}}
\begin{pmatrix}
\bf{0}_2 & \tau_3\\
-\tau_3 & \bf{0}_2
\end{pmatrix}\,, 
& 
\quad  P_2= 
\frac{1}{\sqrt{2}}
\begin{pmatrix}
\bf{0}_2 & \tau^-\\
-\tau^+ & \bf{0}_2
\end{pmatrix}\,,
\end{split}\ee
\be \overline P_1=\frac{1}{\sqrt{2}}
\begin{pmatrix}
\bf{0}_2 & \tau^+\\
-\tau^- & \bf{0}_2
\end{pmatrix}
\,,
\quad \overline  P_2= 
\frac{1}{\sqrt{2}}
\begin{pmatrix}
\bf{0}_2 & \overline \tau_3\\
- \overline \tau_3\ & \bf{0}_2
\end{pmatrix}\,, \\
\ee
with
\be \tau^\pm=\frac{\sigma_1\pm i\, \sigma_2}{2}, \,\, \tau_3=\frac{\bf{1}_2+\sigma_3}{2}, \quad\mbox{and} \quad \overline \tau_3=\frac{\bf{1}_2- \sigma_3}{2}\,. \ee

Then, in terms of $P_{1,2}$ and $\overline P_{1,2}$ we write the Yukawa couplings for the SM fermions that preserve the $SU(2)_L$ gauge symmetry:
 \begin{eqnarray}
-\mathcal{L}^\text{Yukawa} &= & \frac{Y^u_{i j}}{\sqrt{2}} \,\left(\mathbf{Q}^T_{i} \, P_a \, \mathbf{Q}_{j} \right)^\dagger Tr\left[P_a \, M\right] + 
 \frac{Y^d_{i j}}{\sqrt{2}}\, \left(\mathbf{Q}^T_{i}  \,\overline P_a\,  \mathbf{Q}_{j} \right)^\dagger Tr\left[ \overline P_a\, M\right]\nonumber\\ 
         &+& \frac{Y^\nu_{\alpha\beta}}{\sqrt{2}} \,\left(\mathbf{L}^T_{\alpha} \, P_a \, \mathbf{L}_{\beta} \right)^\dagger Tr\left[P_a \, M\right] + 
 \frac{Y^\ell_{\alpha\beta}}{\sqrt{2}}\, \left(\mathbf{L}^T_{\alpha}  \,\overline P_a\,  \mathbf{L}_{\beta} \right)^\dagger Tr\left[ \overline P_a\, M\right]\,+\,\text{h.c.}\label{LYuk1}
\end{eqnarray}
with the Yukawa matrices of quarks and leptons chosen in agreement with  experiments.
\mathversion{normal}
This Lagrangian explicitly breaks the $SU(4)$ global symmetry.  In fact, in terms of the SM quark and lepton fields, eq.~(\ref{LYuk1}) can be written as
 \begin{eqnarray}
-\mathcal{L}^\text{Yukawa} &= & Y^u_{i j} \left( Q_{i L} \,  \tilde q^u_{j L} \right)^\dagger_a\ Tr\left[P_a \, M\right] + 
  Y^d_{i j} \, \left( Q_{i L} \,  \tilde q^d_{j L} \right)^\dagger_a Tr\left[ \overline P_a\, M\right]\nonumber\\ 
         &+& Y^\nu_{\alpha\beta} \,\left(L_{\alpha L} \,  \tilde \nu_{\beta L} \right)^\dagger_a Tr\left[P_a \, M\right] + 
  Y^\ell_{\alpha\beta}\, \left(L_{\alpha L} \,  \tilde \ell_{\beta L}  \right)^\dagger_a Tr\left[ \overline P_a\, M\right]\,+\,\text{h.c.}\label{LYuk2}
\end{eqnarray}
Then, after EW symmetry breaking, we predict for the SM fermion masses
\be m_F = y_F \frac{f \sin \theta}{\sqrt{2}}\,, \ee
$y_F$ being the SM Yukawa coupling of quarks and leptons in the fermion mass basis. 
Notice, in particular, that a Dirac mass for neutrinos is generated as well.
In principle, one can add a Majorana mass term for the RH neutrino fields, which provides an explicit breaking of the $SU(4)$ symmetry, but preserves the
EW gauge group. In this case, the most general mass Lagrangian for leptons is
 \be -\mathcal{L}^{\text{lep}} =Y^\ell_{\alpha\beta}\, \frac{f\sin\theta}{\sqrt{2}} \,\,\overline \ell_{\alpha L}  \ell_{\beta R}+Y^\nu_{\alpha j} \frac{f\sin\theta}{\sqrt{2}}  \,\,\overline \nu_{\alpha L}  \nu_{j R}+ \frac{1}{2}\,(M_R)_{j k} \,  \overline \nu_{j R} \,(\nu_{k R})^c\,+\,\text{h.c.} \label{LYuk3}\ee
where $M_R$ is the Majorana mass term of the three RH neutrinos. We require  $M_R$ to generate a small breaking of $SU(4)$, that is
\begin{equation}
	\left| (M_R)_{j k}  \right|\;\ll\; f\,.
\end{equation}
The couplings in eq.~(\ref{LYuk3}) allow to generate at tree-level a Majorana mass term for the LH neutrinos, in a manner similar to the  standard type I seesaw extension of the SM \cite{seesaw}.  This yields
\begin{equation}
 \mathcal{L}^\nu_{\text{mass}}\;=\;-\frac 1 2 \left( m_\nu \right)_{\alpha\beta}\,\overline\nu_{\alpha L}\,(\nu_{\beta L})^c\;+\;\text{h.c.}
\end{equation}
with
\begin{equation}
	m_\nu\;=\; - m_D\,M_R^{-1}\,m_D^T\quad \text{and}\quad m_D\;=\; Y^\nu\, \frac{f\sin\theta}{\sqrt{2}} \;=\; Y^\nu\, \frac{v_{\rm EW}}{\sqrt{2}}\,.
\end{equation}

\section{ Electroweak scale from radiative corrections}
\label{radiative}

A non-zero mass term for the EGH field $\Pi_4$ is generated at quantum level from those operators in the 
Lagrangian that explicitly break the global symmetry $SU(4)$, i.e. the gauge and Yukawa interactions. 
In this section we will compute  the one-loop Coleman-Weinberg effective potential \cite{Coleman:1973jx} of the model and study 
the new vacuum alignment conditions, which determine the spectrum of the theory.

\subsection{Coleman-Weinberg potential}

The one-loop correction $\delta V(\Phi)$ of the scalar potential $V$ given in (\ref{VPhi}) takes the general expression
	    \begin{equation}
		\label{eq:deltaV}
		\delta V(\Phi)=\frac{1}{64\pi^2}\mathrm{Str}\left[{\cal M}^4 (\Phi) \left(\log\frac{{\cal M}^2(\Phi)}
		    {\mu_0^2}-C\right)\right]+V_{\mathrm{GB}},
	    \end{equation}
where in this case $\Phi\equiv(\sigma,\,\Pi_4)$ denotes  the background scalar fields that we expect to lead to the correct vacuum alignment of the theory
and  ${\cal M}(\Phi) $ is the corresponding  tree-level mass matrix. The  supertrace, $\mathrm{Str}$, is defined as
\begin{equation}
\mathrm{Str} = \sum_{\text{scalars}}-2\sum_{\text{fermions}}+3\sum_{\text{vectors}}.
\end{equation}
and we have $C = 3/2$ for scalars and fermions and $C = 5/6$ for the gauge bosons,
whereas $\mu_0$ is a reference renormalization scale. 
The terms related to the massless Goldstone bosons are described by a separate potential, $V_{GB}$, since these terms lead to infrared divergences due to their vanishing masses. There are several ways of dealing with this issue, for example adding some characteristic mass scale as an infrared regulator. However, since the massive scalars give the dominant contribution to the vacuum structure of the theory, we will simply neglect $V_{GB}$ in the following discussion.

In terms of the background fields $\sigma$ and $\Pi_4$, we can write the first term in eq.~(\ref{eq:deltaV}) as
\be \delta V(\sigma,\Pi_4)\;=\; \delta V_{\mathrm{EW}}(\sigma,\Pi_4)\;+\;\delta V_{\mathrm{top}}(\sigma,\Pi_4)\;+\;\delta V_{\mathrm{sc}}(\sigma,\Pi_4), \qquad \mbox{with}\ee
\begin{align}
    \label{eq:corrEWtop}
    \begin{split}
        \delta V_{\mathrm{EW}}(\sigma,\Pi_4)\;=\;&\frac{3}{1024\pi^2}\, \phi^4\left[2g^4\left(
	    \log\frac{g^2\,\phi^2}{4\mu_0^2}-\frac{5}{6}\right)  +(g^2+g^{\prime\, 2})^2\left(\log\frac{(g^2+g^{\prime\, 2})\,\phi^2}{4\mu_0^2}
	    -\frac{5}{6}\right)\right]\,,
    \end{split}\\
        \delta V_{\mathrm{top}}(\sigma,\Pi_4)\;=\;&-\frac{3}{64\pi^2}\,\phi^4 y_t^4\left(
	    \log\frac{y_t^2\,\phi^2}{2\mu_0^2}-\frac{3}{2}\right)\,,
\end{align}
where $\phi\equiv  \sigma\sin\theta+\Pi_4\cos\theta$. We consider for simplicity only the fermion contribution in the one-loop potential
arising from the virtual top quark. Notice that both $\delta V_{\rm EW}$ and $\delta V_{\rm top}$ introduce an explicit dependence on $\theta$
in the full scalar potential of the theory.

The  quantum correction originated from the scalar sector reads
\begin{equation}
\begin{split}
\delta V_{\mathrm{sc}}(\sigma,\Pi_4)\;=\;& \frac{1}{64 \pi^2} \left[-\frac{3}{2} \left(  m_{\sigma}^4 (\sigma,\Pi_4) +m_\Theta^4(\sigma,\Pi_4) +m_{\tilde\Pi_i}^4(\sigma,\Pi_4) +m_{\tilde \Pi_5}^4(\sigma,\Pi_4) +m_{\Pi_5}^4(\sigma,\Pi_4) \right)  \right. \\
& \left. + m_{\sigma}^4 (\sigma,\Pi_4) \log\left( \frac{m_{\sigma}^2 (\sigma,\Pi_4)}{\mu_0^2} \right)  +
m_{\Theta}^4 (\sigma,\Pi_4) \log\left( \frac{m_{\Theta}^2 (\sigma,\Pi_4)}{\mu_0^2} \right)  \right.\\ 
& \left. +4 m_{\tilde\Pi_i}^4 (\sigma,\Pi_4) \log\left( \frac{m_{\tilde\Pi_i}^2 (\sigma,\Pi_4)}{\mu_0^2} \right) +
m_{\tilde \Pi_5}^4 (\sigma,\Pi_4) \log\left( \frac{m_{\tilde \Pi_5}^2 (\sigma,\Pi_4)}{\mu_0^2} \right)\right. \\
& \left.+m_{\Pi_5}^4 (\sigma,\Pi_4) \log\left(\frac{m_{\Pi_5}^2 (\sigma,\Pi_4)}{\mu_0^2} \right)    \right]\,,\\
\end{split}	
\end{equation}
where the background dependent masses of the scalar fields are  
\begin{eqnarray}
\begin{split}
	m_{\sigma}^2 (\sigma,\Pi_4) &=\frac{1}{2 f^2 }M_\sigma^2 \left(3\,\sigma^2 +\Pi_4^2-f^2\right)\,, \quad 
	m_\Theta^2(\sigma,\Pi_4) = M_\Theta^2+\tilde\lambda \left(\Pi_4^2+\sigma ^2-f^2\right)\,,\\
	m_{\tilde\Pi_i}^2(\sigma,\Pi_4) &= M_\Theta^2+\tilde\lambda \left(\Pi_4^2+\sigma ^2-f^2\right)
	+2 \lambda_f \left(\Pi_4^2+\sigma ^2\right)\,,\\
	m_{\tilde \Pi_5}^2(\sigma,\Pi_4) &= m_{\Theta}^2(\sigma,\Pi_4)+\mu_M^2+2\lambda_f (\Pi_4^2+\sigma ^2)\,,\\
	m_{\Pi_5}^2(\sigma,\Pi_4) &=\frac{1}{2 f^2}M_\sigma^2 \left(\sigma^2 +\Pi_4^2-f^2\right)+\mu_M^2\,.
\end{split}\label{eq:bckfields}
\end{eqnarray}
Notice that these expressions reduce to the tree-level scalar masses (\ref{treemasses}) and (\ref{treemasses2}) 
when we evaluate them for $\langle \Phi \rangle=(f,0)$.

\subsection{Determining the vacuum of the theory}\label{sec:vev}

At the quantum level a new parameter emerges which is the renormalization scale $\mu_0$. Following \cite{Alanne:2014kea} we fix this parameter in such a way that the quantum corrected potential has still an extremum in the $\sigma $ direction by  imposing
\begin{equation}
\label{eq:notads}
\left.\frac{\partial\delta V(\sigma,\Pi_4)}{\partial\sigma}\right|_{\sigma=f,\Pi_4=0}=\;0 \quad \Rightarrow \quad
\log\mu_0^2=\left.\frac{\partial_\sigma\mathrm{Str}\left[\mathcal M^4(\sigma,\Pi_4)\left(\log \mathcal M^2(\sigma,\Pi_4)
-C\right)\right]}{\partial_\sigma\mathrm{Str}\left[\mathcal M^4(\sigma,\Pi_4)\right]}\right|_{\sigma=f,\Pi_4=0} \ .
\end{equation}
This condition induces in $\mu_0$ a dependence  on $f$, $\theta$, $M_\sigma$, $M_\Theta$, $\mu_M$ as well as the effective couplings  $\lambda_f$ and $\tilde\lambda$. 

We are now able to determine the value of $\theta$ that minimises the full (tree-level\footnote{The tree-level potential is $\theta$ independent.} plus one-loop) scalar potential,
by imposing the conditions
\be
 \left. \frac{d \delta V(\sigma,\Pi_4)}{d\theta}\right|_{\sigma=f,\Pi_4=0}=\;  
 \left.  \frac{\partial \delta V(\sigma,\Pi_4)}{\partial \theta}\right|_{\sigma=f,\Pi_4=0}\, +\, 
 \left. \frac{\partial \delta V(\sigma,\Pi_4)}{\partial \mu_0} \frac{\partial \mu_0}{\partial \theta}\right|_{\sigma=f,\Pi_4=0}\;=\;0\label{Vmin1}
\ee
and
\be
	\left. \frac{d^2\delta V(\sigma,\Pi_4)}{d\theta^2} \right|_{\sigma=f,\Pi_4=0}\;>\;0\,.\label{Vmin2}
\ee
This shows that the specific value of $\theta$ that minimises the overall potential is a dynamical issue. We will shortly discover that the theory prefers very small values of $\theta$  without having to fine-tune the parameters of the theory.

\section{Phenomenological Constraints}\label{pheno}

According to the discussion reported in the previous sections, the set of parameters that fully describes the scalar sector of the theory is the following:
\begin{equation}
	f\,,\quad \theta\,,\quad M_\sigma\,,\quad M_\Theta\,,\quad \mu_M\,, \quad\tilde \lambda\,,\quad \lambda_f\,.\label{setpar}
\end{equation}
Several phenomenological constraints allow to relate and reduce the number of free parameters in (\ref{setpar}). 
In fact, as already pointed out in subsection~\ref{sec:gauge}, in order to reproduce the weak gauge boson masses, $f$ and $\theta$ must satisfy the condition given in eq.~(\ref{eq:ew}). The  minimisation of the scalar potential requires also the conditions given in (\ref{Vmin1}) and (\ref{Vmin2}). Furthermore,  important constraints are provided by the Higgs phenomenology, starting with 
the experimental value of the Higgs mass  \cite{PDG}:
 \be m_{h}=125.7\pm 0.4 \mbox{ GeV}\,.\label{Hmass}\ee
The Higgs is one of the two linear combinations of $\sigma$ and $\Pi_4$, that is 
\begin{eqnarray}
	\begin{pmatrix}
		\sigma\\
		\Pi_4
	\end{pmatrix}&=&
	\begin{pmatrix}
		\cos\alpha &-\sin\alpha\\
		\sin\alpha & \cos \alpha
	\end{pmatrix}
	\begin{pmatrix}
		H_1\\
		H_2
	\end{pmatrix}\,,\label{eq:Higgs}
\end{eqnarray}
where $H_1$ and $H_2$ are the mass eigenstates and  $\alpha$ the scalar mixing angle,  chosen in the interval $[0,\pi/2]$. 
The observed Higgs boson will be the lightest eigenstate.  

To extract the couplings of the scalars to the SM fermions we recall that they are proportional to 
 \be -Tr[P_1 M ] \;=\;-Tr[\overline P_2 M ]\;=\; \frac{\cos\theta\, \Pi_4+\sin\theta\,\sigma}{\sqrt{2}}  =\frac{\cos\left(\alpha+\theta\right)\, H_2+\sin\left(\alpha+\theta\right)\, H_1}{\sqrt{2}}\,.  \label{condF}\ee
Hence, from Eqs.~(\ref{LYuk2}) and (\ref{condF}), the SM fermion couplings to the mass eigenstates $H_1$ and $H_2$ read 
\be  \lambda_{H_1 FF}= \frac{1}{\sqrt{2} } y_F  \sin\left(\alpha+\theta\right), \qquad  \lambda_{H_2 FF}= \frac{1}{\sqrt{2} } y_F  \cos\left(\alpha+\theta\right)\,. \ee
 We can now define the SM normalised coupling strength 
\be \begin{split}
 C_F &\;\equiv\;\frac{\lambda_{H_{1\left[2\right]} FF}}{\lambda_{h FF}^{SM}}\;=\;   \sin\left(\alpha+\theta\right) \,\,\,\left[\cos\left(\alpha+\theta\right) \right] \,,
\label{eq:cvcf}\end{split}
\ee
where $\lambda_{h FF}^{SM}\equiv y_F/\sqrt{2}$ is the SM coupling and the index between square brackets refers to $H_2$. Similarly, for the couplings to the gauge bosons, normalised to the SM one $\lambda_{h VV}^{SM}$, we have
\be \begin{split}
C_V &\;\equiv\;\frac{\lambda_{H_{1\left[2\right]} VV}}{\lambda_{hVV}^{SM}} =   \sin\left(\alpha+\theta\right)\,\,\,\, 
\left[  \cos\left(\alpha+\theta\right)\right] \,.
\label{eq:cvcf}\end{split}
\ee
The parameters $C_F$ and $C_V$ must satisfy the  experimental constraints
 \cite{CMS:2014ega}
\be C_V=1.01^{+0.07}_{-0.07}, \quad C_F=0.89 ^{+0.14}_{-0.13} \qquad \mbox{at 68\% C.L.} \label{eq:excvcf}\ee

Finally, we report the trilinear self-coupling of $H_1$ and $H_2$ and confront them with the corresponding SM prediction, 
$\lambda_{hhh}^{\rm SM}=3 \,m_h^2/v_{\rm EW}$. In this case, we have \cite{Alanne:2014kea}
\be  \frac{\lambda_{H_1 H_1 H_1}}{\lambda_{hhh}^{SM}}= v_{\rm EW}\frac{ M_\sigma^2 \cos\alpha}{f m^2_{h} }  , \qquad 
 \frac{\lambda_{H_2 H_2 H_2}}{\lambda_{hhh}^{SM}}= v_{\rm EW}\frac{ M_\sigma^2 \sin\alpha}{f m^2_{h} }\,.\label{trilinear} \ee
 
 We can now investigate two limiting cases, the one (case A) in which $\theta \ll1$ and the other (case B) where $\theta \simeq \pi/2$. 

Using \eqref{eq:excvcf}, for case A, i.e. for  a vacuum aligned mostly along the Goldstone Higgs direction we have 
\be\begin{split}\label{eq:CaseA}
\mbox{Case A:} & \qquad (\alpha, \theta)\approx (\pi/2, 0) \Rightarrow (\sigma, \Pi_4)\approx(H_2, H_1) \Rightarrow h\approx H_1 \approx \Pi_4\,,\\
 \end{split}\ee
where the SM Higgs ($h$) is mostly the pNGB $\Pi_4$.  Instead  for  $\theta \approx \pi/2$, we find
\be\begin{split}\label{eq:CaseB}
\mbox{Case B:} & \qquad (\alpha, \theta)\approx (0, \pi/2) \Rightarrow (\sigma, \Pi_4)\approx(H_1, H_2)  \Rightarrow h\approx H_1 \approx \sigma \ ,\\
\end{split}\ee
for which $\sigma$ is nearly identified with the SM Higgs. 

Radiative corrections will choose the appropriate vacuum alignment as we shall see in the following section.

\section{ Parameter Space Analysis at the Quantum Level}

We are now ready to embark in the numerical analysis of the theory by minimising the quantum corrected potential, following the procedure outlined in the subsection \ref{sec:vev}. This will allow, for a given set of the theory parameters, to establish the specific embedding angle $\theta$ that it is crucial to determine whether or not the Higgs is mostly a pNGB state.

We will learn that the model prefers very small values of $\theta$ without fine-tuning. This is a very welcome feature demonstrating that the EGH paradigm is not a forced feature but rather a natural outcome. It also sets the EGH apart from its  composite counterpart that requires instead a large fine-tuning to accommodate the pNGB nature of the Higgs \cite{Cacciapaglia:2014uja}. 

To better appreciate the dependence of the results on the various parameters of the theory we considered different cases in the numerical analysis, namely
\begin{equation}
\begin{array}{ccccc}
	\textbf{Case 1:} &&&& M_\sigma\;=\;M_\Theta\;\equiv\;M_S\,,\quad \lambda_f  \;=\;0\,,\\
	\textbf{Case 2:} &&&& M_\sigma\;\neq\;M_\Theta\,,\quad \lambda_f\;=\;0\,, \\
	\textbf{Case 3:} &&&& M_\sigma\;\neq\;M_\Theta\,,\quad \lambda_f\;\neq\;0\,,\label{3regimes} 
\end{array}
\end{equation}
where $\lambda_f$ controls the difference, at tree-level, between the $\Theta$ and $\tilde{\Pi}_i$ masses as shown in eq.~\eqref{treemasses}.  
In case 1 all the massive states have a common mass at tree-level (see eq.~(\ref{treemasses})) and the potential will depend on: $\tilde\lambda$, $M_S$, $\mu_M$ and
$\sin\theta$. Imposing the experimental constraints introduced in the previous section further reduces the parameter space. 
  In the other two cases,  we relax the assumption of degenerate tree-level scalar masses. 
  
 \begin{figure}[t!]
  \subfigure
  {\includegraphics[width=5.7cm]{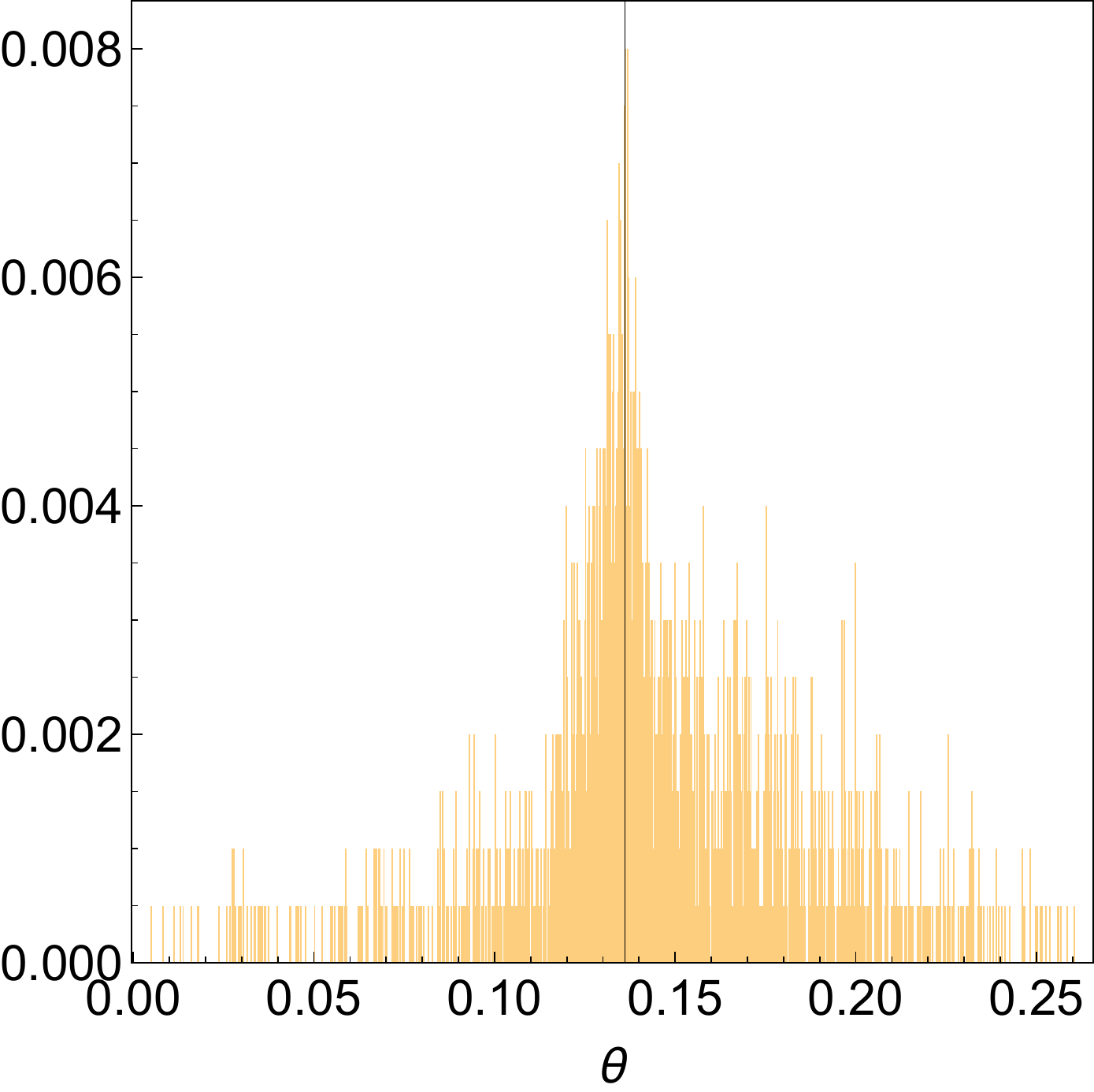}}
  \subfigure
  {\includegraphics[width=8cm]{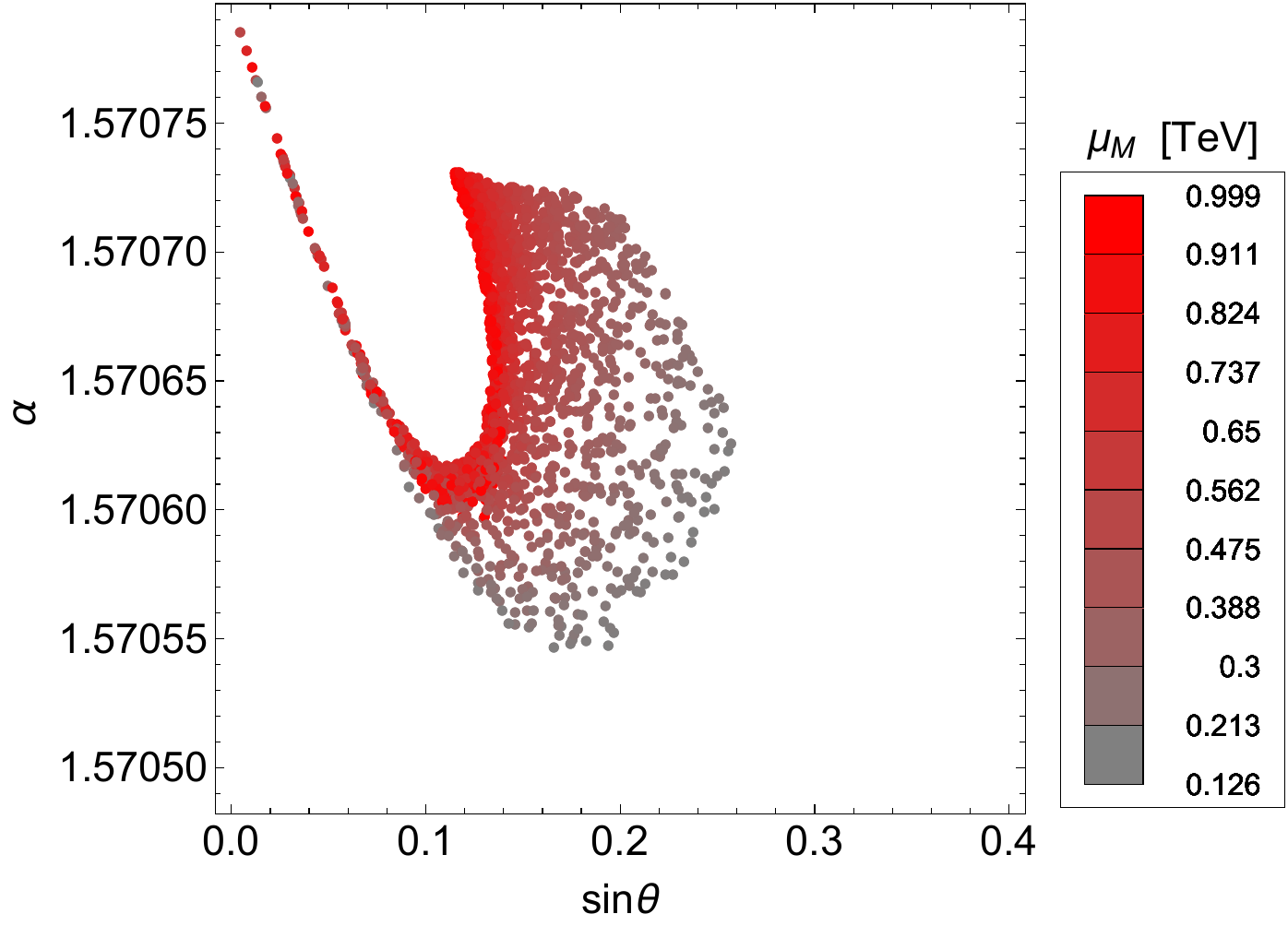}}
      \vspace*{-10pt}\caption{\label{fig:thetaalpha} \textbf{Case 1.} Left panel: Distribution of the vacuum alignment angle $\theta$. The
       mode  is  $\overline\theta\;=\;0.136^{+0.006}_{-0.012}$ (solid vertical line), which corresponds to an average   symmetry breaking scale $\overline f=1.81^{+0.08}_{-0.15}$ TeV.
       In this case the Higgs is mostly a pNGB. Right panel: Correlation of the scalar mixing angle $\alpha$ and $\sin\theta$. The gradient scale 
       describes  the values the parameter $\mu_M$ (see the text for details).}
 \end{figure}

In the following we will present our results in terms of scatter plots that better illustrates the dependence of the model on the parameter space of the theory. 
For each plot  we  fix the SM Yukawa coupling of the top  $y_t=1$ and we allow for a $3\,\sigma$ uncertainty on the value of the Higgs mass and use the central values of the weak gauge boson masses given in \cite{PDG}. Moreover, in all the three regimes we impose the perturbativity bound on the effective quartic coupling $\tilde\lambda$, that is $|\tilde \lambda|<4\,\pi$. The same constraint is applied to $\lambda_f$ in the most general scenario of case 3. Furthermore, we choose random values of  the parameter $\mu_M$ in the interval 
\be
	m_h\;\leq\; \mu_M \; \leq\; 1~\text{TeV}\,, \label{rangemuM}
\ee
with the additional constraint $\mu_M<f$. The latter ensures that $\mu_M$ introduces only a small explicit breaking of the global
$SU(4)$ symmetry.

\subsection{Case 1}

 \begin{figure}[t!]
  \subfigure
  {\includegraphics[width=7.6cm]{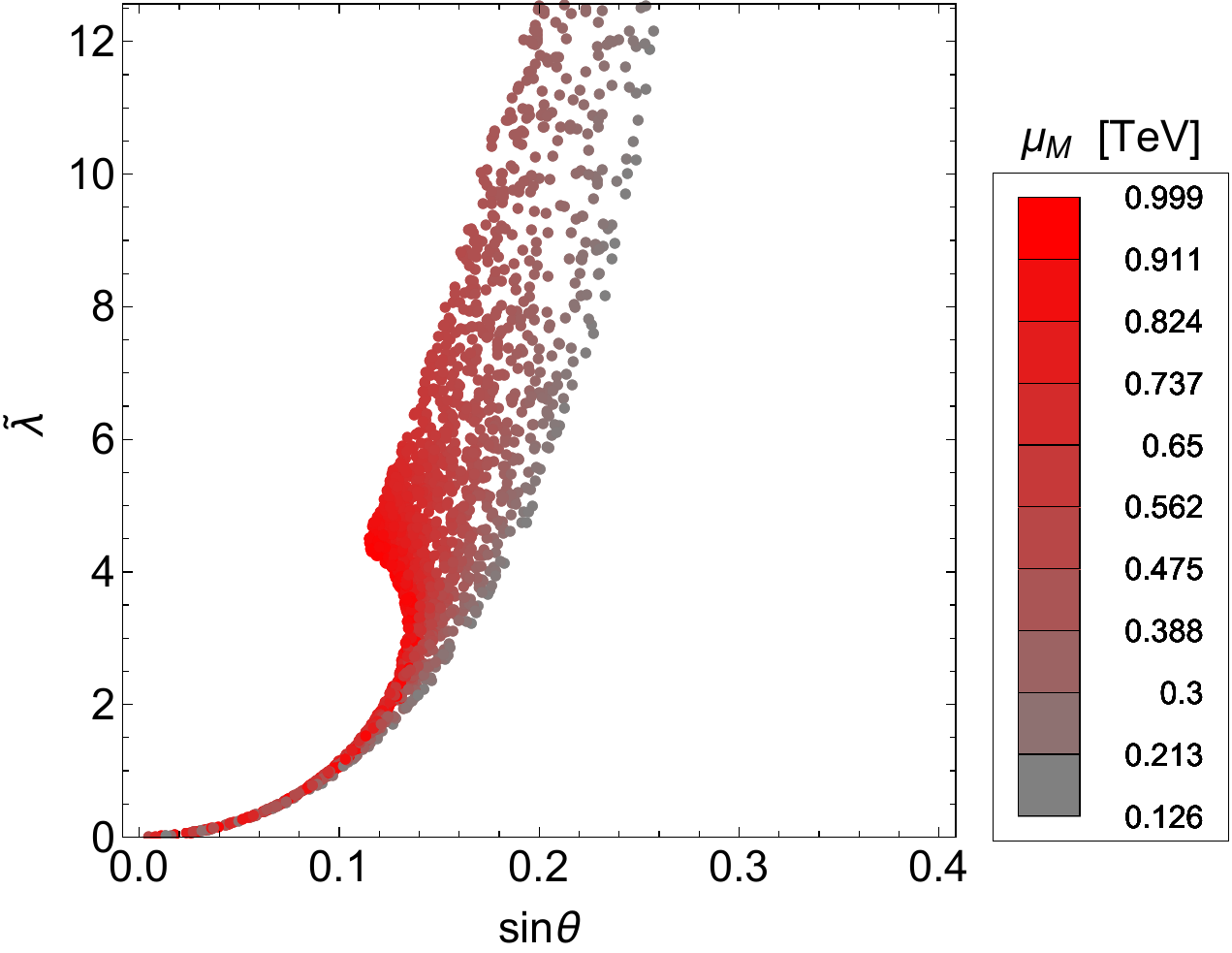}}
  \subfigure
  {\includegraphics[width=7.5cm]{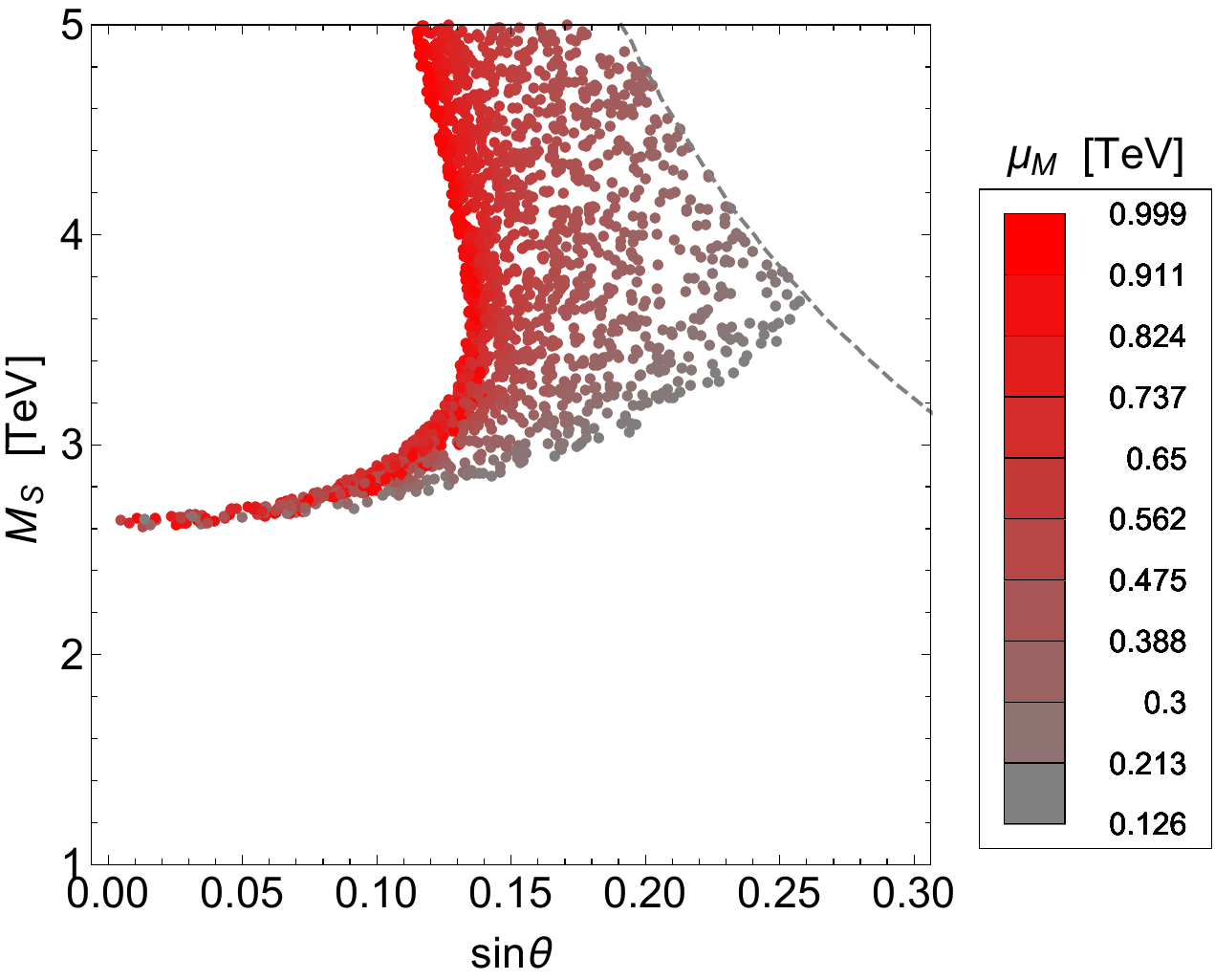}}
      \vspace*{-10pt}\caption{\label{fig:Case1a} \textbf{Case 1.} Left panel: Parameter space projection in the plane  $\sin \theta$ versus $\tilde\lambda$. The  gradient scale of the points describes different intervals of $\mu_M$. For $\sin\theta\lesssim 0.1$ we have $\tilde\lambda\approx K\sin^2\theta$, where $K\approx 90$ is a constant independent of $\mu_M$.  Right panel: Correlation between the effective quartic coupling $\tilde\lambda$ and the scalar mass $M_S$, using an analogous gradient scale for $\mu_M$.  The dashed line corresponds to the perturbative limit 
 $\tilde\lambda=4\,\pi$. For  $\sin\theta\lesssim 0.1$ the scalar mass is approximately constant and approaches the value $M_S\approx 2.6$ TeV (see the text for details). } 
 \end{figure}

Here we vary the common scalar mass  $M_S$, defined in the first line of (\ref{3regimes}), in the interval
\be
	m_h\;\leq\; M_S \; \leq\; 5~\text{TeV}\,.
\ee
 As described in the previous section, for each random value of  $ M_S$ and $\mu_M$, we select the other parameters of the model
  imposing the experimental value of the Higgs mass and the minimisation conditions of the Coleman-Weinberg  potential,
  Eqs.~(\ref{Vmin1}) and (\ref{Vmin2}). In this way we extract the values of the effective quartic coupling $\tilde \lambda$ 
  and the vacuum alignment angle $\theta$,   which are, therefore, implicit functions of the dimensional parameters $M_S$ and $\mu_M$.   
  
 Using this procedure we produce a list of 2000 points satisfying  the constraints above. We plot in the left panel of Fig.~\ref{fig:thetaalpha} the distribution of the  $\theta$ values resulting from this study yielding a mode of:~\footnote{In the following we define the mode as the value that appears most often in a set of data. We report the error on the mode as the width evaluated at half of the mode hight. We use for this statistical analysis histograms with 1000 bins. The error on the scale $f$ of the theory is computed with the standard propagation of errors.}
 \be  \overline\theta\;=\;0.136^{+0.006}_{-0.012} \,, \label{eq:caso1theta}\ee
corresponding to $\overline{\alpha}=1.57$ and the $SU(4)$ spontaneous symmetry breaking scale of 
\be \overline f\;=\; 1.81^{+0.08}_{-0.15}\mbox{ TeV}\,.\ee
Notice that, for a given $\theta$  the scalar mixing angle $\alpha$ is essentially  determined by imposing the experimental constraints  given in eq.~(\ref{eq:excvcf}),  which is reflected in the second panel of Fig.~\ref{fig:thetaalpha}, where the colour gradient  shows the variation of $\mu_M$ within the range specified in  (\ref{rangemuM}). 
The plot clearly shows that the dynamics prefers small values of $\theta$ implying  that the Higgs boson is mostly aligned in the $\Pi_4$ pNGB  direction.

 \begin{figure}[t!]
  \subfigure
  {\includegraphics[width=7.5cm]{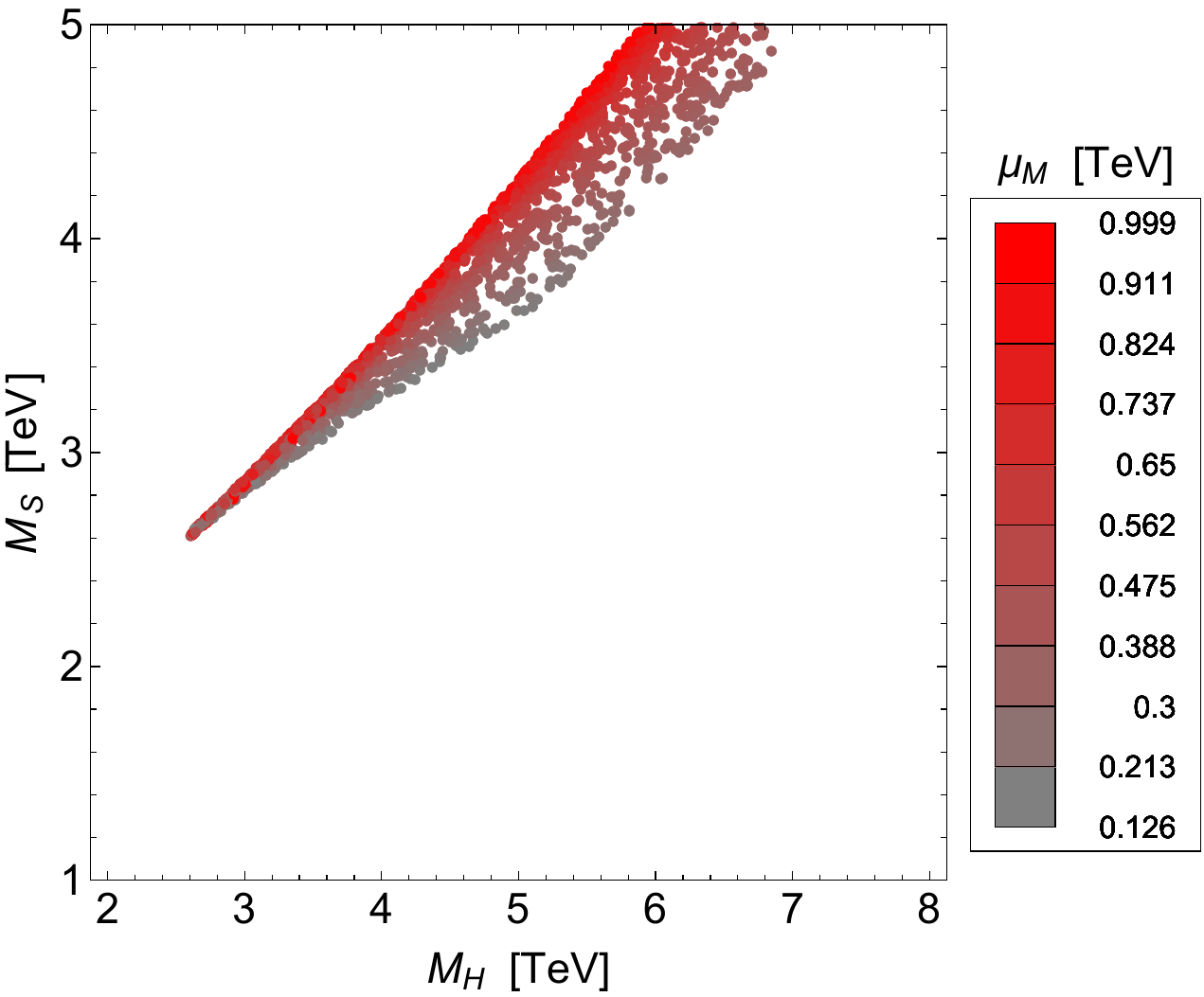}}
  \subfigure
  {\includegraphics[width=8.2cm]{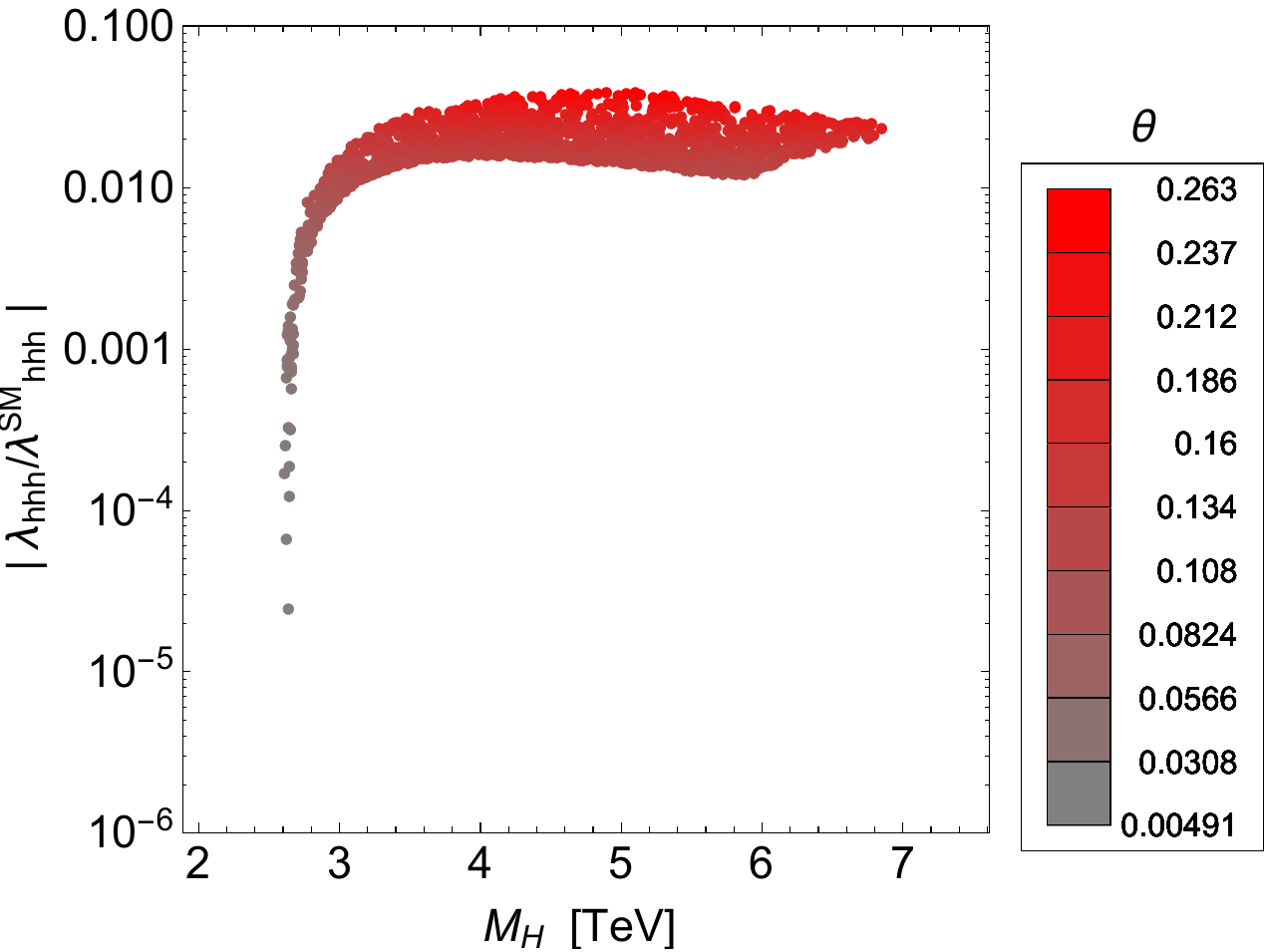}}
      \vspace*{-10pt}\caption{\label{fig:Case1Higgs}  \textbf{Case 1.} Left panel: Correlation between the tree-level mass $M_S$ 
      and the mass of the heaviest scalar mass eigenstate $H$. The lower limit of $M_H$ corresponds to the tree-level value $M_S\approx2.6$ TeV.
      Right panel:The ratio of the trilinear coupling of the light Higgs $h$ and the trilinear coupling in the SM a function of the mass of $M_H$.
      A strong suppression is obtained in the EW conserving limit $\sin\theta=0$ (see the analytic prediction given in~(\ref{trilinear})).} 
 \end{figure}

 In the left panel of Fig.~\ref{fig:Case1a} we show the projection of the allowed parameter space in the plane $\sin\theta$ versus $\tilde\lambda$.
As we can see from this plot, the variation of $\tilde \lambda$ is very simple for $\sin\theta$ close to zero. 
Indeed, from the minimisation condition we find to a very good approximation 
 \be
 		\tilde\lambda\;\approx\; K \sin^2\theta\quad\quad\text{for}\quad\quad \sin\theta\lesssim 0.1\,,
 \ee
 where $K$ depends on $M_S$  and not on $\mu_M$ (for $M_S\approx 2.6 $ GeV, $K\approx90$).
 Henceforth, for $\sin\theta\lesssim 0.1$ the only independent parameter is the tree-level scalar mass $M_S$, which is fixed by the knowledge of the Higgs mass via  
\be m_{h}^2\approx\frac{9}{16\,\pi^2 \,v_{\rm EW}^2} \left[M_Z^4 \,\log \left( \frac{M_Z^2}{M_S^2}\right) \,+
 \,M_W^4 \,\log \left( \frac{M_W^4}{M_S^4}\right)  \,-\, v_{\rm EW}^4\left( \frac 2 3\,+\, \log \left( \frac{v_{\rm EW}^2}{2\, M_S^2}\right)\right)\right]\,.
 \label{eq:analytic}
 \ee
For $m_h = 125$ GeV and $v_{\rm EW}=246$ GeV the previous expression implies
 \be
 	M_S\;\approx\; 2.6\quad\text{TeV}\quad\text{for}\quad\sin\theta\lesssim 0.1\,.\label{boundMS}
 \ee
The general dependence of $M_S$  with respect to $\theta$ is shown in the right panel of Fig.~\ref{fig:Case1a}. The analytic value for small $\theta$ is  confirmed by the numerical analysis.   The dashed line in the plot represents the points  with $\tilde\lambda = 4\,\pi$. Notice that the perturbative bound on $\tilde\lambda$ implicitly sets an upper limit on the vacuum  angle $\theta$. 
The overall spread is due to the dimensional parameter $\mu_M$ for both panels of  Fig.~\ref{fig:Case1a}. 
 
 We turn now to the properties of the heaviest scalar mass eigenstate defined in eq.~(\ref{eq:Higgs}), which here corresponds to  $H\equiv H_2\sim \sigma$. It is clear from the left panel of   Fig.~\ref{fig:Case1Higgs} that the physical mass $M_H$ and the tree-level mass $M_S$ are still close to each other once the quantum corrections are taken into account with the difference due mostly to the effects of $\mu_M$. 
 The mass of the heavy Higgs $H$ also affects the  ratio between the trilinear Higgs coupling $\lambda_{hhh}$ and the corresponding SM one given in (\ref{trilinear}). The latter is shown in the right panel of Fig.~\ref{fig:Case1Higgs}  as a function of $M_H$. As expected from the analytic expression, there is a strong suppression for $\theta \lesssim 0.1$ corresponding to $2.6~\text{TeV}\lesssim M_H\lesssim 3$ TeV.

 \begin{figure}[t!]
  \subfigure
  {\includegraphics[width=8cm]{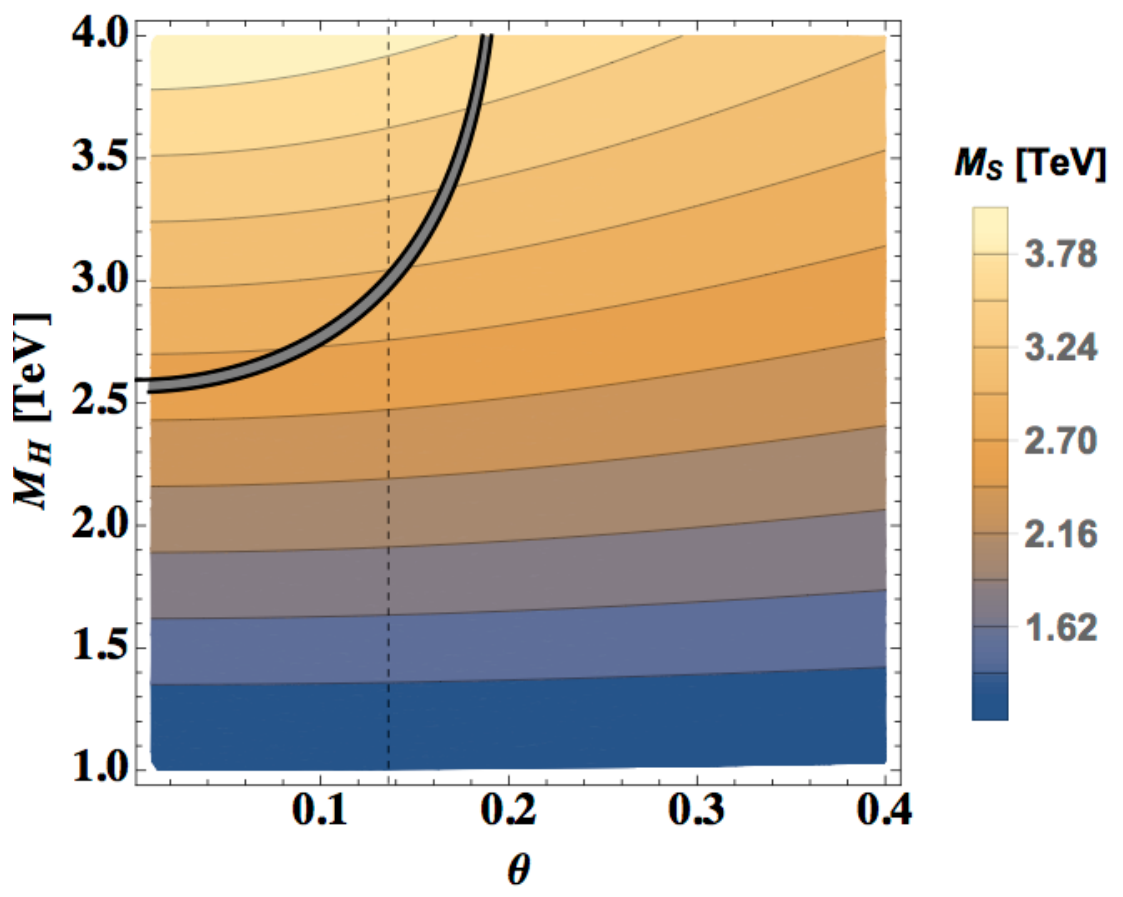}}
    \subfigure
  {\includegraphics[width=8.4cm]{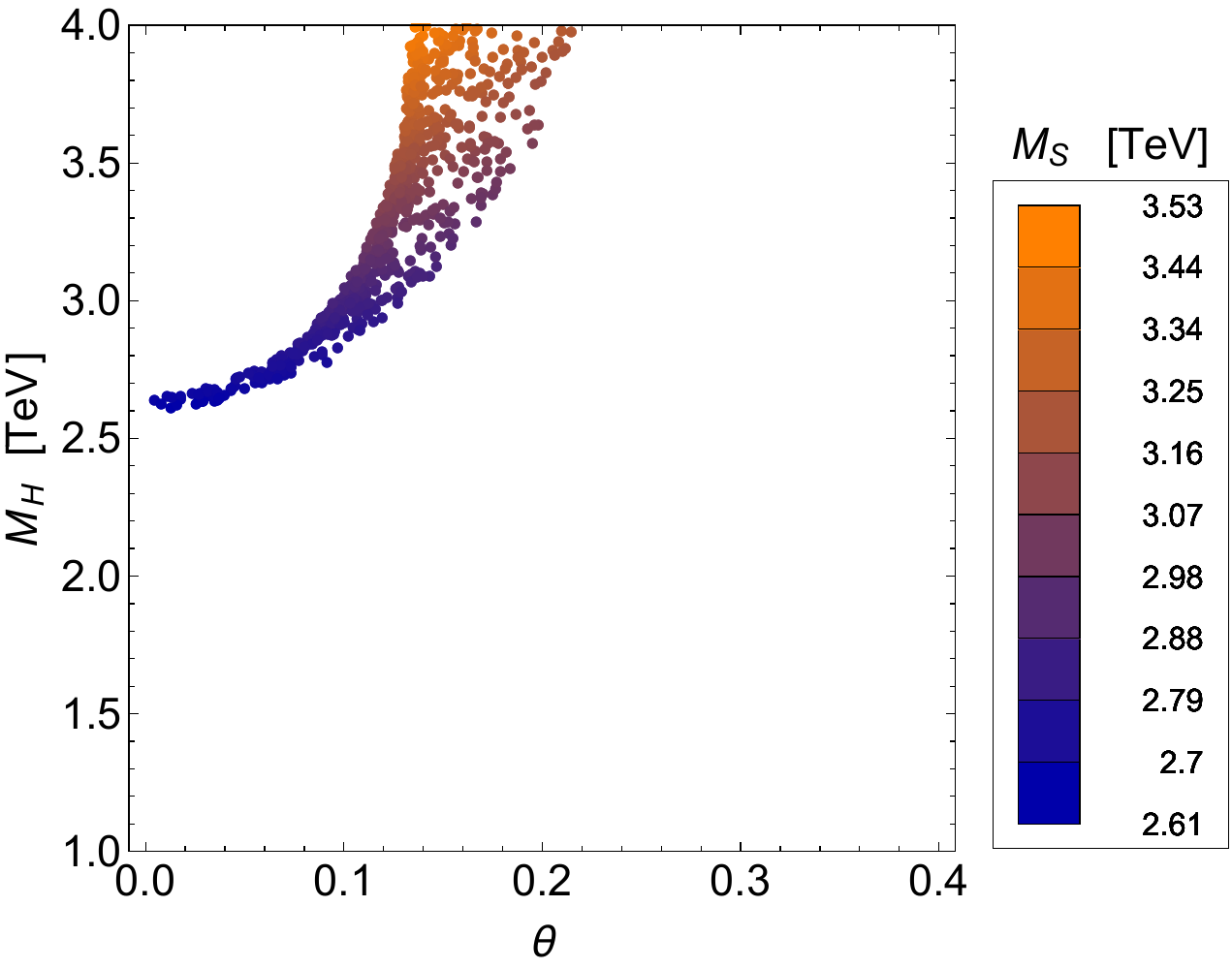}}
        \vspace*{-10pt}\caption{\label{fig:Case1Higgs2} \textbf{Case 1.} Dependence of the heavier scalar mass $M_H$ on the angle $\theta$. Left panel:  The plot has been realised using analytic expressions and expanding in $\sin\theta$. We set here $\mu_M=v_\text{EW}$. The grey band corresponds to the 3$\sigma$ uncertainty on the Higgs mass while the dashed line marks the value of the vacuum angle $\overline \theta$ in  eq. \eqref{eq:caso1theta}.  Right panel: Projection of the parameter space in the plane $M_H$ versus $\theta$. The parameter $\mu_M$ now takes random values in the interval $m_h\leq\mu_M\leq 1$ TeV.} 
 \end{figure}

Further, we want to show in Fig.~\ref{fig:Case1Higgs2} how the $3\sigma$ uncertainty in the mass of the observed Higgs  affects $M_H$.   In the left panel of Fig.~\ref{fig:Case1Higgs2}, we show an \textit{analytic} rendering of this dependence. The dark grey band (reflecting the 3$\sigma$ uncertainty) shows this correlation for fixed values of $M_S$, $\mu_M$ and $\theta$.  This is superimposed on the dependence of the $M_H$ with respect to the unconstrained $\theta$ and $M_S$ values, but still having fixed $\mu_M  = v_{EW}$. 
The numerical analysis confirms the analytical expectation as it is shown in the right panel  of Fig.~\ref{fig:Case1Higgs2}   where, however, we have also allowed $\mu_M$ to vary as in (\ref{rangemuM}), with the points coloured for different values of $M_S$.   

 \begin{figure}[t!]
  {\includegraphics[width=7.5cm]{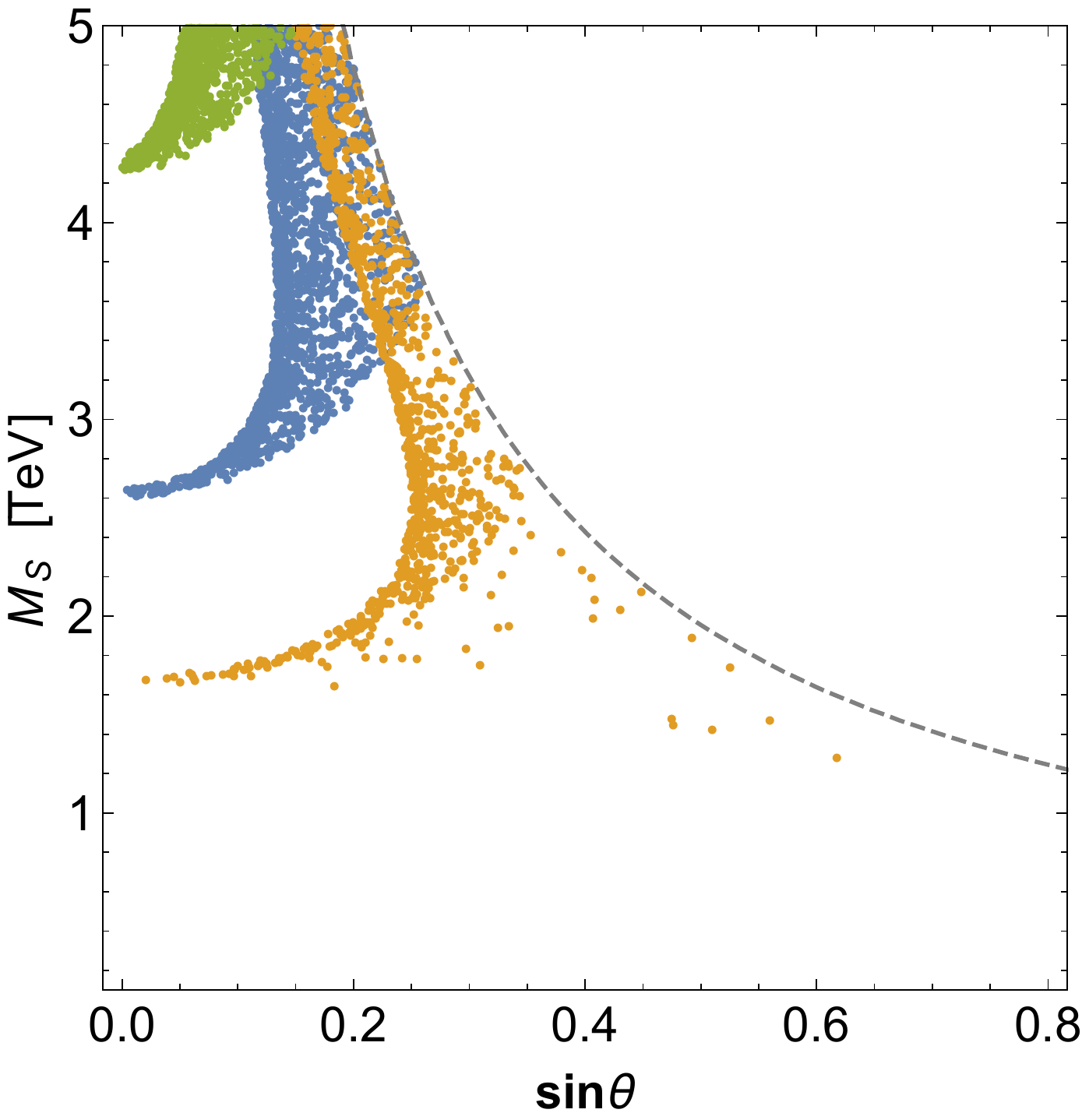}}
        \vspace*{-10pt}\caption{\label{fig:Case1Higgs3} \textbf{Case 1.} 
        The scatter plot shows  allowed regions in the parameter space fixing the
        Higgs mass  respectively at 10\% less/more than the observed Higgs mass  (yellow/green points respectively). 
       We show as well the results using  the exact 3$\sigma$ uncertainty on the Higgs mass (blue points). The dashed line corresponds to $\tilde\lambda=4\pi$.   } 
 \end{figure}

Next, we quantify the impact of the {experimental value of the} Higgs mass  (\ref{Hmass}) in the selection of the vacuum alignment angle $\theta$. This is shown  in Fig.~\ref{fig:Case1Higgs3} by reducing/increasing the physical Higgs mass by 10\% (yellow/green). The blue region corresponds to the observed Higgs mass. 
 Interestingly the lower the mass of the Higgs, the larger would have been the range of the allowed $\theta$ values, while $M_S$ would be lighter.  Again, the dashed line corresponds to $\tilde\lambda = 4 \pi$. The spread is given again by varying $\mu_M$ within the interval (\ref{rangemuM}).

\subsection{Case 2}

 \begin{figure}[t!]
  \subfigure
  {\includegraphics[width=5.3cm]{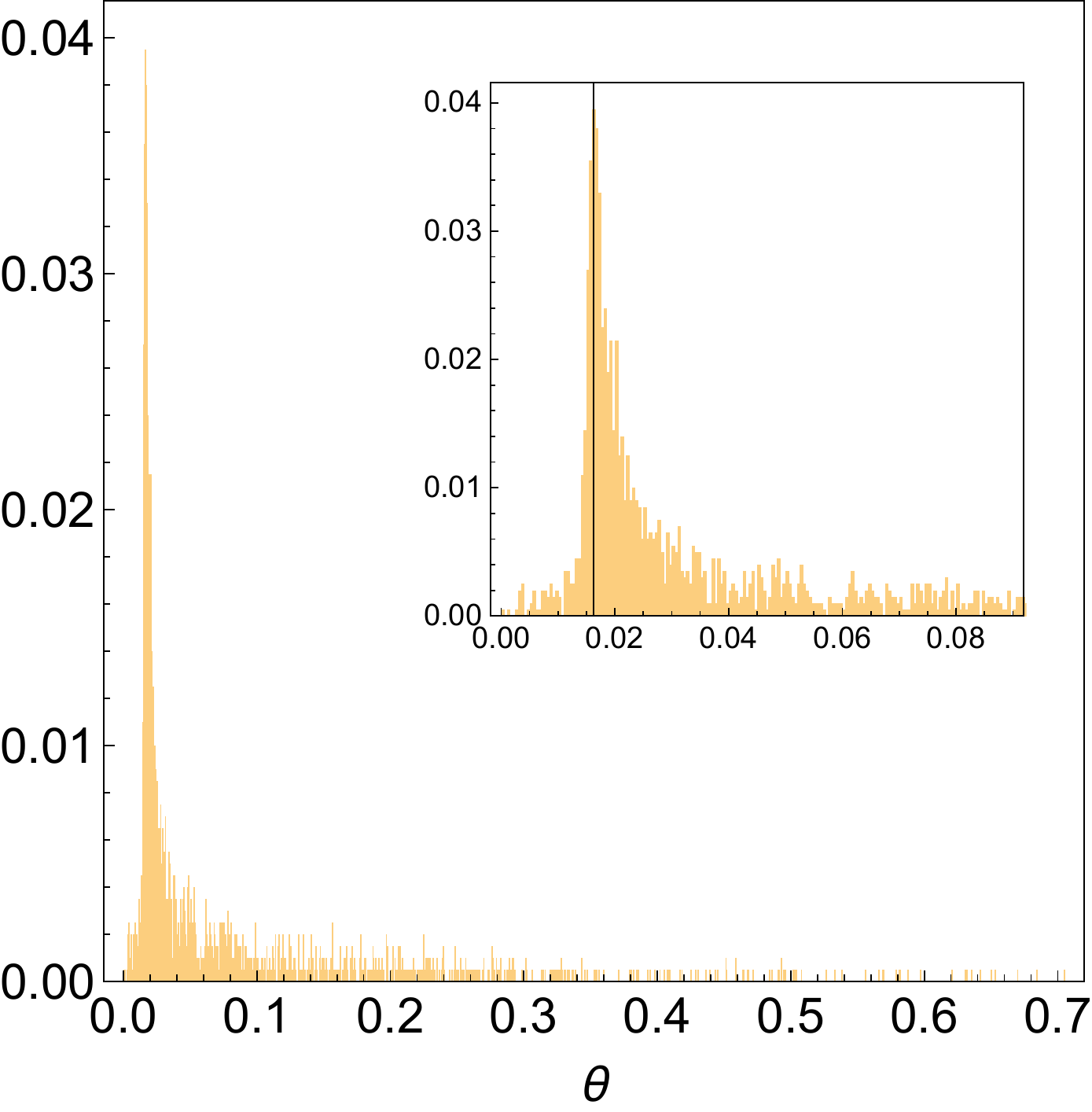}}
  \subfigure
  {\includegraphics[width=7.5cm]{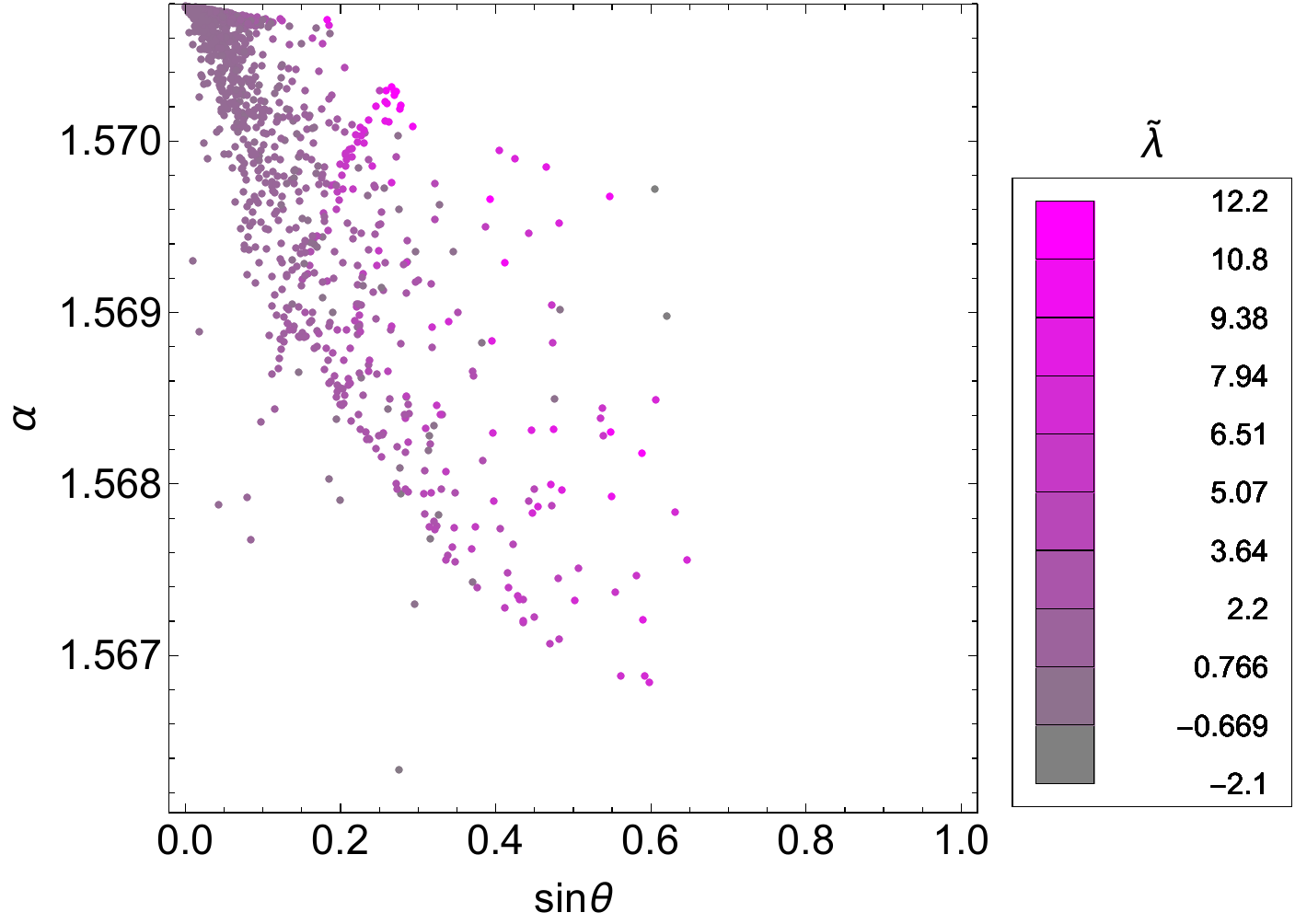}}
      \vspace*{-10pt}\caption{\label{fig:thetaalpha2} \textbf{Case 2.} Left Panel: Distribution of the vacuum alignment angle $\theta$. In the upper right side of the plot
      we show the distribution for small $\theta$. The vertical solid line corresponds to the mode at $\overline\theta=0.016^{+0.004}_{-0.002}$.  Right panel: Correlation of the scalar mixing angle $\alpha$ and $\sin\theta$. The colour gradient gives the value of the effective quartic coupling $\tilde\lambda$. See the text for details.      } 
 \end{figure}

In this case we have that $ M_\sigma \neq M_\Theta$ and $ M_\Theta =  M_{\tilde{\Pi}} = M_S$. Therefore, the parameters characterising the model are  $\tilde\lambda$, $M_\sigma$, $M_S$, $\mu_M$ and $\sin\theta$. We consider  values of the scalar masses within the interval
\be m_h\;\leq \;M_S\,, M_\sigma \;\leq \;5\, \mbox{ TeV}\,.\ee 
The procedure used in this section is the same as  described above.
Thus the randomised parameters are $M_S, M_\sigma$  and $\mu_M$.
As in the previous scenario, we construct a list of 2000 points satisfying all the experimental constraints. 
We found that the mode of the vacuum angle $\theta$ is
 \be  \overline\theta\;=\;0.016^{+0.004}_{-0.002}\,. \label{thetaavcase2}\ee
The average scalar mixing is also in this case $\overline\alpha=1.57$. The associated $SU(4)$ breaking scale reads
\be \overline f\;=\; 15.2^{+3.9}_{-1.4}\mbox{ TeV}\,.\label{favcase2}\ee
We notice that the central value of $\theta$ is smaller than in the first case, correspondingly we have a higher central value for the spontaneous symmetry breaking scale which is now around 10 TeV. Because the parameter space is larger than in the first case we also observe a larger deviation from the central value.  
 
  In  Fig.~\ref{fig:thetaalpha2} we show in the left panel the resulting distribution of the values of the vacuum alignment angle $\theta$, 
  while in  the  plot on the right side  we display the correlation between $\alpha$ and $\sin\theta$, with
 the colour gradient corresponding to fixed values of $\tilde\lambda$.
  Notice that very few points  with  $\theta>0.4$    are found in the numerical analysis, while
  in most of the parameter space the dynamics of the theory  favours small values of $\theta$.     
Therefore, we conclude that also in this regime the physical Higgs particle emerges as a pNGB, that is $h\equiv H_1\approx \Pi_4$ (see eq.~\eqref{eq:CaseA}).

 \begin{figure}[t!]
  \subfigure
 {\includegraphics[width=7.4cm]{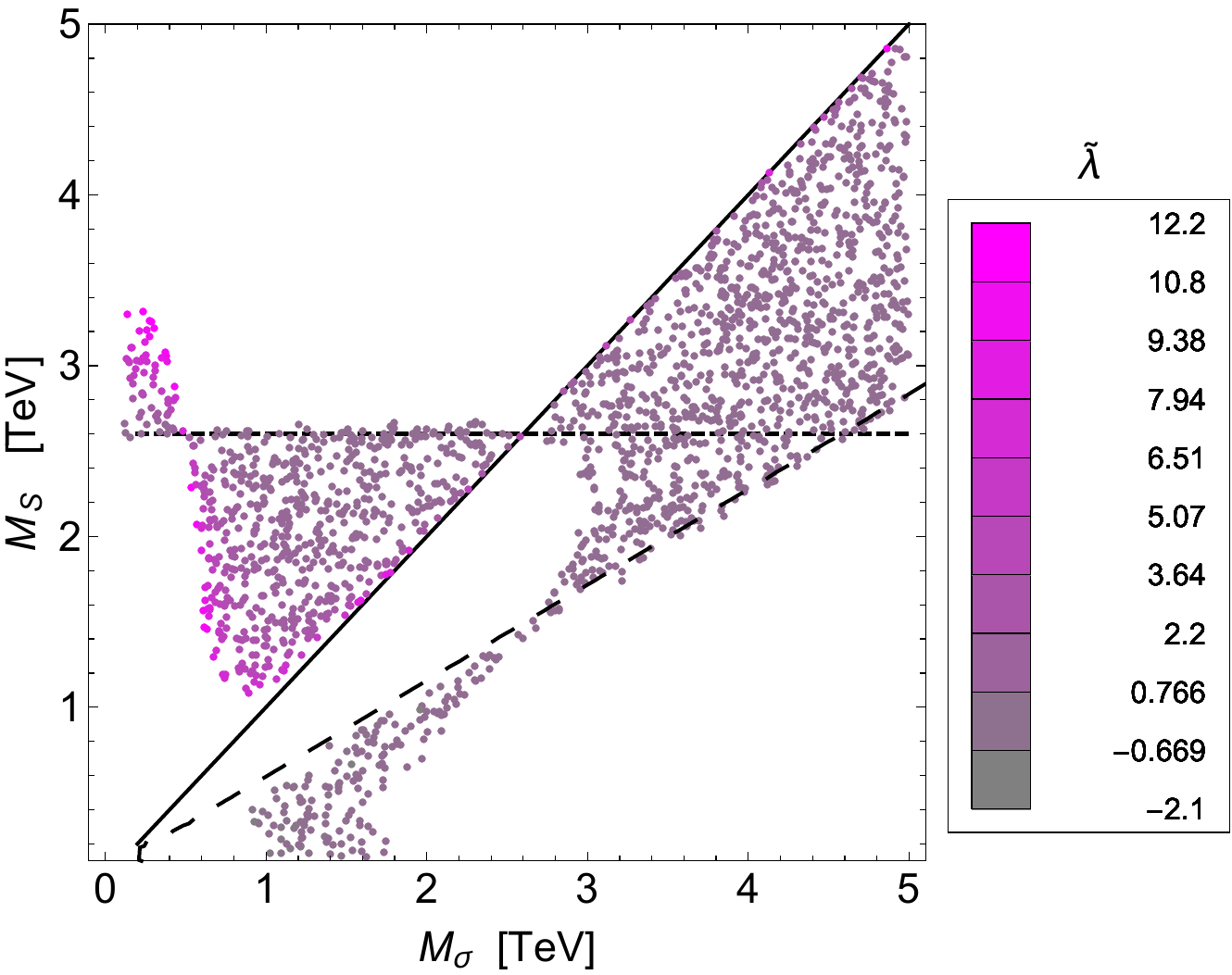}}
  \subfigure
   {\includegraphics[width=7.5cm]{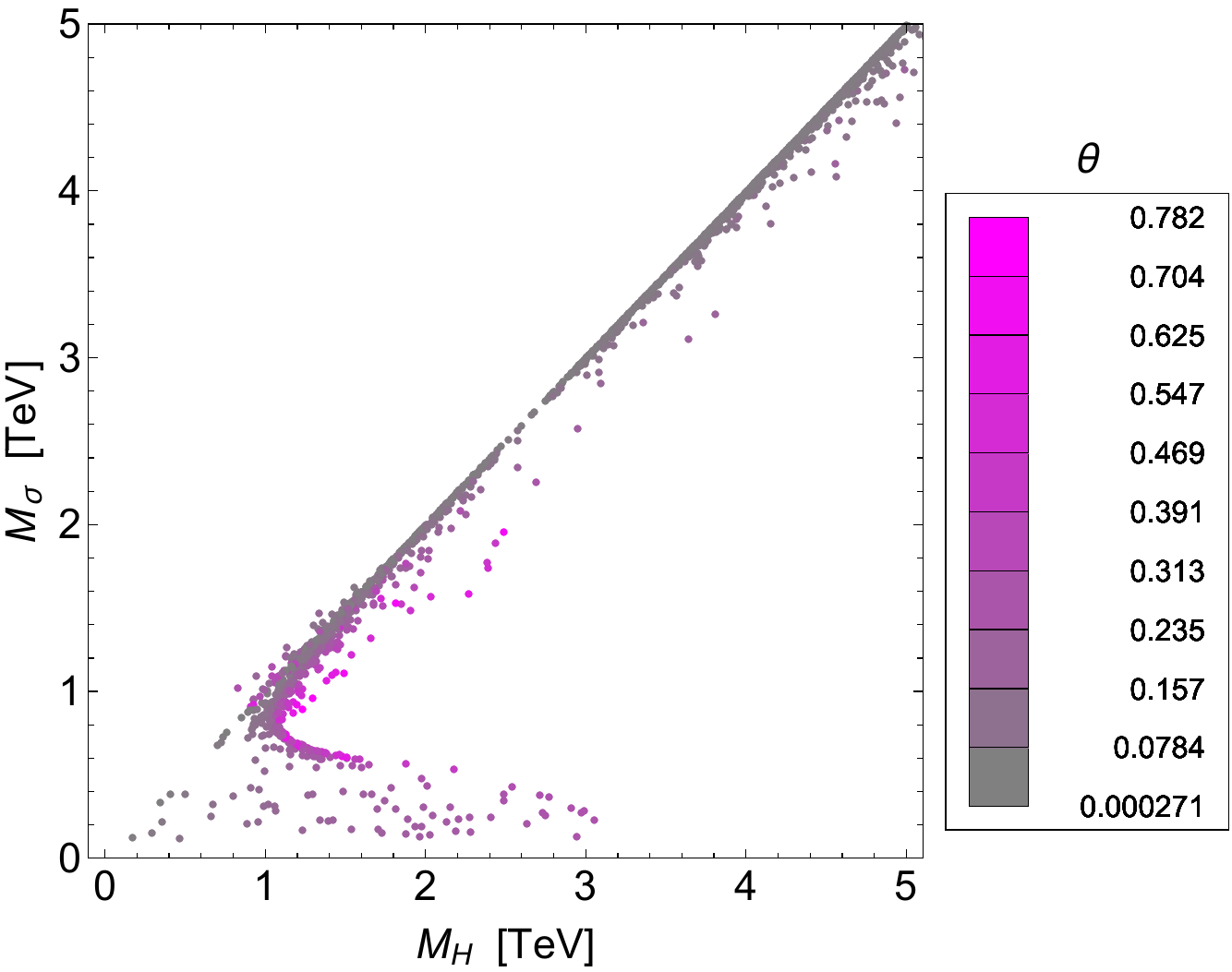}}
      \vspace*{-10pt}\caption{\label{fig:Case2b} \textbf{Case 2.}  Left panel: Correlation between the mass parameters  $M_\sigma$ and $M_S$. 
       Right panel: Dependence of the heaviest scalar mass, $M_H$, with respect to its tree-level  value, 
      $M_\sigma$.  See the text for details.} 
 \end{figure}

In Fig.~\eqref{fig:Case2b} (left panel) we show the correlation between the tree-level scalar masses $M_\sigma$ and $M_S$.   We mark with a solid line the particular case  $M_\sigma=M_S$ corresponding to the case 1 limit studied before.
Considering only the scalar and the top corrections, and neglecting in first approximation the parameter 
$\mu_M$, we get from the minimisation condition (\ref{Vmin1}) the relation
\be  
\tilde\lambda= \sin^2\theta\,\, g(M_\sigma, M_S) \,,
\ee
where $g(M_\sigma, M_S)$ is a complicated function that depends on the scalar masses. In the limit $\tilde\lambda\approx 0$
one has
\be M_S^4\approx \frac{M_\sigma^4 \left(\log\left[ \frac{m_t^2}{M_\sigma^2}\right]-1\right)}{6\left(\log\left[ \frac{m_t^2}{M_\sigma^2}\right]+1\right)}\,,\ee
where $m_t$ indicates the top mass.
This condition corresponds  to the long-dashed line in left panel of Fig.~\ref{fig:Case2b}. 
In the right panel of the same figure, instead, we report the correlation between  $M_\sigma$ and $M_H$, 
with the colour gradient indicating the values of $\theta$. We can see that for masses $M_\sigma\gtrsim 1$ TeV the mass of the heavier scalar is almost
entirely given by the tree-level term and most of the points correspond to very small values of  $\theta$.

\begin{figure}[t!]
{\includegraphics[width=8cm]{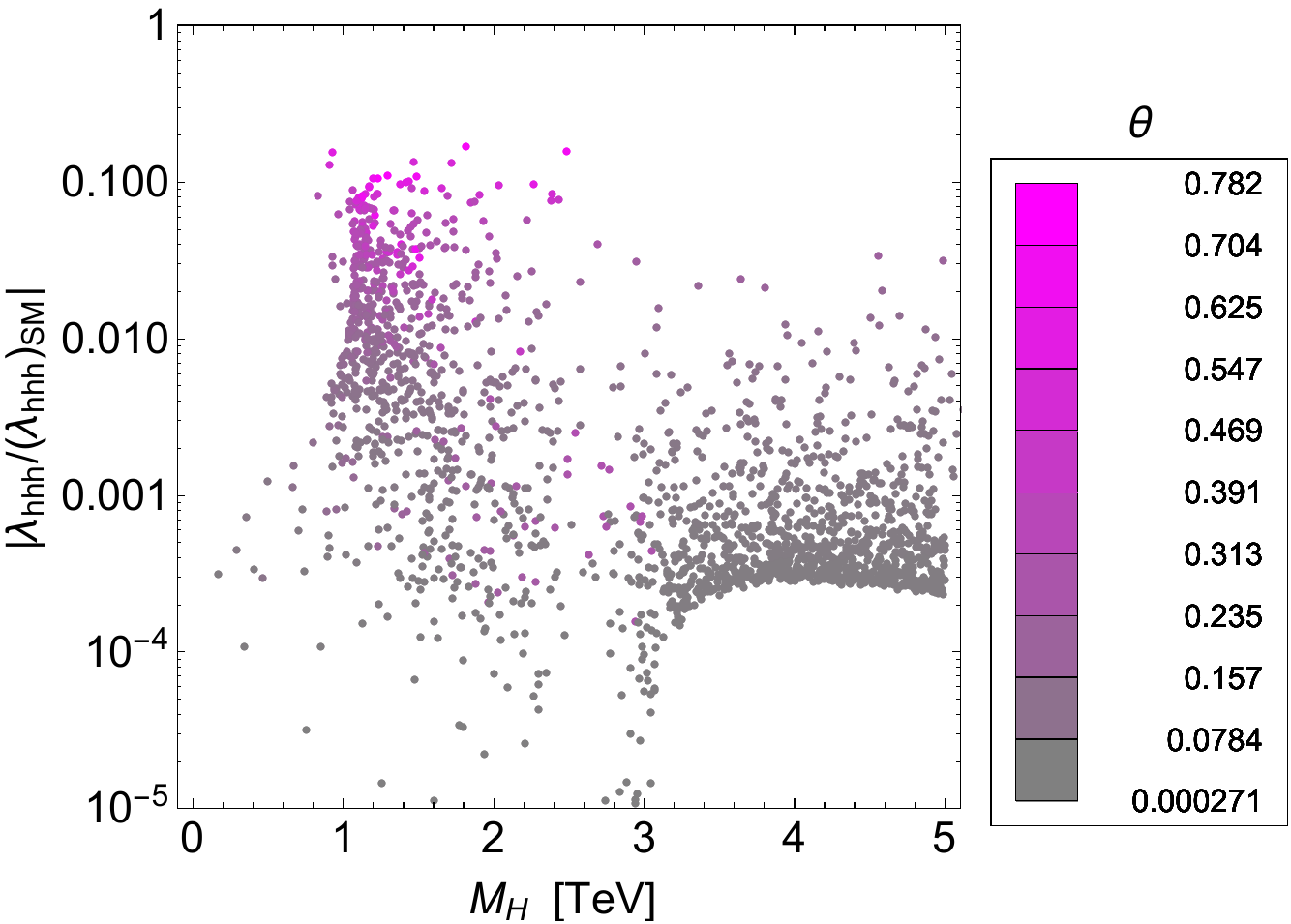}}
      \vspace*{-10pt}\caption{\label{fig:Case2c} \textbf{Case 2.}  The ratio between  the Higgs  trilinear coupling $\lambda_{hhh}$
       and the SM one as function of $M_H$.  }  
 \end{figure}

Finally, the scatter plot in Fig. \ref{fig:Case2c} displays the ratio of the  trilinear coupling with respect to the SM one  as function of $M_H$, each point, again, corresponding to a certain $\theta$. We can see that larger values of $\theta$ give a non-negligible ratio, while if $\theta$ is very small, like the average value of the sample obtained in the analysis suggests, eq.~(\ref{thetaavcase2}), there is a strong suppression.

\subsection{Case 3}\label{case3sub}

Finally we consider the most general possible spectrum of the theory, that is  $M_\sigma\neq M_\Theta \neq M_{\tilde \Pi_i}$. 
The parameters used in the analysis are $\tilde\lambda$, $\lambda_f$, $M_\sigma$, $M_\Theta$, $\mu_M$ and $\sin\theta$. As in the previous cases, we randomly generate the scalar masses and extract the values of $\theta$ and $\tilde\lambda$ that satisfy all the phenomenological constraints.
In particular,  the scalar masses are varied within the interval
\be m_h \; \leq\; M_\sigma\,,\,  M_\Theta\,, M_{\tilde \Pi_i}\; \leq\; 5\, \mbox{ TeV}\,.\ee 

In this more general regime the correlations between the parameters of the theory are very similar to  the ones obtained in the previous case.
However, we can notice that  the points are a bit more ``smeared" out due to $\lambda_f$ being in general different from zero. 
We find again a scalar mixing angle of $\overline\alpha=1.570$ and the mode of the distribution to be
 \be  \overline\theta\;=\;0.018^{+0.004}_{-0.003}	\,.\ee
The average $SU(4)$ breaking scale  associate to $\overline{\theta}$ is
\be \overline f\;=\;13.9^{+2.9}_{-2.1}\mbox{ TeV}\,,\ee
which is very similar to the results obtained in the previous case, Eqs.~(\ref{thetaavcase2}) and (\ref{favcase2}).

\begin{figure}[t!]
  \subfigure
  {\includegraphics[width=4.7cm]{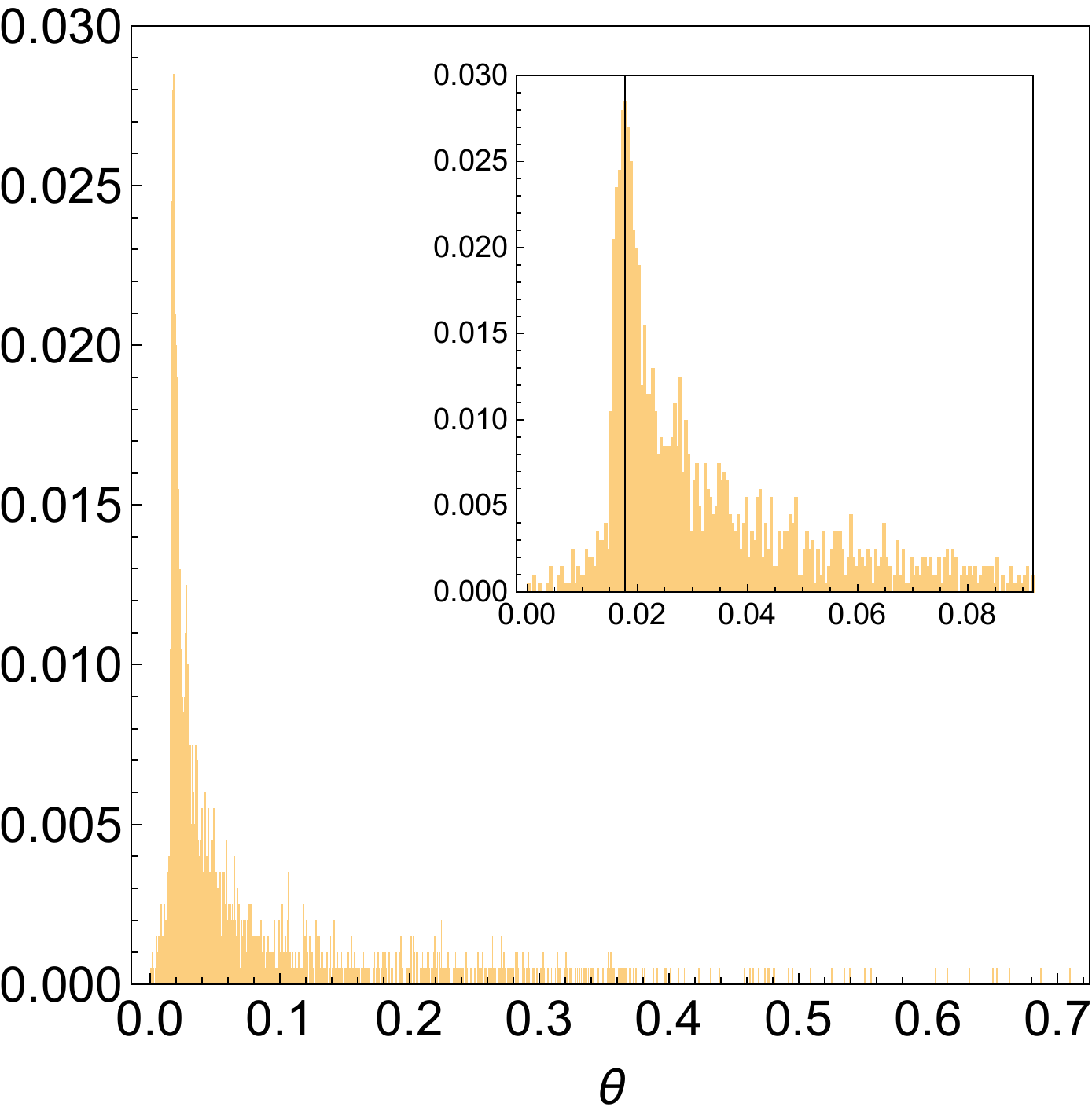}}
  \subfigure
  {\includegraphics[width=6.5cm]{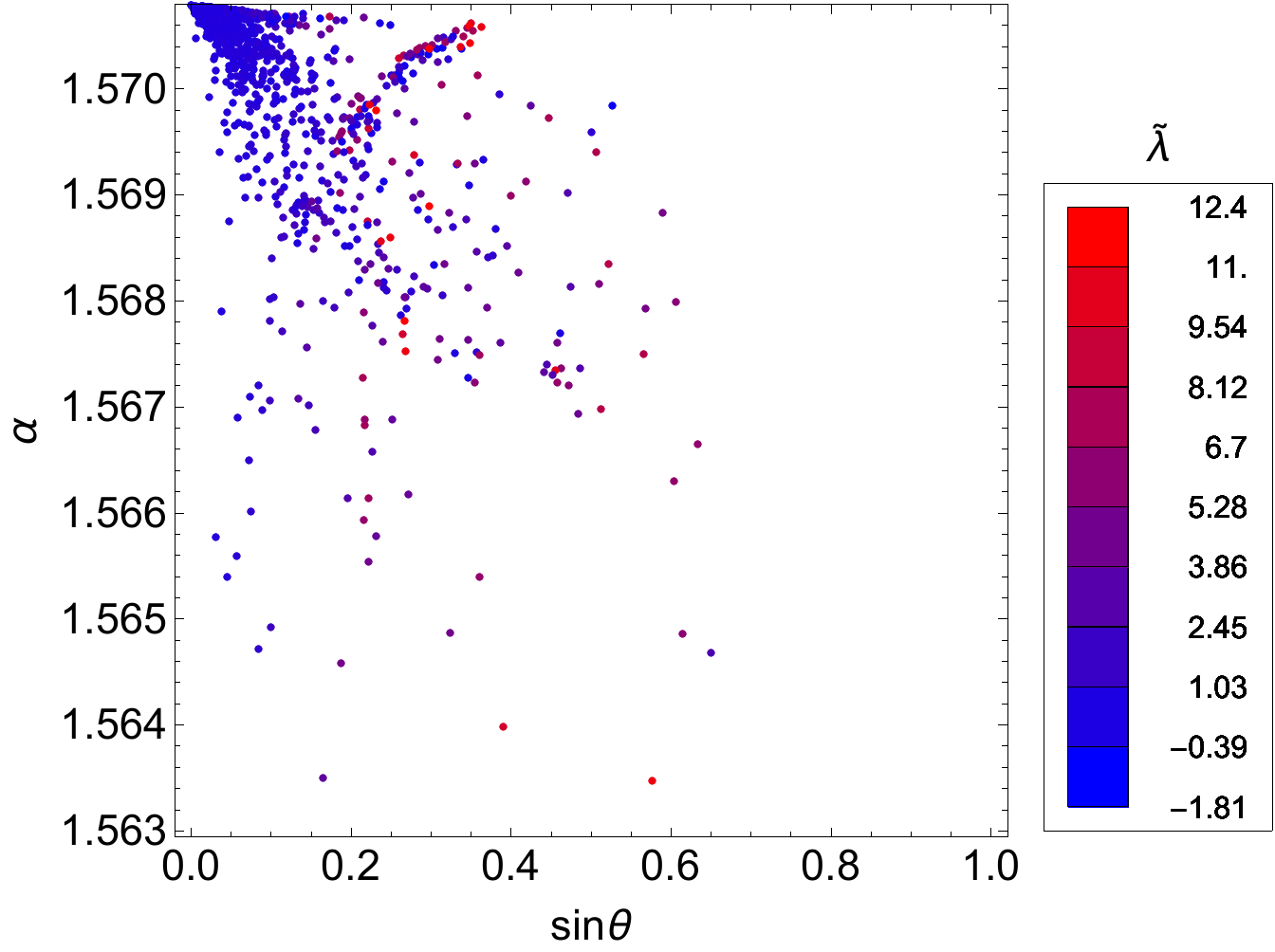}}
      \vspace*{-10pt}\caption{\label{fig:thetaalpha3} \textbf{Case 3.} Left Panel:  Distribution of  $\theta$.   In the upper right side of the plot
      we show the distribution for small $\theta$. The vertical solid line corresponds to the mode with  value of $ \overline\theta\;=0.018^{+0.004}_{-0.003}$. Right panel: 
      Correlation of the scalar mixing angle $\alpha$ and $\sin\theta$. The colour gradient 
       describes  the value of the parameter $\tilde\lambda$.} 
 \end{figure}

The distribution of $\theta$ as well as the correlation between $\alpha$ and $\theta$ are reported in Fig.~\ref{fig:thetaalpha3}.
Also in this case we deduce that the Higgs particle is mostly  the pNGB $\Pi_4$.  
We do not show here additional scatter plots since the results are very similar to the previous case.

\subsection{Electroweak precision observables}

The presence of new massive neutral and charged scalar fields can lead to sizeable corrections to the $Z$ and $W^\pm$ self energies.
In the model under study is therefore interesting to evaluate the couplings of the massive scalar fields contained in $M$ coupled to the gauge bosons.
The deviations with respect to the SM contributions can be described through the so called oblique parameters, $S$, $T$ and $U$   
 \cite{Peskin:1990zt, Peskin:1991sw}. 
The relevant diagrams are shown in Fig. \ref{fig:Feyn}. In Appendix~\ref{app:STU} we list all the relevant Feynman rules  and the main integrals used in the computation.
 We will use as experimental values for the oblique parameters, the values reported in   \cite{Baak:2011ze}
 (defined for the reference masses $m_{t,ref}=173$ GeV and $m_{h,ref}=125$ GeV):
\be S = 0.05 \pm 0.11,	\quad T = 0.09 \pm 0.13,	\quad U = 0.01 \pm 0.11  \,\,\,.\label{eq:STU}\ee

 The oblique parameters are defined through the vacuum polarisation amplitudes  as follows:
\begin{eqnarray}
	i\Pi_{XY}^{\mu\nu}(q^2)&=&ig^{\mu\nu}\Pi_{XY}(q^2)+(q^\mu q^\nu\mbox{terms})  \equiv \int  d^4xe^{-iqx}\langle J_X^\mu(x)J_Y^\nu(0)\rangle\;,\label{def-vacuum-pol}
\end{eqnarray}
where $(XY)=(11), (22), (33), (3Q),\mbox{and} (QQ)$ and 
\begin{eqnarray}
	\Pi_{XY}(q^2)\equiv \Pi_{XY}(0)+q^2\Pi_{XY}^\prime(q^2)\;.\label{def-vacuum-pol2}
\end{eqnarray}
Then, the three oblique   parameters $S$, $T$ and $U$ are  
\begin{eqnarray}
	&&\alpha S\equiv 4e^2[\Pi_{33}^\prime(0)-\Pi_{3Q}^\prime(0)]\;,\nonumber\\
	&&\alpha T\equiv\frac{e^2}{s_W^2c_W^2m_Z^2}[\Pi_{11}(0)-\Pi_{33}(0)]\;,\label{def-STU}\\
	&&\alpha U\equiv 4e^2[\Pi_{11}^\prime(0)-\Pi_{33}^\prime(0)]\;,\nonumber
\end{eqnarray}
where $\alpha\equiv e^2/(4\pi)$ is the fine-structure constant. 
The analytic expressions for $S$, $T$ and $U$ are listed in the Appendix~\ref{app:STU} in terms of the physical masses.

The parameters $T$ and $U$ are proportional to  $\cos^2(\theta+\alpha)$, which, due to the low value of $\theta$ and  $\alpha\sim\pi/2$, 
strongly suppresses any contribution coming from the heavier scalar $H$. On the other hand,
the $S$ parameter features two distinct contributions one still proportional to  $\cos^2(\theta+\alpha)$ and the other to $\sin^2\theta$. 
The function multiplying this latter term   depends explicitly on  $M_\Theta$ and $ M_{\tilde\Pi_i}$. 
 
\begin{figure}[t!]
  \subfigure
  {\includegraphics[width=5.2cm]{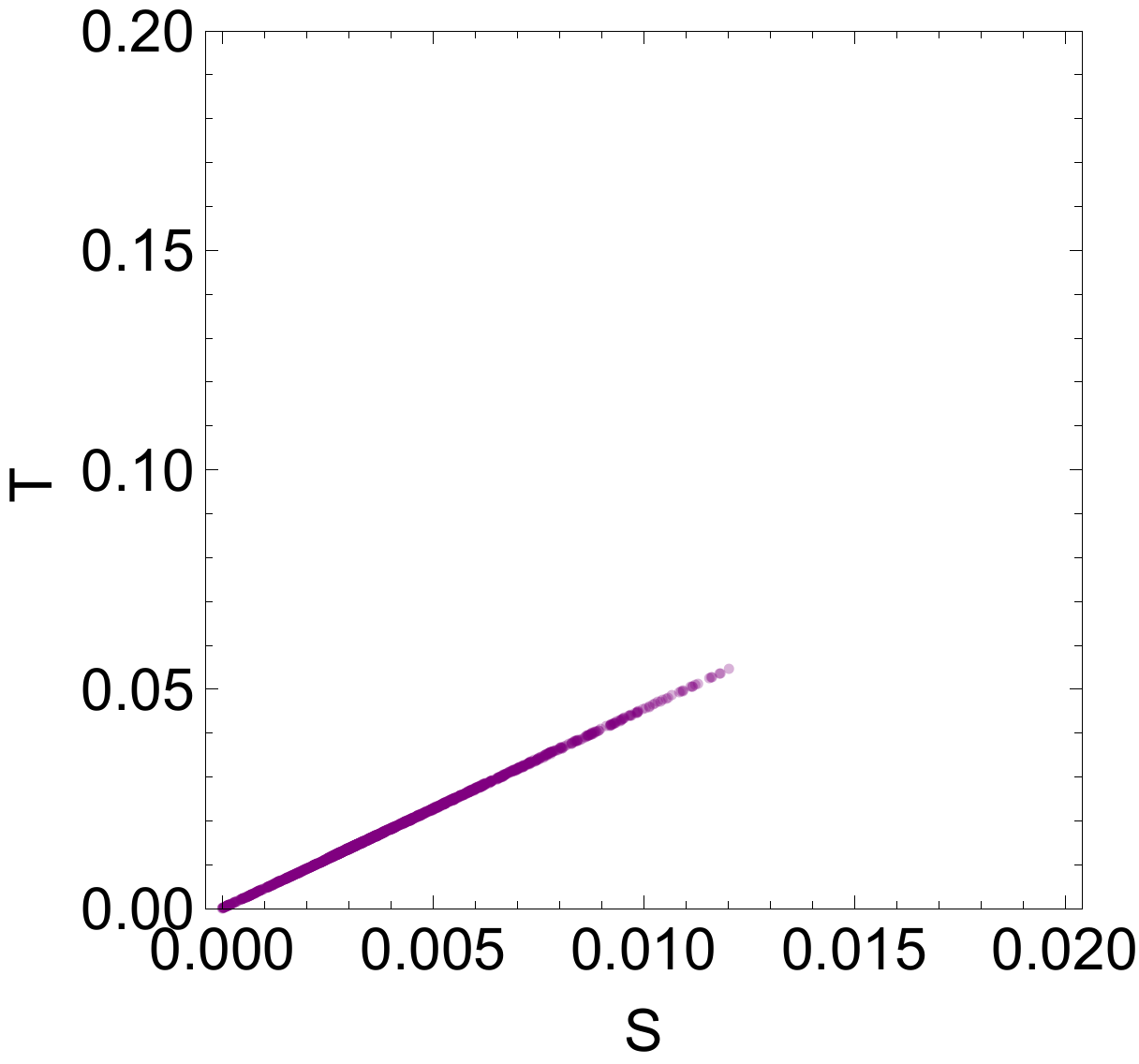}}
  \subfigure
  {\includegraphics[width=5cm]{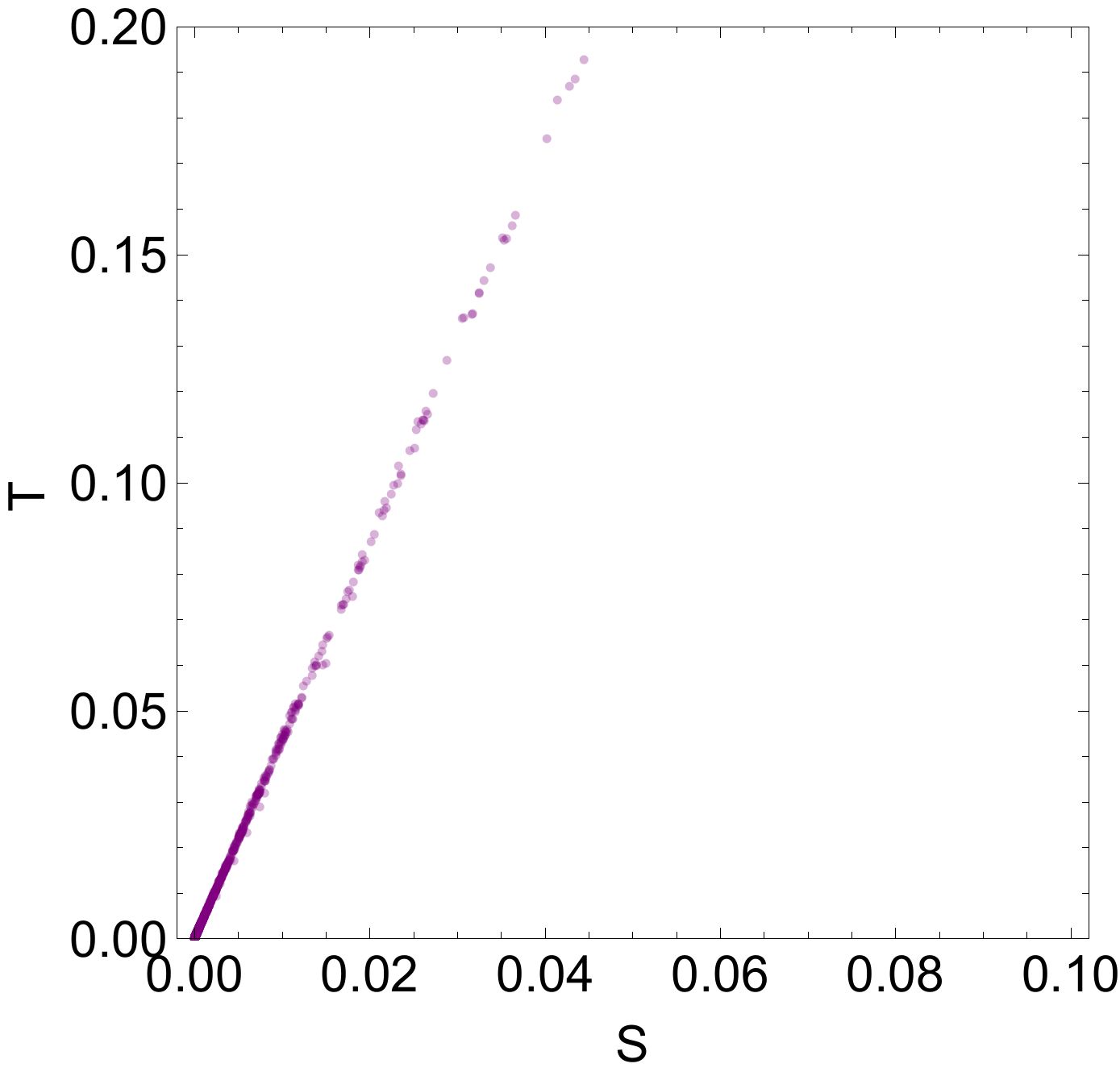}}
   \subfigure
  {\includegraphics[width=5cm]{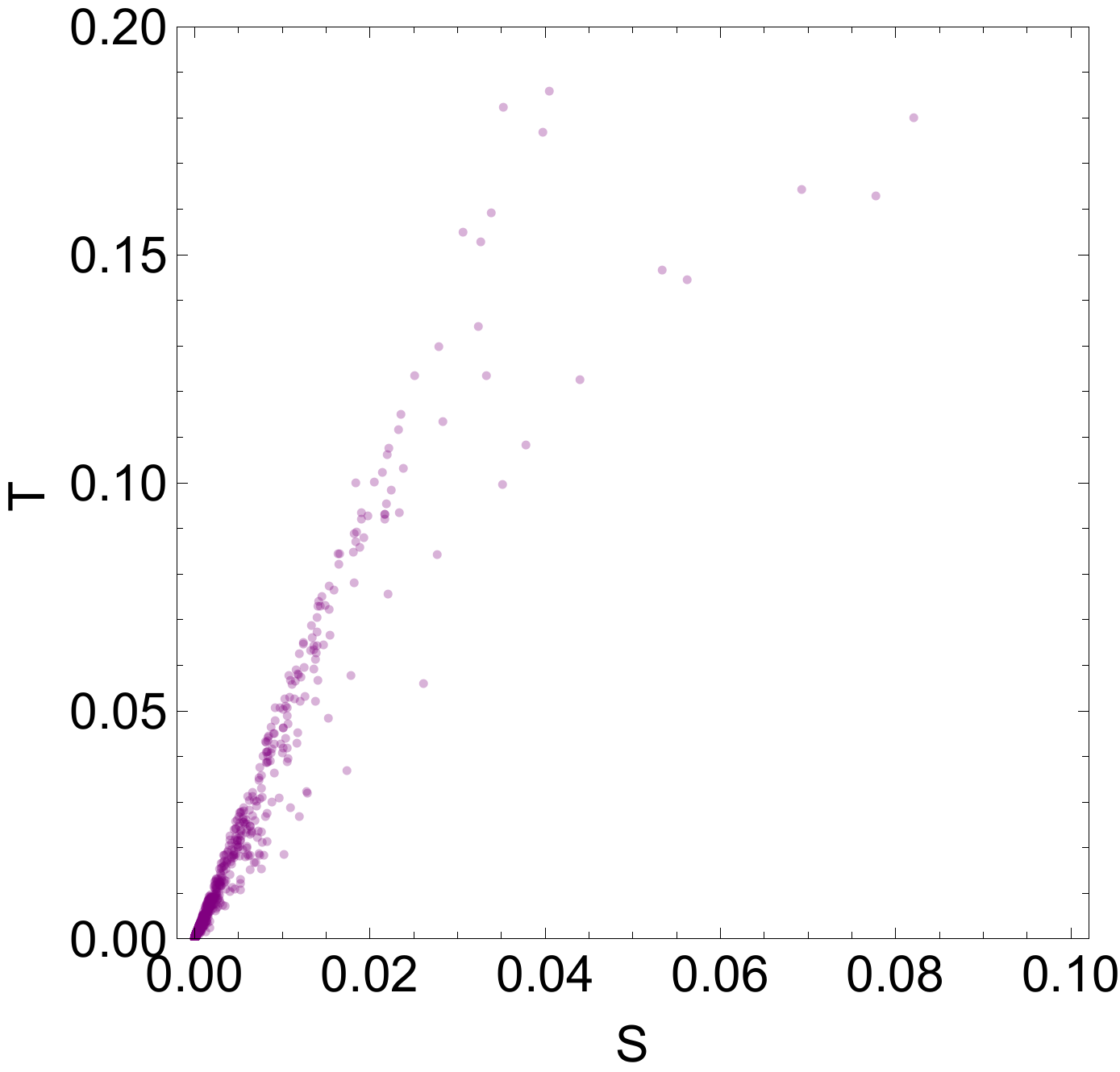}}
      \vspace*{-10pt}\caption{\label{fig:ST} {Correlation between S and T} 
      in the three regimes studied in in  Section~\ref{pheno}:  case 1 (left panel), case 2 (middle panel) and case 3 (right panel).  } 
 \end{figure}

\begin{figure}[t!]
  \subfigure
  {\includegraphics[width=5.1cm]{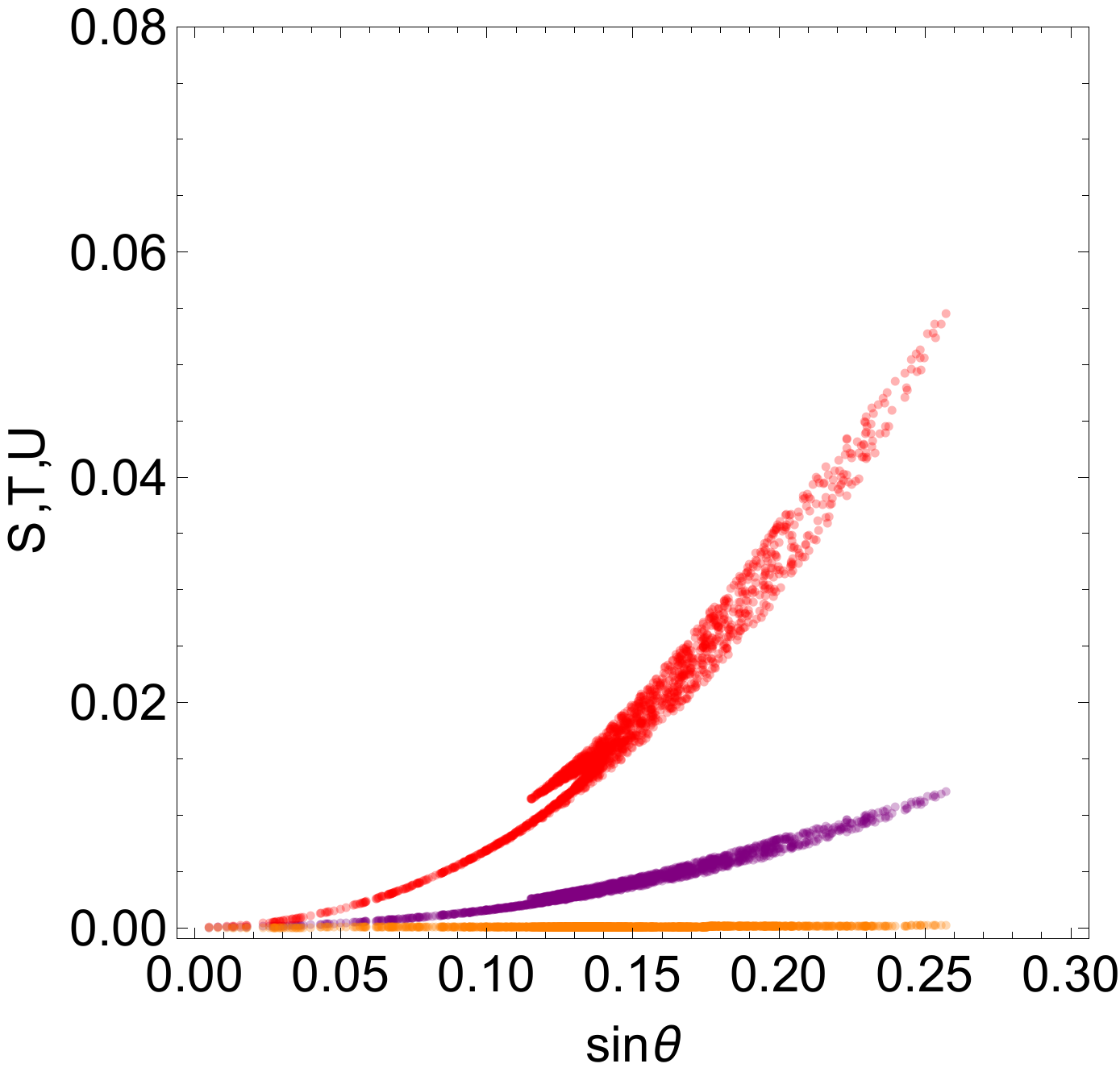}}
  \subfigure
  {\includegraphics[width=5cm]{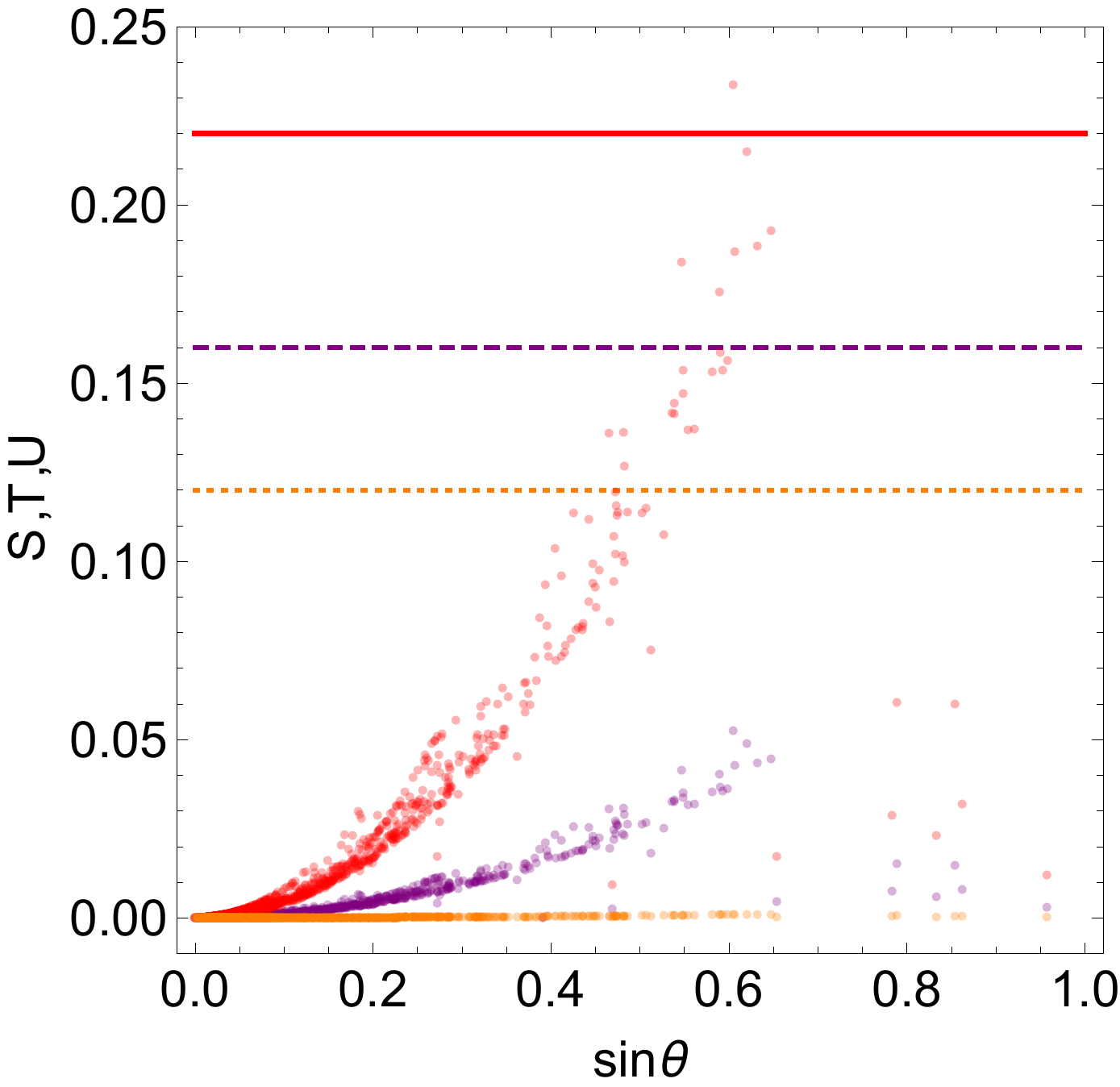}}
  \subfigure
  {\includegraphics[width=5cm]{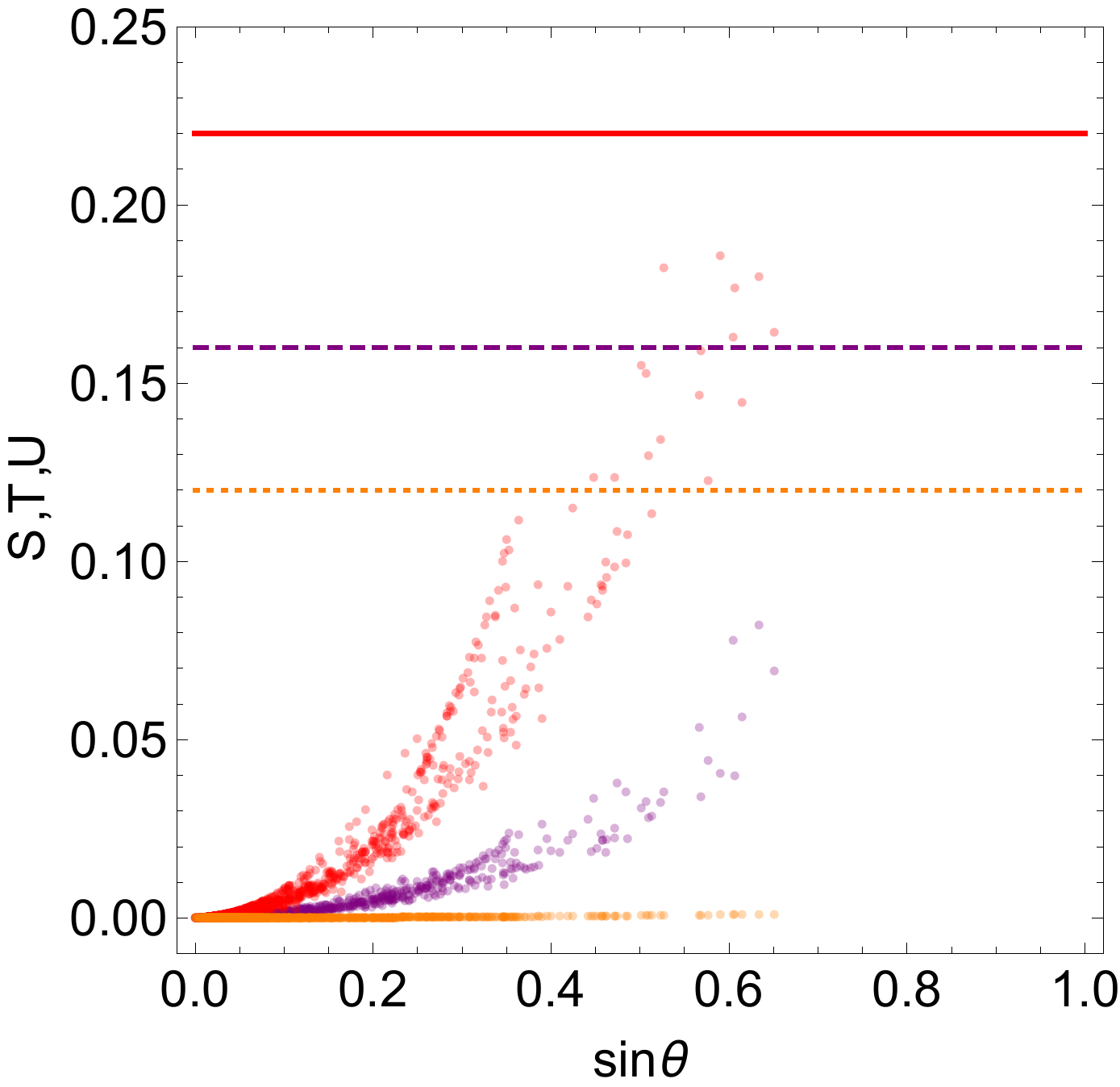}}
      \vspace*{-10pt}\caption{\label{fig:STUtheta} The parameters $S$ (purple) , $T$ (red)  and $U$ (orange) as function of $\sin\theta$, the angle defining the vacuum alignment of the model. The scatter plots  correspond to the scalar spectra studied in the Section~\ref{pheno}, i.e.,  case 1 (left panel),  case 2 
      (middle panel) and  case 3 (right panel).  
      As expected, the larger is the value of $\sin\theta$ the larger are the contributions to the oblique parameters.
      The solid horizontal lines represents the bounds from the fit reported in~\eqref{eq:STU}.} 
 \end{figure}

 In Fig.~\ref{fig:ST} {we report the correlation between} 
 $S$ and $T$ for all the three cases studied  in  Section~\ref{pheno}. 
 We show {for completeness} 
 the scatter plots of $S$ versus $U$  and $T$ versus $U$  in Appendix~\ref{app:STU},  respectively in Figs. \ref{fig:SU} and \ref{fig:TU}.  
 {The variation of $S$, $T$ and $U$ with respect to $\sin\theta$ is shown in Fig.~\ref{fig:STUtheta}}. 
 
 It is clear from this analysis, and scatter plots in Figs.~\ref{fig:ST} and \ref{fig:STUtheta}, that the model is phenomenologically viable and that only few points can be affected by the EW precision constraints. The results also show which level of precision must be reached 
 in order to constrain the  {parameter space of the} theory.

\section{Conclusions}

In this paper we analysed in detail the paradigm, put forward in the exploratory work \cite{Alanne:2014kea}, according to which the elementary Higgs sector of the Standard Model is enhanced to an $SU(4)$ symmetry that breaks spontaneously to $Sp(4)$.
The embedding of the electroweak gauge sector is parametrised by an angle  {$0\leq\theta\leq \pi/2$}. 
In this scenario the observed Higgs particle is shown to emerge as a pNGB  with its mass arising via radiative corrections.
 
Via detailed analytical and numerical analyses we have demonstrated that not only a pNGB of the Higgs is possible but that it is, indeed, naturally embodied in the elementary realisation. 
 We analysed several parameter space scenarios. For the most general one {(case 3 in subsection~\ref{case3sub})} we demonstrated that the preferred electroweak alignment angle is centred around $\theta \simeq 0.02 $, corresponding to the Higgs chiral symmetry breaking scale  $ f \simeq 14~$TeV.  This is almost 60 times higher than the SM electroweak scale. Due to the perturbative nature of the theory the  new scalars remain in the few TeV energy range.

We stress that it is the structure of the quantum corrections that is responsible for this intriguing result.  Crucially, the origin and structure of the quantum corrections here, dictating the specific electroweak vacuum alignment physics, is  dramatically different from the composite cousin models. The reason being that for the composite case the final vacuum alignment is dictated by cut-off contributions that require new operators rendering the pNGB nature fine-tuned \cite{Cacciapaglia:2014uja}. Furthermore, in the composite case the new resonances are of the order of $4\pi f$ and therefore typically much harder to discover at present and future colliders. 

 We have also determined the electroweak precision parameters and shown that the model is phenomenological viable.  Several technical details are summarised in the appendices including the Feynman rules of the model. 
 
The elementary nature of the pNGB Higgs permits to investigate different aspects of the ultraviolet physics such as a potential asymptotically safe nature \cite{Litim:2014uca,Litim:2015iea} or lead to new theories of grand unification.

\section{Acknowledgments}
We thank T. Alanne  and G. Cacciapaglia  for relevant  discussions.
 The CP$^3$-Origins center is partially supported by the DNRF:90 grant.

\appendix

\section{Group Generators}\label{app:generators}
We use the following  parametrization of  the  vacuum  of the theory:
    \begin{equation}
	\label{eq:vacuum}
	E_{\theta}=\cos\theta \,E_B+\sin\theta\, E_H\,.
    \end{equation}
    We give here the representation of the unbroken and broken generators of $SU(4)$, respectively $S_\theta^i$ with $i=1,...,10$ and $X_\theta^i$ with $i=1,...,5$,  associated with $E_{\theta}$:

    \begin{equation}
	\label{eq:}
	\begin{split}
	    &S_{\theta}^{1,2,3}=
	    \frac{1}{2\sqrt{2}}
	    \left(
		\begin{array}{cc}
		\sigma_{i}   & 0 \\
		0		    & -\sigma_{i}^{T}
		\end{array}
	    \right),\quad
	    S_{\theta}^{4}=\frac{1}{2\sqrt{2}}\left(
		\begin{array}{cc}
		0   & i\sigma_{1} \\
		-i\sigma_{1}		    & 0
		\end{array}
	    \right),\quad
	    S_{\theta}^{5}=\frac{1}{2\sqrt{2}}\left(
		\begin{array}{cc}
		0   & i\sigma_{3} \\
		-i\sigma_{3}		    & 0
		\end{array}
	    \right),\\
	    &S_{\theta}^{6}=\frac{1}{2\sqrt{2}}\left(
		\begin{array}{cc}
		0   & 1 \\
		1		    & 0
		\end{array}
	    \right),\\
   	    &S_{\theta}^{7}=\frac{\cos\theta}{2\sqrt2}\left(
		\begin{array}{cc}
		\sigma_{1} &0\\
		0& \sigma_{1}^{T}
		\end{array}
	    \right)
	    +\frac{ \sin\theta  }{2\sqrt{2}}\left(
		\begin{array}{cc}
		0   & \sigma_{3} \\
		\sigma_{3}		    & 0
		\end{array}
	    \right),\quad
	    S_{\theta}^{8}=\frac{\cos\theta}{2\sqrt2}\left(
		\begin{array}{cc}
		\sigma_{2} &0\\
		0& \sigma_{2}^{T}
		\end{array}
	    \right)
	    -\frac{ \sin\theta  }{2\sqrt{2}}\left(
		\begin{array}{cc}
		0   & i \\
		-i		    & 0
		\end{array}
	    \right),\\
	    &S_{\theta}^{9}=\frac{\cos\theta}{2\sqrt2}\left(
		\begin{array}{cc}
		\sigma_{3} &0\\
		0& \sigma_{3}^{T}
		\end{array}
	    \right)
	    -\frac{ \sin\theta  }{2\sqrt{2}}\left(
		\begin{array}{cc}
		0   & \sigma_{1} \\
		\sigma_{1}		    & 0
		\end{array}
	    \right),\quad
	    S_{\theta}^{10}=\frac{\cos\theta}{2\sqrt2}\left(
		\begin{array}{cc}
		0 & i\sigma_{2}\\
		-i\sigma_{2} & 0
		\end{array}
	    \right)
	    +\frac{ \sin\theta  }{2\sqrt{2}}\left(
		\begin{array}{cc}
		1   & 0 \\
		0    & -1
		\end{array}
	    \right).
	\end{split}
    \end{equation}
    and
    \begin{equation}
	\begin{split}
	    \label{eq:}
	    &X_{\theta}^{1}=\frac{\cos\theta}{2\sqrt{2}} \left(
		\begin{array}{cc}
		0   &  \sigma_{3}\\
		\sigma_{3}    & 0
		\end{array}
	    \right)-\frac{\sin\theta}{2\sqrt{2}}\left(
		\begin{array}{cc}
		\sigma_{1}   & 0 \\
		0    & \sigma_{1}^{T}
		\end{array}
	    \right),\quad
	    X_{\theta}^{2}=\frac{\cos\theta}{2\sqrt{2}} \left(
		\begin{array}{cc}
		0   &  i\\
		-i    & 0
		\end{array}
	    \right)+\frac{\sin\theta}{2\sqrt{2}}\left(
		\begin{array}{cc}
		\sigma_{2}   & 0 \\
		0    & \sigma_{2}^{T}
		\end{array}
	    \right)\\
	    &X_{\theta}^{3}=\frac{\cos\theta}{2\sqrt{2}} \left(
		\begin{array}{cc}
		0   &  \sigma_{1}\\
		\sigma_{1}    & 0
		\end{array}
	    \right)+\frac{\sin\theta}{2\sqrt{2}}\left(
		\begin{array}{cc}
		\sigma_{3}   & 0 \\
		0    & \sigma_{3}^{T}
		\end{array}
	    \right),\quad
	    X_{\theta}^{4}=\frac{1}{2\sqrt{2}} \left(
		\begin{array}{cc}
		0   &  \sigma_{2}\\
		\sigma_{2}    & 0
		\end{array}
	    \right),\\
   	    &X_{\theta}^{5}=\frac{\cos\theta}{2\sqrt{2}} \left(
		\begin{array}{cc}
		1  &  0\\
		0    & -1
		\end{array}
	    \right)-\frac{\sin\theta}{2\sqrt{2}}\left(
		\begin{array}{cc}
		0 & i\sigma_{2}   \\
		-i \sigma_{2} & 0
		\end{array}
	    \right),
	\end{split}
    \end{equation}
where we use the following normalisation: 
    \begin{equation}
	\label{eq:genNorm}
	\mathrm{Tr}[S_{\theta}^aS_{\theta}^b]=\frac{1}{2}\delta^{ab},\quad \mathrm{Tr}[X_{\theta}^iX_{\theta}^j]
	    =\frac{1}{2}\delta^{ij}.
    \end{equation}
    Following \cite{Alanne:2014kea}, we choose to embed in $SU(4)$  the full custodial symmetry group of the SM, $SU(2)_L\otimes SU(2)_R$, identifying the left and right generators as:
    
    \be T^i_L=\frac{1}{2}\begin{pmatrix}
    \sigma_i & 0\\
    0 & 0
    \end{pmatrix}, \quad T^i_R=\frac{1}{2}\begin{pmatrix}
  0   & 0\\
    0 & -\sigma_i^T
    \end{pmatrix}\ee
where $\sigma_i$ with $i=1,2,3$ are the Pauli matrices. The generator of the hypercharge is  identified with the third generator of the $SU(2)_R$ group, $T_Y = T_{R}^3$ .

\section{Stability of the potential}\label{app:stability}

We study in this appendix the stability of the tree-level scalar potential given in (\ref{eq:pot}).
The latter  can be recast using  the following sextuplets:
    \begin{equation}
	\label{eq:sext}
	\varphi_1=(\sigma,i \mathbf{\Pi}),\quad\text{and}\quad\varphi_2=(\Theta,- i  \mathbf{\tilde{\Pi}}),
    \end{equation}
    which allow  to express the following real quantities:
    \begin{equation}
	\label{eq:phis}
	\varphi_1^{\dagger}\varphi_1=|\varphi_1|^2=\sigma^2+\mathbf{\Pi}^2,\quad \varphi_2^{\dagger}\varphi_2=|\varphi_2|^2=\Theta^2+\mathbf{\tilde{\Pi}}^2,\quad
	    \text{and}\quad\varphi_1^{\dagger}\varphi_2=\rho\,|\varphi_1|\, |\varphi_2| =\sigma\Theta-\mathbf{\Pi}\cdot\mathbf{\tilde\Pi}.
    \end{equation}
   In the last expression we have $|\rho|\in [0,1]$, because of the Cauchy inequality, i.e. $0\leq| \varphi_1^\dagger \varphi_2| \leq|\varphi_1|\, |\varphi_2| $.  
   In terms of the two sextuplets, the potential reads
    \begin{equation}
	\label{eq:potential2}
	\begin{split}
	    V_M=&\frac{1}{2}m_M^2(|\varphi_1|^2+|\varphi_2|^2)+\frac{1}{2}c_{MR}(-|\varphi_1|^2
		+|\varphi_2|^2)
		+c_{MI}\rho\,|\varphi_1|\, |\varphi_2| 
	    +\frac{\lambda}{4}(|\varphi_1|^2+|\varphi_2|^2)^2\\
	    &+\lambda_1\left[\frac{1}{4}(|\varphi_1|^2+|\varphi_2|^2)^2+(|\varphi_1|^2)
		(|\varphi_2|^2)-(\rho\,|\varphi_1|\, |\varphi_2| )^2\right]\\
	    &-\lambda_{2R}\left[\frac{1}{4}(|\varphi_1|^2-|\varphi_2|^2)^2
		-(\rho\,|\varphi_1|\, |\varphi_2| )^2\right]-\lambda_{2I}(-|\varphi_1|^2+|\varphi_2|^2)
		\rho\,|\varphi_1|\, |\varphi_2| \\
	    &\left[\frac{\lambda_{3R}}{4}(-|\varphi_1|^2+|\varphi_2|^2)
		+\frac{\lambda_{3I}}{2}\rho\,|\varphi_1|\, |\varphi_2| \right](|\varphi_1|^2+|\varphi_2|^2)\,.
	\end{split}
    \end{equation}
   Restricting ourselves only to real parameters, i.e. setting $c_{MI}=\lambda_{2I}=\lambda_{3I}=0$, the potential becomes    
\begin{equation}
	\label{eq:potential3}
	\begin{split}
	    V_M=&\frac{1}{2}m_M^2(|\varphi_1|^2 + |\varphi_2|^2)-\frac{1}{2}c_{MR}(|\varphi_1|^2 - |\varphi_2|^2)
		+\lambda_{11} |\varphi_1|^4 +\lambda_{22} |\varphi_2|^4 +2 \lambda_{12} |\varphi_1|^2  |\varphi_2|^2\,,
	\end{split}
    \end{equation}  
    where we introduce the effective couplings
    \be 
    \begin{split}
   & \lambda_{11} \;=\; \frac 1 4\, \left(\lambda+ \lambda_1-\lambda_{2}
   -\lambda_{3}\right)\,,\\
    &\lambda_{22} \;=\; \frac 1 4 \left(\lambda+ \lambda_1-\lambda_{2}
    +\lambda_{3}\right)\,,\\
    & \lambda_{12}\;=\; \frac 1 4 \left(\lambda+3\lambda_1+\lambda_{2}
    \right) \;+\; \frac 1 2 \rho^2 \,\left(-\lambda_1 
    +\lambda_{2}\right)\,.
    \end{split}
    \ee
Then, the minimum of the potential is  reached if the following conditions are satisfied:
    \be \label{mina}\begin{cases}
    \rho=0 & \mbox{if}\quad (-\lambda_1 
    +\lambda_{2})\geq 0\\
    \rho=\pm1 & \mbox{if}\quad (-\lambda_1 
    +\lambda_{2})\leq 0
    \end{cases}\,.\ee
In order to have a stable vacuum configuration, a scalar potential of the form $\lambda_{ab}\,\phi_a^2\,\phi_b^2$ has to be bounded from below (the quartic potential indeed is a biquadratic form of real fields). This in particular requires that the effective quartic coupling of the scalar potential in eq.~(\ref{eq:potential3})  must be positive  for all values of the fields and for all scales. 
The copositivity conditions are less restrictive conditions with respect to the Sylvester criterium since the biquadratic form has its domain in the 
non-negative orthant 
$\mathbb{R}^n_{+}$ and not in the whole $\mathbb{R}^n$. The conditions of copositivities  \cite{Kannike:2012pe} can be derived as follows, taking into account the minima conditions in eq.~(\ref{mina}):
\be 
\begin{split}
 & \lambda_{11} \geq 0\quad \wedge \quad  \lambda_{22}\geq 0 \quad   \wedge \quad  \lambda_{12}+\sqrt{ \lambda_{11} \lambda_{22} } \geq 0\,. \label{listconds}\\
 \end{split}
\ee
So either $\lambda_{12}$ is positive or it is negative, with  $|\lambda_{12}|\in \left[0,\sqrt{\lambda_{11}\lambda_{22}}\right]$. 

The condition that $\lambda_{11}\geq 0$ is always satisfied given the relation reported in eq.~(\ref{eq:ren}). 
Conversely, the condition  $\lambda_{22}\geq 0$ implies the relation
\begin{eqnarray}
	\frac{ \tilde\lambda}{2}-2\lambda_{2}-\frac{M_\sigma ^2}{8\, f^2}
	&\geq &0\label{eq:cond2}\,,
\end{eqnarray}
with $\tilde \lambda$ defined in eq.~(\ref{tildeL}). 
Finally, the last condition  in (\ref{listconds})  gives
\begin{eqnarray}
	2 \tilde\lambda-4 \lambda_f \left( \rho ^2-1\right)+ \frac{M_\sigma}{f^2} \sqrt{ 4 \,f^2 (\tilde\lambda-4\lambda_{2})  -  M_\sigma^2 }&\geq &0\label{eq:cond3}\,,
\end{eqnarray}
where $\lambda_f$ is given by eq.~(\ref{lambdaF}), while the request of positivity of the radicand corresponds to the condition~\eqref{eq:cond2}. 
Therefore, if  $\rho=0~(\pm1)$, then $\lambda_f < 0$ ($>0$) and eq.~\eqref{eq:cond3} gives a lower bound
\be \label{mina2}\begin{cases}
    \rho=0 & \mbox{then}\quad2 \,\tilde\lambda-4 |\lambda_f|\geq - C \\
    \rho=\pm 1& \mbox{then}\quad2\, \tilde\lambda\geq -C \\    \end{cases}\,,\ee
$C\equiv\frac{M_\sigma}{f^2} \sqrt{ 4 f^2 (\tilde\lambda-4\lambda_{2})  -  M_\sigma^2 }$ being a  positive number.

\section{Electroweak Precision Observables}\label{app:STU}
When we gauge  the electroweak group, the three  degrees of freedom associated with the generators $S^{1,2,3}$, namely  the three NGBs, $\Pi_{1,2,3}$, become the longitudinal components of the massive gauge bosons $W^{\pm}$ and $Z$.  We use the following notation for the $W^\pm$ bosons:
\be W^{\pm}=\frac{\left(W_1 \mp i W_2 \right)}{\sqrt{2}}.\ee 
We are interested to identify the scalar degrees of freedom which are charged under $U(1)_{EM}$. In particular we find convenient to redefine  $\Pi_{1,2}$  and  $\tilde\Pi_{1,2}$  in the following way:
\be \Pi^{\pm}=\frac{\left(\Pi_2 \mp i \Pi_1 \right)}{\sqrt{2}},\qquad \tilde\Pi^{\pm}=\frac{\left(\tilde\Pi_2 \mp i \tilde\Pi_1 \right)}{\sqrt{2}}.\ee
The latter contribute at one loop  to the self energy of the gauge bosons. The topology of the relevant diagrams is depicted in Fig. \ref{fig:Feyn} and the explicit computation is reported below.
 \begin{figure}[t!]
	\centering
\subfigure{
	\begin{fmffile}{PiOmega} 
	\begin{fmfgraph*}(100,70)
		\fmfleft{i1,i2} 
		\fmfright{o1,o2} 
		
		\fmf{wiggly}{i1,v,o1}
		\fmf{dashes,tension=0.5}{v,v}
	\end{fmfgraph*}
	\end{fmffile}}
\subfigure{
		\begin{fmffile}{PiPhi} 
		\begin{fmfgraph*}(100,70)
			\fmfleft{i1} 
			\fmfright{o1} 
		
			\fmf{wiggly}{i1,v1}
			\fmf{dashes,left,tension=0.5}{v1,v2}
			\fmf{dashes,left,tension=0.5}{v2,v1}
			\fmf{wiggly}{v2,o1}
		\end{fmfgraph*}
		\end{fmffile}}
\subfigure{
		\begin{fmffile}{PiPsi} 
		\begin{fmfgraph*}(100,70)
			\fmfleft{i1} 
			\fmfright{o1} 	
			\fmf{wiggly}{i1,v1}
			\fmf{dashes,left,tension=0.5}{v1,v2}
			\fmf{wiggly,left,tension=0.5}{v2,v1}
			\fmf{wiggly}{v2,o1}
		\end{fmfgraph*}
		\end{fmffile}}
\caption{\label{fig:Feyn}Feynman diagrams contributing to the vacuum polarisation amplitude.  The contributions from charged and neutral scalars appearing in the theory, together with their couplings, are given in the appendix. }
\end{figure}

\begin{figure}[t!]
  \subfigure
  {\includegraphics[width=5.2cm]{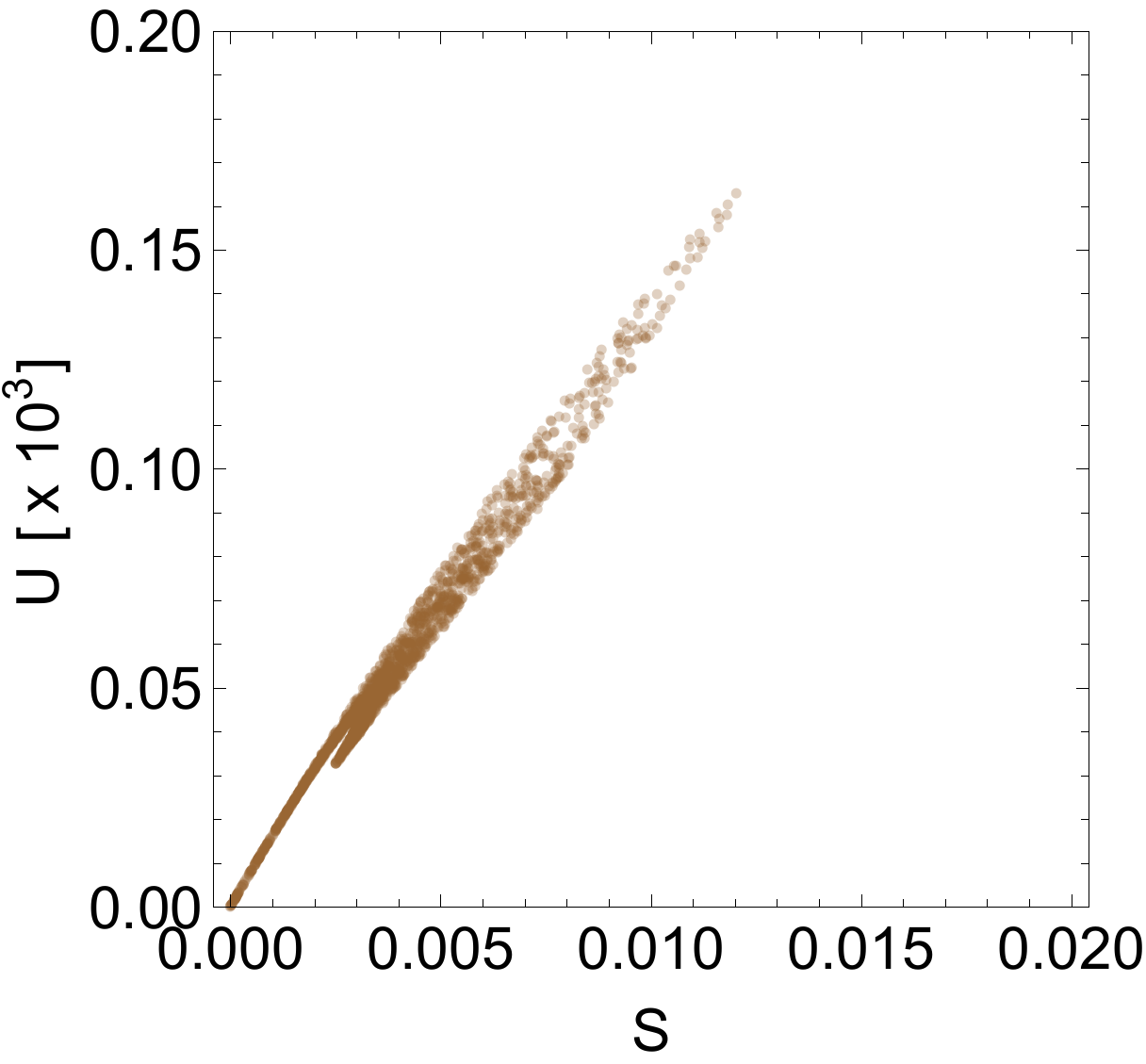}}
  \subfigure
  {\includegraphics[width=5cm]{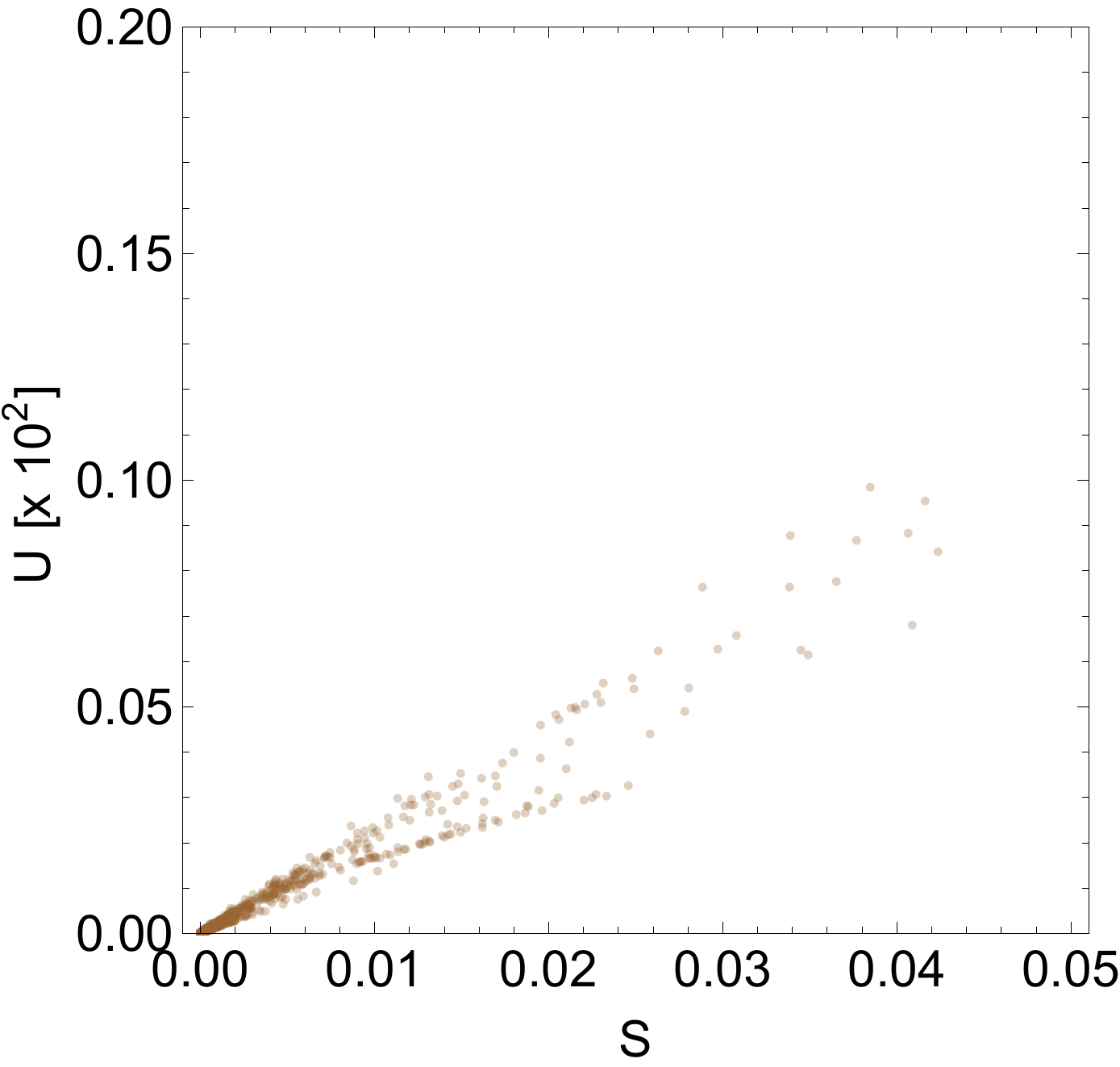}}
   \subfigure
  {\includegraphics[width=5cm]{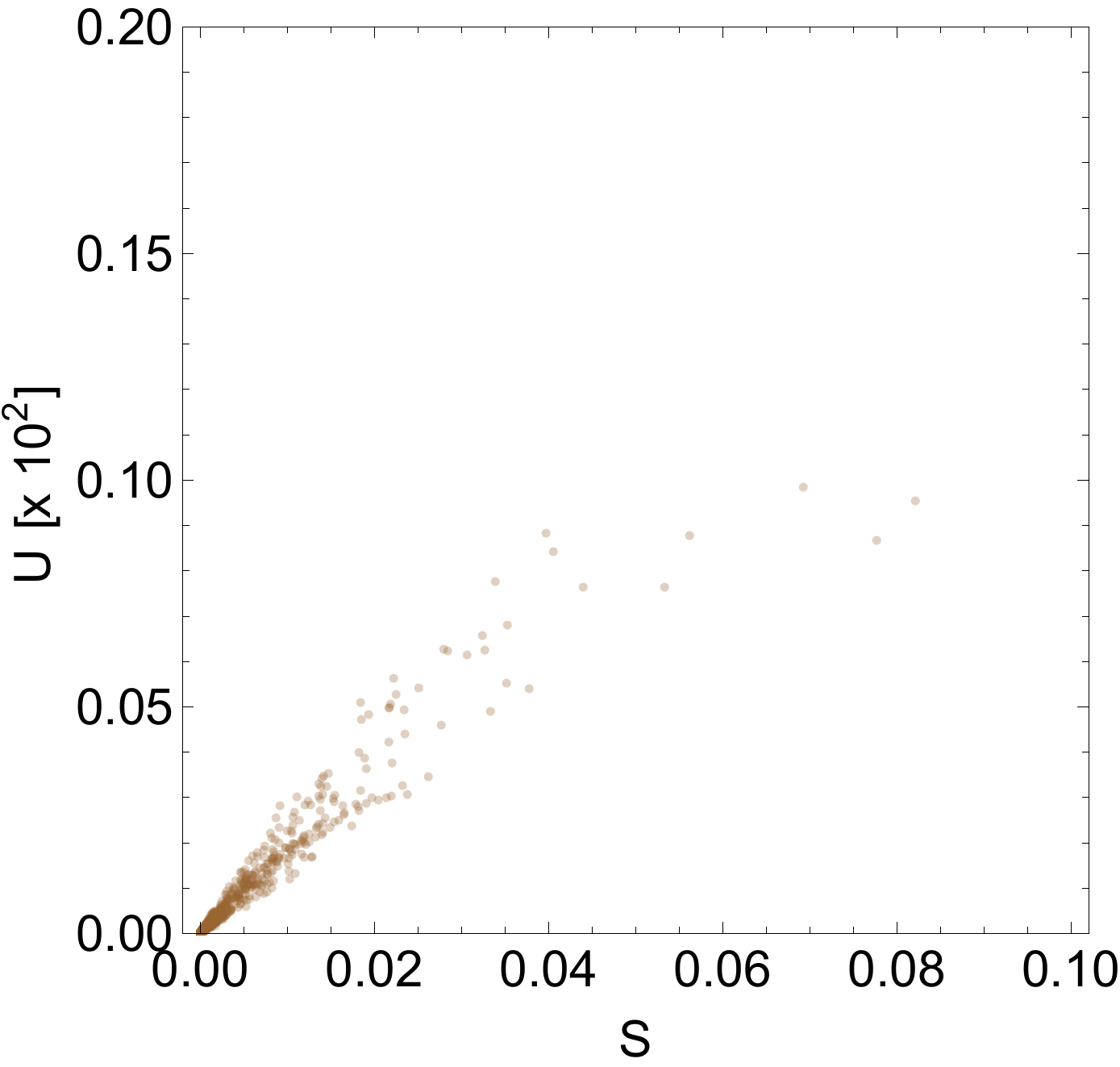}}
      \vspace*{-10pt}\caption{\label{fig:SU}  Correlation between  $S$ and $U$ in the three regimes studied in the Section~\ref{pheno}:   case 1
      (left panel), case 2 (middle panel) and  case 3 (right panel).  } 
 \end{figure}

\begin{figure}[t!]
  \subfigure
  {\includegraphics[width=5.1cm]{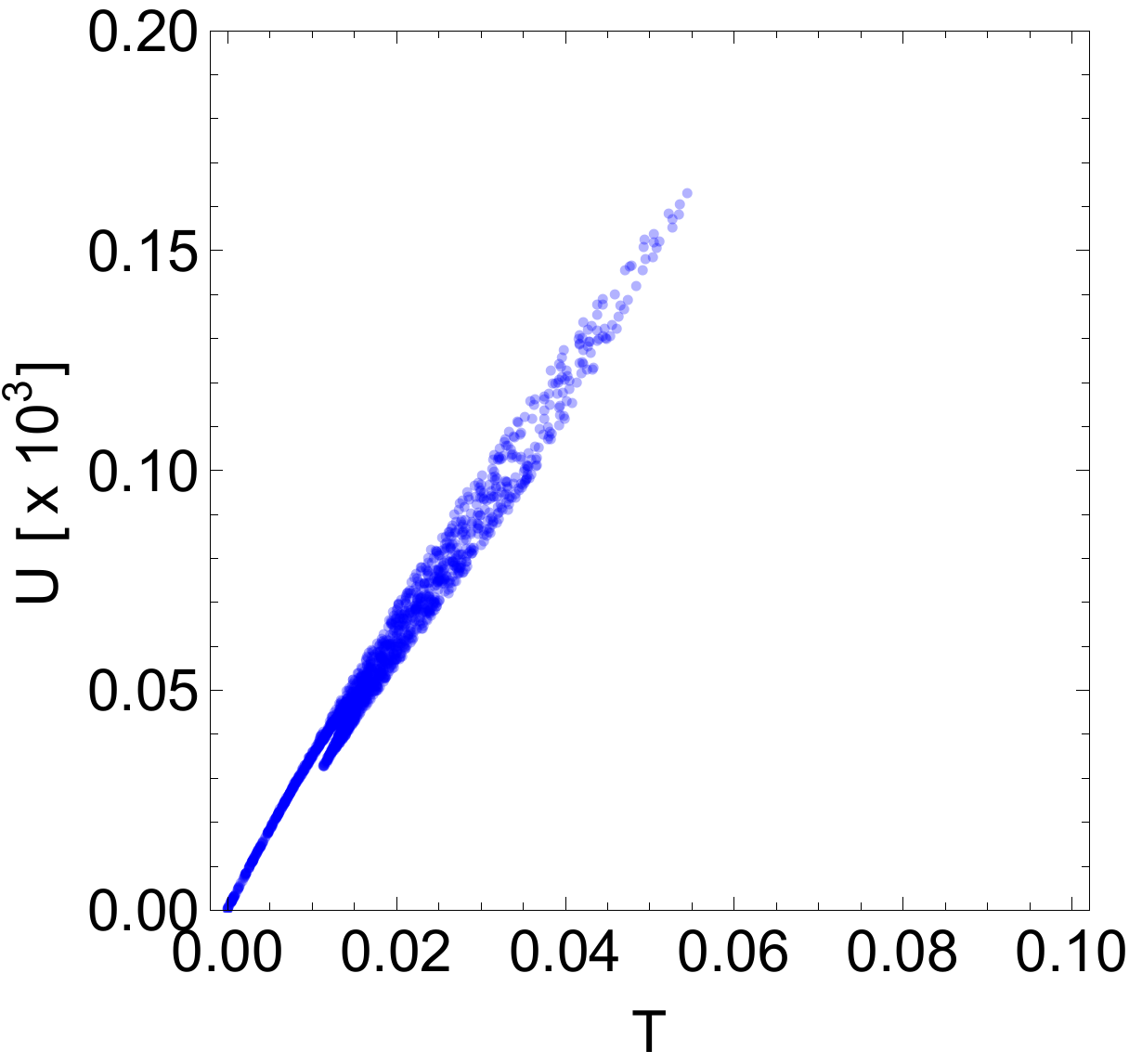}}
  \subfigure
  {\includegraphics[width=5cm]{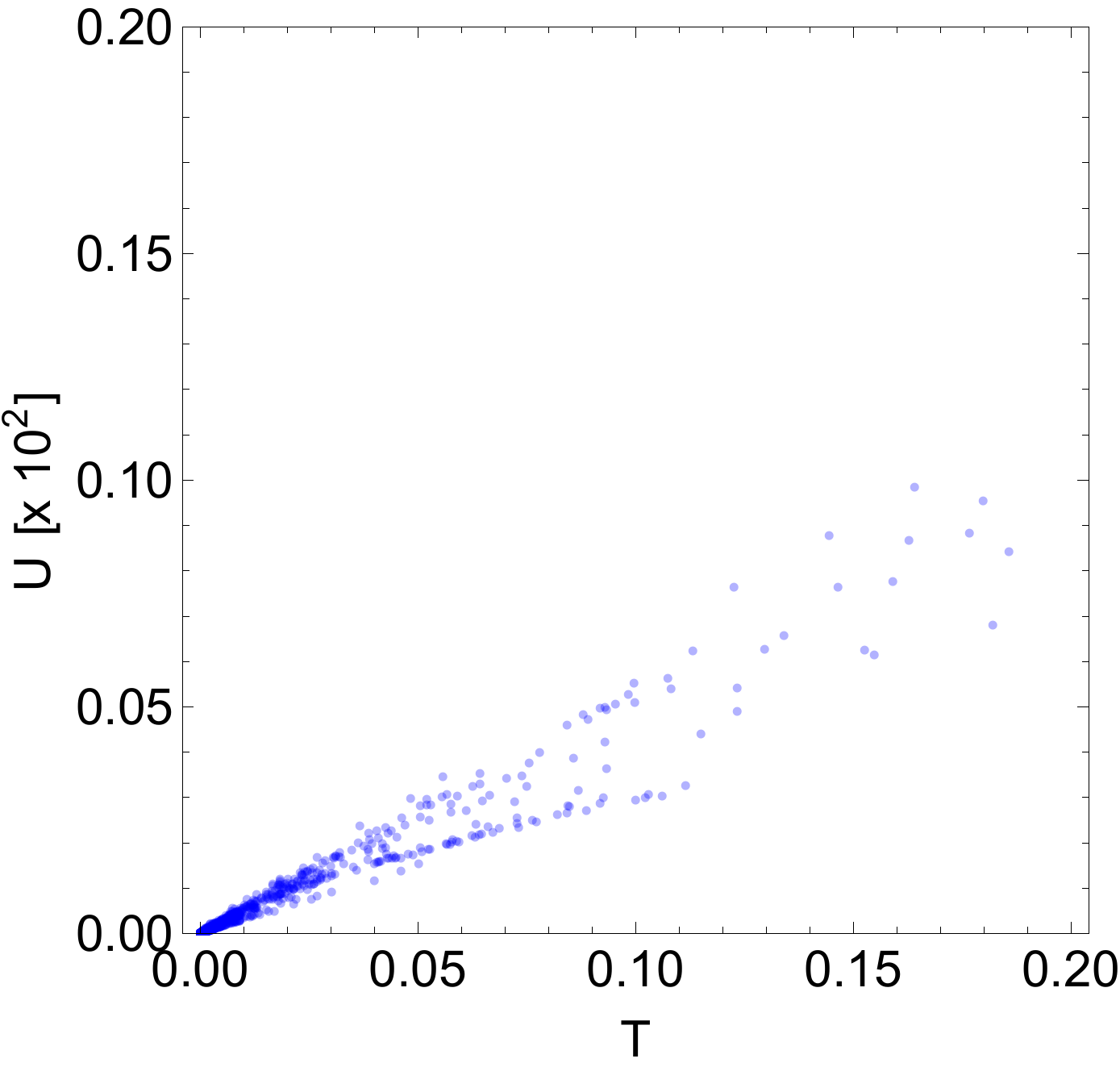}}
   \subfigure
  {\includegraphics[width=5cm]{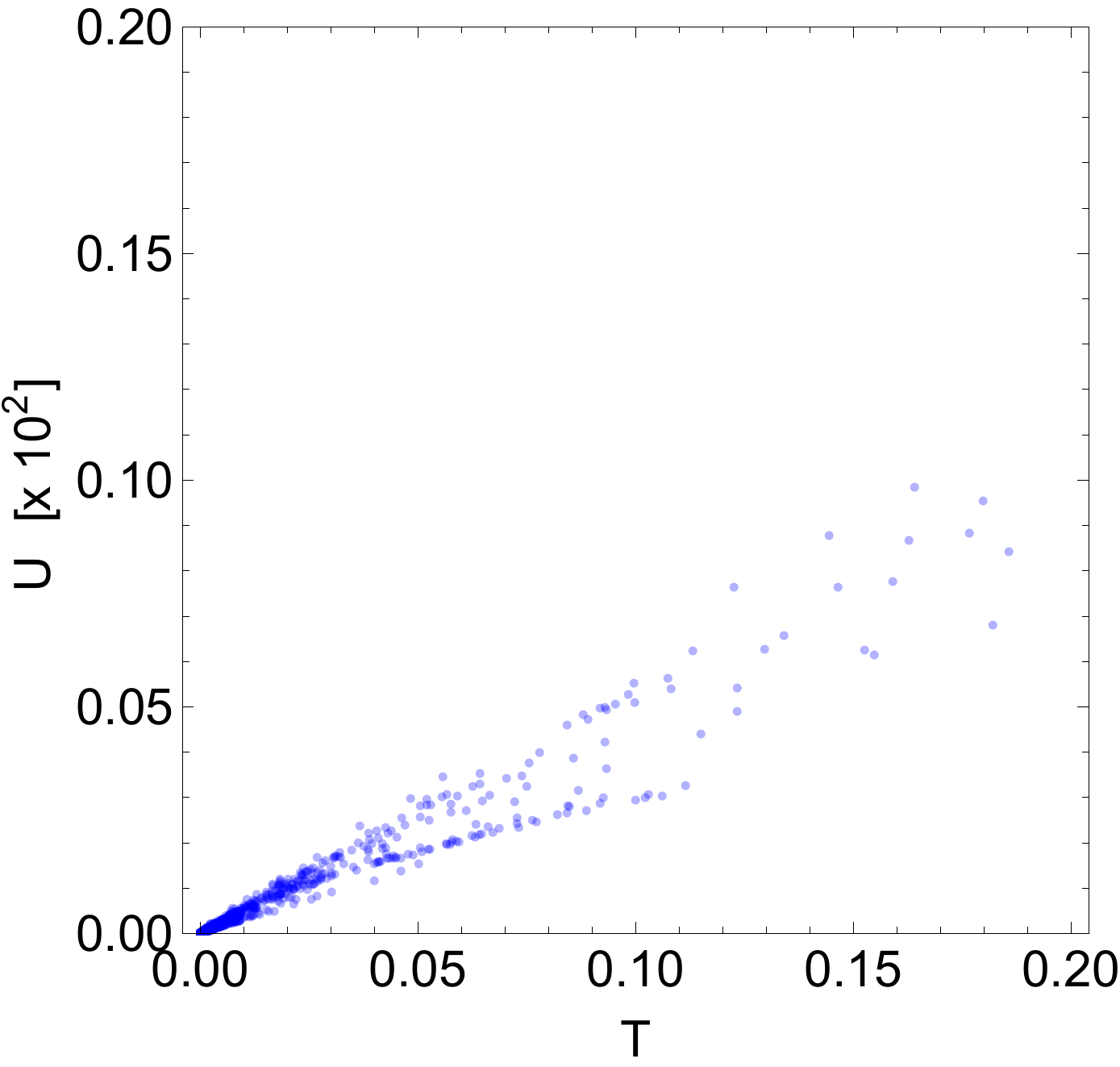}}
      \vspace*{-10pt}\caption{\label{fig:TU} Correlation between  $T$ and $U$ in the three regimes studied in the Section~\ref{pheno}:   case 1
      (left panel), case 2 (middle panel) and  case 3 (right panel).  } 
 \end{figure}

Using dimensional regularisation we have:
\begin{eqnarray}
i\,\Pi_\Omega^{\mu\nu}(q^2,m^2)&=&ig^{\mu\nu}\int\frac{d^4k}{(2\pi)^4} \frac{i}{k^2-m^2}\rightarrow ig^{\mu\nu}\mu^{4-d}\int\frac{d^dk}{(2\pi)^d} \frac{i}{k^2-m^2}\nonumber\\
	&=&-ig^{\mu\nu}\frac{m^2}{(4\pi)^2}(\Upsilon+1-\log\frac{m^2}{\mu^2})\;,
	\label{iPi-Omega}\\
i\,\Pi_\Phi^{\mu\nu}(q^2,m_1^2,m_2^2)&=&\int\frac{d^4k}{(2\pi)^4} \frac{(2k+q)^\mu(2k+q)^\nu}{(k^2-m_1^2)[(k+q)^2-m_2^2]}\rightarrow\mu^{4-d}\int\frac{d^dk}{(2\pi)^d} \frac{(2k+q)^\mu(2k+q)^\nu}{(k^2-m_1^2)[(k+q)^2-m_2^2]}\nonumber\\
	&=&\frac{ig^{\mu\nu}}{(4\pi)^2}\bigg[ (m_1^2+m_2^2)(\Upsilon+1) -2f_2(m_1^2,m_2^2)+q^2[-\frac{1}{3}(\Upsilon+1)+2f_1(m_1^2,m_2^2)]\bigg]\nonumber\\
	&&+(q^\mu q^\nu\mbox{~terms})+\mathcal{O}(q^4)\;,\label{iPi-Phi}\\
i\,\Pi_\Psi^{\mu\nu}(q^2,m_S^2,m_G^2)&=&g^{\mu\nu}\int\frac{d^4k}{(2\pi)^4} \frac{i}{(k^2-m_S^2)}\frac{-i}{[(k-q)^2-m_G^2)]}\rightarrow g^{\mu\nu} \mu^{4-d}\int\frac{d^4k}{(2\pi)^4} \frac{i }{(k^2-m_S^2)}\frac{-i}{[(k-q)^2-m_G^2)]}\nonumber\\
	&=&\frac{ig^{\mu\nu} }{(4\pi)^2}\bigg[\Upsilon- f_3 (m_S^2,m_G^2)     \bigg] \;,\label{iPi-Psi}
\end{eqnarray}
where $\mu$ is an arbitrary mass scale parameter and 
$\Upsilon\equiv2/(4-d)-\gamma+\log(4\pi)$. The functions
$f_1(m_1^2,m_2^2)$ and $f_2(m_1^2,m_2^2)$ are as
\begin{eqnarray}
f_1(m_1^2,m_2^2)&\equiv &\int_0^1dx\,x(1-x)\log\bigg[\frac{xm_1^2 +(1-x)(m_2^2-q^2 x)}{\mu^2}\bigg]\;,\label{f1}\\
f_2(m_1^2,m_2^2)&\equiv &\int_0^1dx[xm_1^2+(1-x)m_2^2] \log\bigg[\frac{xm_1^2+(1-x)(m_2^2-q^2 x)}{\mu^2}\bigg]\;,\label{f2}\\
f_3(q^2,m_S^2,m_G^2)&\equiv &\int_0^1dx \log\bigg[\frac{(1-x)(m_S^2- x q^2) + m_G^2 x }{\mu^2}\bigg]\;.\label{f3}
\end{eqnarray}
In particular, if $m_1=m_2=m$ in eqs. \eqref{iPi-Phi} and \eqref{iPi-Psi} we have
\be
\begin{split}
	i\,\Pi_\Phi^{\mu\nu}(q^2,m^2,m^2)&=\frac{ig^{\mu\nu}}{(4\pi)^2}\bigg[2m^2(\Upsilon+1-\log\frac{m^2}{\mu^2})+\frac{q^2}{3}(-\Upsilon-1+\log\frac{m^2}{\mu^2})\bigg]+(q^\mu q^\nu\mbox{~terms})+\mathcal{O}(q^4)\;,\label{iPi-Phi-2}\\
	i\,\Pi_\Psi^{\mu\nu}(q^2,m^2,m^2)&= \frac{ig^{\mu\nu} }{(4\pi)^2}\bigg[\Upsilon- \log\frac{m^2}{\mu^2}    \bigg] \;.
\end{split}
\ee
The parameters $S$ $T$ and $U$ are defined in eq.~({\ref{def-STU}}) and take the explicit form
\begin{eqnarray}
S&=&    \frac{\cos^2(\theta+\alpha)}{72 \pi} 
\left(\frac{ -5 m_H^4+22 m_H^2 m_Z^2-5 m_Z^4}{\left(m_H^2-m_Z^2\right)^2}
+\frac{5 m_h^4-22 m_h^2 m_Z^2+5 m_Z^4}{\left(m_h^2-m_Z^2\right)^2}
\right. \nonumber\\
& &\left. -\frac{6 m_h^4  \left(m_h^2-3 m_Z^2\right) \log \left(\frac{m_h^2}{m_Z^2}\right)}{\left(m_h^2-m_Z^2\right)^3}
 +\frac{6 m_H^4  \left(m_H^2-3 m_Z^2\right) \log \left(\frac{m_H^2}{m_Z^2}\right)}{\left(m_H^2-m_Z^2\right)^3}\right)\nonumber\\
 & &  +\frac{ \sin^2\theta}{72\, \pi} \left(
 -\frac{6 \left(M_\Theta^6-3 M_\Theta^4 M_{\tilde\Pi}^2\right) \log \left(\frac{M_{\tilde\Pi}^2}{M_\Theta^2}\right)}{\left(M_{\Theta}^2-M_{\tilde\Pi}^2\right)^3}
 +\frac{ -5 M_\Theta^4+22 M_\Theta^2 M_{\tilde\Pi}^2-5 M_{\tilde\Pi}^4}{\left(M_\Theta^2-M_{\tilde\Pi}^2\right)^2}\right)\,,
\end{eqnarray}
\begin{eqnarray}
T	&=&  \frac{\cos^2(\theta+\alpha)}{16 \pi} 
 \left(\frac{\log \left(\frac{m_H^2}{m_h^2}\right)}{c_W^2}
 -\frac{\left(4 m_h^2+m_Z^2\right) \log \left(\frac{m_h^2}{m_Z^2}\right)}{c_W^2 s_W^2 \left(m_h^2-m_Z^2\right)}
 +\frac{\left(4 m_H^2+m_Z^2\right) \log \left(\frac{m_H^2}{m_Z^2}\right)}{c_W^2 s_W^2 \left(m_H^2-m_W^2\right)}\right.\nonumber\\
 & & \left.+\frac{\left(4 m_h^2+m_W^2\right) \log \left(\frac{m_h^2}{m_W^2}\right)}{s_W^2 \left(m_Z^2-m_W^2\right)}-\frac{\left(4 m_H^2+m_W^2\right) \log \left(\frac{m_H^2}{m_W^2}\right)}{c_W^2 \left(m_h^2-s_W^2\right)}\right)\,,
\end{eqnarray}
\begin{eqnarray}
U&=&     -\frac{\cos^2(\theta+\alpha)}{12 \pi}\left(
2 \left(m_W^2-m_Z^2\right) \left(\frac{m_h^2 \left(m_h^4-m_W^2 m_Z^2\right)}{\left(m_h^2-m_W^2\right)^2 \left(m_h^2-m_Z^2\right)^2}-\frac{m_H^2 \left(m_H^4-m_W^2 m_Z^2\right)}{\left(m_H^2-m_W^2\right)^2 \left(m_H^2-m_Z^2\right)^2}\right)    \right.\nonumber\\
& &\left.   
+\frac{m_W^4 \left(m_W^2-3 m_h^2\right) \log \left(\frac{m_h^2}{m_W^2}\right)}{\left(m_h^2-m_W^2\right)^3}+\frac{m_Z^4 \left(m_Z^2-3 m_h^2\right) \log \left(\frac{m_h^2}{m_Z^2}\right)}{\left(m_Z^2-m_h^2\right)^3}\right.\nonumber\\
& &\left.    +\frac{m_W^4 \left(m_W^2-3 m_H^2\right) \log \left(\frac{m_H^2}{m_W^2}\right)}{\left(m_W^2-m_H^2\right)^3}+\frac{m_Z^4 \left(m_Z^2-3 m_H^2\right) \log \left(\frac{m_H^2}{m_Z^2}\right)}{\left(m_H^2-m_Z^2\right)^3}\right)\,.
\end{eqnarray}
As we we can read from these expressions the only oblique parameter depending on the masses of $M_{\tilde\Pi}$  and $M_\Theta$ is $S$. 
We show the correlations  $S$ versus $U$  and $T$ versus $U$ in  Figs.~\ref{fig:SU} and \ref{fig:TU}, respectively, for the three regimes studied in 
Section~\ref{pheno}. 

\section{Feynman rules for vacuum polarisation amplitudes}\label{app:Feynrules}
We report here the Feynman rules for the non-standard couplings of the neutral and charged scalar fields with the gauge bosons $W^\pm$ and $Z$.
In particular, we list below the trilinear  and quadrilinear couplings between  the vector bosons (indicated by $V$)  and the neutral and charged scalar components (indicated respectively  by $S^0$ and $S^\pm$) appearing in $M$
in eq. \eqref{eq:MM}. 

 The trilinear couplings between the $W^\pm$ vector bosons and charged/neutral scalars are the following:
\small{
\be\begin{split}
W_\mu^+(-p-q)S^0(p)S^-(q)
	:&\,\, \frac{ i\,e }{2 s_W} W^+_\mu\left[  (\sin\theta \Theta-\cos\theta\tilde\Pi_4 + i \tilde \Pi_3) \pd\tilde \Pi^-    
	- (h \,\sin(\theta+\alpha) +H\, \cos(\theta+\alpha) - i \Pi_3) \pd\Pi^- 
	\right]\\
	\mbox{Feynman Rules} \, \, :  & \hspace{20pt}\bullet \hspace{20pt}W^+_\mu  \Theta  \tilde \Pi^-\rightarrow
	-i \frac{e}{2 s_W}  \sin\theta (p^\mu-q^\mu)\\
	 & \hspace{20pt}\bullet \hspace{20pt} W^+_\mu   \tilde \Pi_4\tilde \Pi^-\rightarrow 
	i \frac{e}{2 s_W}  \cos\theta (p^\mu-q^\mu)\\
	 &\hspace{20pt}\bullet \hspace{20pt} W^+_\mu  \tilde \Pi_3\tilde \Pi^-\rightarrow 
	\frac{e}{2 s_W}   (p^\mu-q^\mu)\\  
	 &\hspace{20pt}\bullet \hspace{20pt} W^+_\mu  h  \Pi^-\rightarrow
	i \frac{e}{2 s_W}  \sin(\theta+\alpha) (p^\mu-q^\mu)\\
	 &\hspace{20pt}\bullet \hspace{20pt} W^+_\mu   H  \Pi^-\rightarrow
	i \frac{e}{2 s_W}  \cos(\theta+\alpha) (p^\mu-q^\mu)\\
	 &\hspace{20pt}\bullet \hspace{20pt} W^+_\mu   \Pi_3 \Pi^-\rightarrow
	\frac{e}{2 s_W}   (p^\mu-q^\mu)\\
	\end{split}\ee
}
The trilinear couplings with the $Z$ boson are given by:
\be\begin{split}	
Z_\mu(-p-q)S^+(p) S^-(q):
	 & -i \frac{e (1-2 s_W^2)}{2 s_W c_W}  Z \left[ \tilde\Pi^+ \pd \tilde \Pi^-  +  \Pi^+ \pd  \Pi^-     \right]  \\
\mbox{Feynman Rules} \, \, :   &  \hspace{20pt}\bullet \hspace{20pt}  Z \, \tilde\Pi^+  \tilde \Pi^-   \rightarrow
	 i \frac{e (1-2 s_W^2)}{2 s_W c_W}  (p^\mu- q^\mu) \\
	 & \hspace{20pt}\bullet \hspace{20pt} Z \, \Pi^+  \Pi^-   \rightarrow
	 i \frac{e (1-2 s_W^2)}{2 s_W c_W}  (p^\mu- q^\mu) \\                                                      
Z_\mu(-p-q)S^0(p) S^{0 \dag}(q):
	 & \frac{e }{2  s_W c_W}  Z \left[   (\sin\theta\Theta  - \cos\theta\tilde\Pi_4)\pd \tilde\Pi_3  +  \Pi_3 \pd \left(\sin(\theta+\alpha) h  +\cos(\theta+\alpha) H\right) \right]\\
\mbox{Feynman Rules} \, \, : & \hspace{20pt}\bullet \hspace{20pt}Z \tilde\Pi_3 \Theta \rightarrow
\frac{e }{2  s_W c_W}  \sin\theta  (p^\mu- q^\mu) \\
	 & \hspace{20pt}\bullet \hspace{20pt}Z \tilde\Pi_3 \tilde\Pi_4 \rightarrow
	  - \frac{e }{2  s_W c_W}  \cos\theta   (p^\mu- q^\mu) \\
	 & \hspace{20pt}\bullet \hspace{20pt}Z \Pi_3 h \rightarrow
	 -  \frac{e }{2  s_W c_W}  \sin(\theta +\alpha) (p^\mu- q^\mu) \\
	 & \hspace{20pt}\bullet \hspace{20pt}Z \Pi_3 H  \rightarrow
	\frac{e }{2  s_W c_W}  \cos(\theta +\alpha)   (p^\mu- q^\mu) \\
\end{split}\ee
The vertices with two gauge bosons can be of two types: \\1) $V V^\prime S$ type, which couplings read:
\small{
\be\begin{split}
W_\mu^-(-p-q)A_{\nu}(p)  S^+(q):           &   \,\, \, e\, m_W   W_\mu^-A_{\nu} \Pi^+     \\
								\mbox{Feynman Rules} \, \, :  	& \hspace{20pt}\bullet \hspace{20pt} W_\mu^-A_{\nu} \Pi^+  \rightarrow i g^{\mu\nu}\,e\, m_W    \\
W_\mu^-(-p-q)Z_{\nu}(p)  S^+(q):             &   \,\, \, -e\, s_W m_Z   W_\mu^-Z_{\nu} \Pi^+     \\
							\mbox{Feynman Rules} \, \, :  	& \hspace{20pt}\bullet \hspace{20pt} W_\mu^-Z_{\nu} \Pi^+  \rightarrow -i g^{\mu\nu}\, \, e\, s_W m_Z       \\
\end{split}\ee
}
2)  $VVS$ type, which are given by the following couplings: 
\small{
\be\begin{split}
W_\mu^-(-p-q)W_{\nu}^+(p)  S^0(q):            &  \,\,\frac{e^2}{2 s_W^2} f\sin\theta\sin(\theta+\alpha) W_\mu^-W_{\nu}^+h     \\
								\mbox{Feynman Rules} \, \, :  	& \hspace{20pt}\bullet \hspace{20pt}   W_\mu^-W_{\nu}^+ h  \rightarrow i g^{\mu\nu}\,  m_W \frac{e}{ s_W} \sin(\theta+\alpha)  \\
W_\mu^-(-p-q)W_{\nu}^+(p) S^0(q):               &  \,\, \frac{e^2}{2 s_W^2} f\sin\theta\cos(\theta+\alpha)  W_\mu^-W_{\nu}^+H   \\
								\mbox{Feynman Rules} \, \, :  	& \hspace{20pt}\bullet \hspace{20pt}  W_\mu^-W_{\nu}^+H \rightarrow i g^{\mu\nu}\,  m_W \frac{e}{ s_W} \cos(\theta+\alpha)  \\
Z_\mu(-p-q)Z_{\nu}(p) S^0(q):                  &  \,\,\frac{e^2}{4 s_W^2 c_W^2} f\sin\theta\sin(\theta+\alpha)  Z_\mu Z_{\nu}h    \\
								\mbox{Feynman Rules} \, \, :  	& \hspace{20pt}\bullet \hspace{20pt}   Z_\mu Z_{\nu}h  \rightarrow i g^{\mu\nu} \,  m_Z \frac{e}{s_W c_W} \sin(\theta+\alpha)  \\
Z_\mu(-p-q)Z_{\nu}(p) S^0(q) :             &  \,\,\frac{e^2}{4 s_W^2 c_W^2} f\sin\theta\cos(\theta+\alpha)  Z_\mu Z_{\nu}H   \\
								\mbox{Feynman Rules} \, \, :  	& \hspace{20pt}\bullet \hspace{20pt}   Z_\mu Z_{\nu} H \rightarrow i g^{\mu\nu} \,  m_Z \frac{e}{s_W c_W} \cos(\theta+\alpha)  \\
\end{split}\ee
}

\newpage
The quadrilinear couplings of type   $V V S S $ for the $W^\pm$ boson read:
\small{
\be\begin{split}
W_\mu^+W_\nu^- S^+ S^-
	:&~\frac{e^2}{2s^2_W} g^{\mu\nu}W_\mu^+W_\nu^- ( \tilde \Pi^+ \tilde \Pi^-+ \Pi^+  \Pi^-) \\
		\\\mbox{Feynman Rules} \, \, :  &\hspace{20pt}\bullet \hspace{20pt} W_\mu^+W_\nu^-  \tilde \Pi^+ \tilde \Pi^-\rightarrow  i   g^{\mu\nu}\frac{e^2}{2s^2_W} , \\
						& \hspace{20pt}\bullet \hspace{20pt} W_\mu^+W_\nu^-   \Pi^+  \Pi^-\rightarrow  i  g^{\mu\nu} \frac{e^2}{2s^2_W}\\
W_\mu^+W_\nu^- S^0 S^{0 \dag}	:&~\frac{e^2}{4s^2_W} g^{\mu\nu}W_\mu^+W_\nu^-\left[  \sin(\alpha+\theta)^2 h^2   + \cos(\theta+\alpha)^2 H^2       + \sin^2\theta\Theta^2   + \cos\theta^2\tilde\Pi_4^2      +\tilde\Pi_3^2  +\Pi_3^2 \right]\\
\mbox{Feynman Rules} \, \, :  	& \hspace{20pt}\bullet \hspace{20pt} W_\mu^+W_\nu^-  h^2     \rightarrow   i g^{\mu\nu}\frac{e^2}{2s_W^2} \sin(\alpha+\theta)^2 \\
				&   \hspace{20pt}\bullet \hspace{20pt}W_\mu^+W_\nu^- H^2     \rightarrow   i g^{\mu\nu}\frac{e^2}{2s^2_W}  \cos(\theta+\alpha)^2 \\							
				&   \hspace{20pt}\bullet \hspace{20pt}W_\mu^+W_\nu^-  \Theta^2      \rightarrow   i g^{\mu\nu}\frac{e^2}{2s^2_W} \sin^2\theta\ \\							
				&   \hspace{20pt}\bullet \hspace{20pt}W_\mu^+W_\nu^- \tilde\Pi_4^2       \rightarrow   i g^{\mu\nu}\frac{e^2}{2s^2_W} \cos\theta^2  \\	
				&  \hspace{20pt}\bullet \hspace{20pt}W_\mu^+W_\nu^-\tilde \Pi_3^2       \rightarrow   i g^{\mu\nu}\frac{e^2}{2s^2_W}  \\			
				&  \hspace{20pt}\bullet \hspace{20pt}W_\mu^+W_\nu^-  \Pi_3^2       \rightarrow   i g^{\mu\nu}\frac{e^2}{2s^2_W}  \\
\end{split}\ee
}

\newpage
Similarly, for the $Z$ boson we have:
\small{
\be\begin{split}																												
Z_\mu Z_\nu S^+ S^- :   &~\frac{e^2 (1-2 s_W^2)^2}{4 s_W^2 c_W^2}  g^{\mu\nu}Z_\mu Z_\nu  (\tilde \Pi^+ \tilde \Pi^- +\Pi^+ \Pi^-)\\
						 \mbox{Feynman Rules} \, \, :    &  \hspace{20pt}\bullet \hspace{20pt}    Z_\mu Z_\nu  \tilde \Pi^+ \tilde \Pi^-\rightarrow~ i g^{\mu\nu} \frac{e^2 (1-2 s_W^2)^2}{ 2 s_W^2 c_W^2}\\
										 & \hspace{20pt}\bullet \hspace{20pt}  	Z_\mu Z_\nu \Pi^+ \Pi^-\rightarrow~ i g^{\mu\nu} \frac{e^2 (1-2 s_W^2)^2}{ 2 s_W^2 c_W^2}  \\
Z_\mu Z_\nu S^0 S^{0 \dag}
						:&~\frac{e^2}{8 s_W^2c_W^2}g^{\mu\nu} Z_\mu Z_\nu \left[  \sin(\alpha+\theta)^2 h^2   + \cos(\theta+\alpha)^2 H^2     + \sin^2\theta\,\Theta^2   + \cos^2\theta\tilde\Pi_4^2     +\tilde\Pi_3^2 +\Pi_3^2\right]\\
					\mbox{Feynman Rules} \, \, :  	& \hspace{20pt}\bullet \hspace{20pt}  Z_\mu Z_\nu h^2 \rightarrow~  i g^{\mu\nu}\frac{e^2}{2s^2_W c^2_W}\sin(\alpha+\theta)^2    \\
						&  \hspace{20pt}\bullet \hspace{20pt}  Z_\mu Z_\nu H^2 \rightarrow~  i g^{\mu\nu}\frac{e^2}{2s^2_W c^2_W} \cos(\alpha+\theta)^2    \\
						& \hspace{20pt}\bullet \hspace{20pt}  Z_\mu Z_\nu \Theta^2 \rightarrow~  i g^{\mu\nu}\frac{e^2}{2s^2_W c^2_W} \sin^2\theta    \\
						& \hspace{20pt}\bullet \hspace{20pt}  Z_\mu Z_\nu \tilde\Pi_4^2 \rightarrow~  i g^{\mu\nu}\frac{e^2}{2s^2_W c^2_W}\cos\theta^2    \\
						& \hspace{20pt}\bullet \hspace{20pt}  Z_\mu Z_\nu \tilde\Pi_3^2 \rightarrow~  i g^{\mu\nu}\frac{e^2}{2s^2_W c^2_W}    \\
						& \hspace{20pt}\bullet \hspace{20pt}  Z_\mu Z_\nu  \Pi_3^2 \rightarrow~  i g^{\mu\nu}\frac{e^2}{2s^2_W c^2_W}    \\
\end{split}\ee
}

\end{document}